\documentclass[a4paper,11pt]{article}
\pdfoutput=1 

\usepackage{jheppub} 
                     
\usepackage{amsmath,amssymb,amsthm,amscd,graphicx}
\usepackage{psfrag}
\input epsf.sty
\newtheorem{theorem}{Theorem}[section]

\theoremstyle{definition}

\newtheorem{example}[theorem]{Example}

\newcommand{\CF}{{\cal F}}

\newcommand{\CJ}{{\cal J}}

\newcommand{\CM}{{\cal M}}
\newcommand{\CN}{{\cal N}}
\newcommand{\CO}{{\cal O}}
\newcommand{\CP}{{\cal P}}

\newcommand{\CR}{{\cal R}}

\newcommand{\CW}{{\cal W}}

\def\IZ{{\mathbb Z}}

\def\IR{{\mathbb R}}
\def\IC{{\mathbb C}}
\def\IP{{\mathbb P}}
\def\IT{{\mathbb T}}

\newcommand{\tr}{{\rm Tr}}
\newcommand{\re}{{\rm e}}
\newcommand{\ri}{{\rm i}}
\newcommand{\rd}{{\rm d}}
\newcommand{\mx}{{\mathsf{x}}}
\newcommand{\my}{{\mathsf{y}}}
\newcommand{\mb}{{\mathsf{b}}}
\newcommand{\map}{{\mathsf{p}}}
\newcommand{\mq}{{\mathsf{q}}}
\newcommand{\im}{{\mathsf{i}}}
\newcommand{\mA}{{\mathsf{A}}}

\newcommand{\mO}{{\mathsf{O}}}
\newcommand{\mP}{{\mathsf{P}}}
\newcommand{\mV}{{\mathsf{V}}}
\newcommand{\mJ}{{\mathsf{J}}}

\newcommand{\fad}{\operatorname{\Phi}_{\mathsf{b}}}

\newcommand{\mypsi}[2]{\operatorname{\Psi}_{#1,#2}}

\newcommand{\be}{\begin{equation}}
\newcommand{\ee}{\end{equation}}
\newcommand{\ba}{\begin{aligned}}
\newcommand{\ea}{\end{aligned}}
\newcommand{\ben}{\begin{eqnarray}\displaystyle}
\newcommand{\een}{\end{eqnarray}}

\newcommand{\sectiono}[1]{\section{#1}\setcounter{equation}{0}}

\newdimen\tableauside\tableauside=1.0ex
\newdimen\tableaurule\tableaurule=0.4pt
\newdimen\tableaustep
\def\phantomhrule#1{\hbox{\vbox to0pt{\hrule height\tableaurule width#1\vss}}}
\def\phantomvrule#1{\vbox{\hbox to0pt{\vrule width\tableaurule height#1\hss}}}
\def\sqr{\vbox{%
  \phantomhrule\tableaustep
  \hbox{\phantomvrule\tableaustep\kern\tableaustep\phantomvrule\tableaustep}%
  \hbox{\vbox{\phantomhrule\tableauside}\kern-\tableaurule}}}
\def\squares#1{\hbox{\count0=#1\noindent\loop\sqr
  \advance\count0 by-1 \ifnum\count0>0\repeat}}
\def\tableau#1{\vcenter{\offinterlineskip
  \tableaustep=\tableauside\advance\tableaustep by-\tableaurule
  \kern\normallineskip\hbox
    {\kern\normallineskip\vbox
      {\gettableau#1 0 }%
     \kern\normallineskip\kern\tableaurule}%
  \kern\normallineskip\kern\tableaurule}}
\def\gettableau#1{\ifnum#1=0\let\next=\null\else
\squares{#1}\let\next=\gettableau\fi\next}

\newcommand{\figref}[1]{Fig.~\protect\ref{#1}}

\title{\boldmath Spectral Theory and Mirror curves of Higher Genus}

\author{Santiago Codesido,}
\author{Alba Grassi}
\author{and Marcos Mari\~no}

\affiliation{D\'epartement de Physique Th\'eorique et Section de Math\'ematiques,\\
Universit\'e de Gen\`eve, Gen\`eve, CH-1211 Switzerland}

\emailAdd{santiago.codesido@unige.ch, alba.grassi@unige.ch, marcos.marino@unige.ch}

\abstract{Recently, a correspondence has been proposed between spectral theory and topological strings on toric Calabi--Yau manifolds. In this paper we develop in detail this 
correspondence for mirror curves of higher genus, which display many new features as compared to the genus one case studied so far. Given a curve of genus $g$, 
our quantization scheme leads to $g$ different trace class operators. Their spectral properties are encoded in a generalized spectral determinant, 
which is an entire function on the Calabi--Yau moduli space. We conjecture 
an exact expression for this spectral determinant in terms of the standard and refined topological string amplitudes. 
This conjecture provides a non-perturbative definition of the topological string on these geometries, in which the genus expansion emerges in a 
suitable 't Hooft limit of the spectral traces of the operators. In contrast to what happens 
in quantum integrable systems, our quantization scheme leads to a single quantization condition, which is 
elegantly encoded by the vanishing of a quantum-deformed theta function on the mirror curve. We illustrate our 
general theory by analyzing in detail the resolved $\IC^3/\IZ_5$ orbifold, which is the simplest toric Calabi--Yau manifold with a genus two mirror curve. By applying our conjecture to this example, we find 
new quantization conditions for quantum mechanical operators, in terms of genus two theta functions, as well as new number-theoretic properties for the periods of this Calabi--Yau.}

\begin{document}
\maketitle
\flushbottom

\sectiono{Introduction}

It has been conjectured in \cite{ghm} that there is a precise correspondence between the spectral theory of certain operators and local mirror symmetry. 
This correspondence postulates that the Weyl quantization of mirror curves to toric Calabi--Yau (CY) threefolds leads to trace class operators on $L^2(\IR)$, and that the 
spectral determinant of these operators is captured by topological string amplitudes on the underlying CY. As a corollary, one 
finds an exact quantization condition for their spectrum, in terms of the 
vanishing of a (deformed) theta function. The correspondence unveiled in \cite{ghm} builds upon previous work on the quantization of mirror curves \cite{adkmv,acdkv} and on the relation between 
supersymmetric gauge theories and quantum integrable systems \cite{ns}. It incorporates in addition 
key ingredients from the study of the ABJM matrix model at large $N$ \cite{dmp,mp,hmo,hmo2,hmmo,km}. These ingredients 
are necessary for a fully non-perturbative treatment, beyond the perturbative WKB approach of \cite{acdkv} and of other recent works on the quantization of spectral curves.

The correspondence of \cite{ghm} between spectral theory and topological strings can be used to give a non-perturbative definition of the standard topological string. 
The (un-refined) topological string amplitudes appear as quantum-mechanical instanton corrections to the spectral problem, and due to their peculiar form, 
they can be singled out by a 't Hooft-like limit of the so-called fermionic spectral traces of the operator. In addition, by using the integral kernel of the operator, which was determined 
explicitly in \cite{kas-mar} in many cases, one can write down a matrix model whose $1/N$ expansion gives exactly the genus expansion of the topological string \cite{mz,kmz}. Therefore, 
one can regard the correspondence of \cite{ghm} as a large $N$ Quantum Mechanics/topological string correspondence, with many features of large $N$ gauge/string dualities. 
In particular, it is a strong/weak duality, since the Planck constant in the quantum-mechanical problem, $\hbar$, is identified as the {\it inverse} string coupling constant. 

All the examples of the correspondence that have been studied so far involve local del Pezzo CYs, and their mirror curve has genus one \cite{ghm,mz,kmz,gkmr}.
 It was pointed out in \cite{ghm} that the relationship between the spectral theory of trace class operators and topological string amplitudes 
 should hold for general toric CYs, i.e. it should hold for mirror curves of arbitrary genus. In this paper we present a compelling picture for the 
 spectral theory/mirror symmetry correspondence in the higher genus case. This generalization involves some new ingredients. 
 In the theory developed in \cite{ghm} for the genus one case, the basic object is the spectral determinant of the trace class 
operator obtained by quantization of the mirror curve. It turns out that a curve of genus $g_\Sigma$ leads to $g_\Sigma$ different operators, which are related by 
explicit transformations\footnote{In this paper, we will denote by $g_\Sigma$ the genus of the mirror curve, which should not be confused with the genus $g$ appearing in the 
genus expansion. The former is a spacetime genus, while the latter is a worldsheet genus.}. As we show in this paper, there is nevertheless a single, 
generalized spectral determinant, which is an {\it entire} function on the moduli space of the CY manifold. The spectra of the different operators associated to a 
higher genus mirror curve are encoded in a {\it single} quantization condition, 
which is given, as in \cite{ghm}, by the vanishing of the generalized spectral determinant. This quantization condition can be formulated in an elegant way as the vanishing of a 
quantum-deformed Riemann theta function on the mirror curve; it determines a family of codimension one submanifolds in moduli space.  

The fact that we obtain a single quantization condition from a curve of genus $g_\Sigma$ might be counter-intutitive to readers familiar with quantum integrable systems, 
like for example the quantum Toda chain and its generalizations. In those systems, the quantization of the spectral curve leads to $g_\Sigma$ quantization conditions. This is of course due to the fact that the 
underlying quantum-mechanical system is $g_\Sigma$-dimensional, and there are $g_\Sigma$ commuting Hamiltonians that 
can (and should) be diagonalized simultaneously. It should be noted, however, that the spectral curve by itself does not carry this additional information. In fact, in the case of 
quantum mechanical problems on the real line it is quite common that the quantization of a higher genus curve leads to a single quantization 
condition. This is what happens, for example, for the Schr\"odinger equation with a confining, polynomial potential of higher degree.

As we have just mentioned, one of the main consequences of the conjecture of \cite{ghm} is that it provides a non-perturbative definition of topological string theory. This can be 
also generalized to the higher genus case: as we show in this paper, the generalized spectral determinant 
leads to fermionic spectral traces $Z({\boldsymbol{N}}, \hbar)$, depending on $g_\Sigma$ non-negative integers ${\boldsymbol{N}}=(N_1, \cdots, N_{g_\Sigma})$. In the 't Hooft limit
\be
\label{thooft}
 \hbar \rightarrow \infty, \qquad N_i \rightarrow \infty, \qquad {N_i \over \hbar} = \lambda_i \, \, \,\, \,  \,\, \, \text{fixed}, \quad i=1, \cdots, g_\Sigma, 
 \ee
these traces have the asymptotic expansion 
\be
\label{logz-exp}
\log \, Z({\boldsymbol{N}}, \hbar) =\sum_{g\ge 0} \CF_g({\boldsymbol{\lambda}}) \hbar^{2-2g}, 
\ee
where $\CF_g({\boldsymbol{\lambda}})$ are the genus $g$ free energies of the topological string, in an appropriate 
conifold frame. In particular, we can regard these fermionic spectral traces, which are completely well defined objects, 
as non-perturbative completions of the topological string partition function. 

The theory of quantum mirror curves of higher genus is relatively intricate, and we develop it in full 
detail for what is probably the simplest genus two mirror curve, namely, the total resolution of the 
$\IC^3/\IZ_5$ orbifold. We perform a detailed study of the associated spectral theory, and in particular 
we determine the vanishing locus of the spectral determinant on the two-dimensional moduli space, 
in the so-called maximally supersymmetric case $\hbar=2\pi$. 
In addition, we give compelling evidence that the 
expansion of the topological string free energies near what we call the maximal conifold locus gives the large $N$ expansion of the fermionic spectral traces. 
This provides a non-perturbative completion of the topological string on this background. As a bonus, we obtain non-trivial identities for 
the values of the periods of this CY at the maximal conifold locus in terms of the dilogarithm, in the spirit of \cite{rv,dk}. 

The organization of this paper is the following. In section 2, we develop the theory of 
quantum operators associated to higher genus mirror curves and we construct the 
appropriate generalization of the spectral determinant. In section 3 we present an explicit, 
conjectural expression for the spectral determinant in terms of topological string amplitudes, and we explain 
how the large $N$ limit of the spectral traces provides a non-perturbative definition of the all-genus topological string free energy. 
In section 4, we test these ideas in detail in the example of the 
resolved $\IC^3/\IZ_5$ orbifold. In section 5, we conclude and present some problems for future research. The Appendix summarizes information about the special geometry of the resolved $\IC^3/\IZ_5$ orbifold which is 
needed in section 4.

\sectiono{Quantizing mirror curves of higher genus}

\subsection{Mirror curves}

In this paper we will consider mirror curves to toric CY threefolds, and we will promote them to quantum operators. Let us first review some 
well-known facts about local mirror symmetry \cite{kkv,ckyz} and the corresponding algebraic curves. 
The toric CY threefolds which we are interested in can be 
described as symplectic quotients, 
\be
X=\mathbb{C}^{k+3}//G,
\ee
where $G=U(1)^k$. The quotient is specified by a matrix of charges $Q_i^\alpha$, $i=0, \cdots, k+2$, $\alpha=1, \cdots, k$. The CY condition 
requires the charges to satisfy \cite{witten-phases}
\begin{equation}  
\sum_{i=0}^{k+2} Q_i^\alpha=0, \qquad \alpha = 1, \ldots, k .
\label{anomaly}
\end{equation}  
The mirrors to these toric CYs were constructed in \cite{kkv, Bat,hv}. They can be written in terms of
$3+k$ complex coordinates $Y^i\in \IC^*$, $i=0, \cdots, k+2$, which satisfy the constraint
\be
\label{dterm-Y}
\sum_{i=0}^{k+2} Q_i^\alpha Y^i=0,  \qquad \alpha = 1, \ldots, k.  
\ee
The mirror CY manifold $\widehat X$ is given by 
\be
w^+w^-= W_X ,
\ee
where
\be
\label{wx}
W_X= \sum_{i=0}^{k+2} x_i \re^{Y_i}. 
\ee
The complex parameters $x_i$, $i=0, \cdots, k+2$, give a redundant parametrization of the moduli space, 
and some of them can be set to one. Equivalently, 
we can consider instead the coordinates
\be
\label{z-moduli}
z_\alpha=\prod_{i=0}^{k+2} x_i^{Q_i^\alpha}, \qquad  \alpha = 1, \ldots, k.
\ee
The constraints (\ref{dterm-Y}) have a three-dimensional family of solutions. One of the parameters corresponds to a translation of all the coordinates:
\be
\label{trans}
Y^i \rightarrow Y^i+c, \qquad i=0, \cdots, k+2, 
\ee
which can be used for example to set one of the $Y^i$s to zero. The remaining coordinates can be expressed in terms of two variables which we will denote by 
$x$, $y$. There is still a group of symmetries left, given by transformations of the form \cite{akv}, 
\be
\label{can-t}
\begin{pmatrix} x \\ y \end{pmatrix}\rightarrow G \begin{pmatrix} x \\ y \end{pmatrix}, \qquad G\in {\rm SL}(2, \IZ). 
\ee
After solving for the variables $Y^i$ in terms of $x$, $y$, one finds a function 
\be
\label{wxxy}
W_X (\re^x, \re^y), 
\ee
which, due to the translation invariance (\ref{trans}) and the symmetry (\ref{can-t}), is only well-defined up to an overall factor of the form $\re^{ \lambda x + \mu y}$, 
$\lambda, \mu \in \IZ$, and a transformation of the form (\ref{can-t}). The equation
\be
\label{riemann}
W_X (\re^x, \re^y)=0  
\ee
defines a Riemann surface $\Sigma$ embedded in $\IC^* \times \IC^*$. 
We will call (\ref{riemann}) the {\it mirror curve} to the toric CY threefold $X$. All the information about the closed string amplitudes 
on $\widehat X$ is encoded in $\Sigma$, as shown in \cite{mmopen,bkmp,eo-proof}. 

The equation of the mirror curve (\ref{riemann}) can be written down in detail, as follows. Given the matrix of charges $Q^\alpha_i$, we introduce the vectors, 
\be
\overline \nu^{(i)}=\left(1, \nu^{(i)}_1, \nu^{(i)}_2\right),\qquad i=0, \cdots, k+2,
\ee
satisfying the relations
\be
\sum_{i=0}^{k+2} Q^\alpha_i \overline \nu^{(i)}=0. 
\ee
In terms of these vectors, the function (\ref{wxxy}) can be written as
\be
\label{coxp}
W_X(\re^x, \re^y)=\sum_{i=0}^{k+2}x_i  \exp\left( \nu^{(i)}_1 x+  \nu^{(i)}_2 y\right). 
 \ee
Clearly, there are many sets of vectors satisfying these constraints, but they differ in reparametrizations 
and overall factors (as we explained above), 
and therefore they define the same Riemann surface. 
The genus of this Riemann surface, $g_\Sigma$, depends on the toric data, encoded in the matrix of charges, or 
equivalently in the vectors $\overline \nu_i$. Among the parameters 
(\ref{z-moduli}), there will be $g_\Sigma$ ``true" moduli of the geometry, and in addition there will be $r_\Sigma$ ``mass parameters", which lead typically to rational mirror maps 
(this distinction has been emphasized in \cite{hkp,kpsw}.)

\subsection{Quantization} 

The quantization of mirror curves studied in \cite{ghm}, building on \cite{adkmv,acdkv}, is simply based on 
Weyl quantization of the function (\ref{wxxy}), i.e. the variables $x$, $y$ are promoted to 
Heisenberg operators $\mx$, $\my$ satisfying
\be
[\mx, \my]= \im \hbar. 
\ee
In the genus one case, when the CY is the canonical bundle over a del Pezzo surface $S$, 
\be
\label{delpezzo}
X= \CO(K_S) \rightarrow S, 
\ee
the function (\ref{wxxy}) can be written in a canonical form, as 
\be
W_S(\re^x, \re^y)= \CO_S(x,y) + u, 
\ee
where $u$ is the modulus of the Riemann surface. The quantum operator associated to the toric CY threefold, $\mO_S$, is obtained by 
Weyl quantization of the function $\CO_S(x,y)$, and $u$ plays the r\^ole of (minus) the exponentiated energy, or the fugacity. 

The higher genus case is much richer, due to the fact that there are $g_\Sigma$ different moduli for the curve. As a consequence, there 
will be $g_\Sigma$ different ``canonical" forms for the curve, which we will write as
\be
\CO_i (x,y) + \kappa_i=0, \qquad i=1, \cdots, g_\Sigma. 
\ee
Here, $\kappa_i$ is a modulus of $\Sigma$, and in practice it is one of the $x_j$s appearing in (\ref{coxp}). 
Of course, the different canonical forms of the curves are related by reparametrizations and overall factors, so we will write 
\be
\label{o-rel}
\CO_i +\kappa_i  = \CP_{ij} \left( \CO_j +\kappa_j\right), \qquad i,j=1, \cdots, g_\Sigma, 
\ee
where $\CP_{ij}$ is a monomial of the form $\re^{ \lambda x + \mu y}$. Equivalently, we can write 
\be
\label{oi-exp}
\CO_i = \CO_i^{(0)}+ \sum_{j \not=i} \kappa_j \CP_{ij}. 
\ee
We can now perform a standard Weyl quantization of the operators $\CO_i(x,y)$. In this way we obtain 
$g_\Sigma$ different operators, which we will denote by $\mO_i$, $i=1, \cdots, g_\Sigma$. These operators are Hermitian. 
The relation (\ref{o-rel}) becomes, 
\be
\label{op-rel}
\mO_i+ \kappa_i =\mP_{ij}^{1/2} \left( \mO_j  + \kappa_j \right) \mP^{1/2}_{ij}, \qquad i,j=1, \cdots, g_\Sigma, 
\ee
where $\mP_{ij}$ is the operator corresponding to the monomial $\CP_{ij}$. In this relation, the ``splitting" of $\mP_{ij}$ in two square roots is due to the fact that we are 
using Weyl quantization, which leads to Hermitian operators. The expression 
(\ref{oi-exp}) becomes, after promoting both sides to operators, 
\be
\mO_i=\mO_i^{(0)}+ \sum_{j \not=i} \kappa_j \mP_{ij}. 
\ee
We can regard the operator $\mO_i^{(0)}$ as an ``unperturbed" operator, while the moduli $\kappa_j$ encode different 
perturbations of it. We will also need, 
\be
\label{unp-inv}
\rho_i^{(0)}=\left( \mO_i^{(0)}\right)^{-1}, \qquad i=1, \cdots, g_\Sigma.
\ee
By comparing the coefficients of $\kappa_j$ in the relation 
(\ref{op-rel}), we find
\be
\mP_{ij}= \mP_{ji}^{-1}
\ee
and
\be
\label{p-rels}
\mP_{ik}=\mP_{ij}^{1/2} \mP_{jk} \mP_{ij}^{1/2}, \qquad i\not= k. 
\ee
Amusingly, these relationships are a sort of non-commutative version of the the relations between transition functions in the theory of bundles. We will set, by convention, 
\be
\mP_{ii}=1, \quad i=1, \cdots, g_\Sigma. 
\ee
We also have
\be
\label{ozero-rels}
\mO_i^{(0)}= \mP_{ij}^{1/2} \mO_j^{(0)}\mP_{ij}^{1/2}. 
\ee
Before proceeding, let us examine some examples to illustrate the considerations above.

 \begin{center}
 \begin{figure} \begin{center}
 {\includegraphics[scale=0.33]{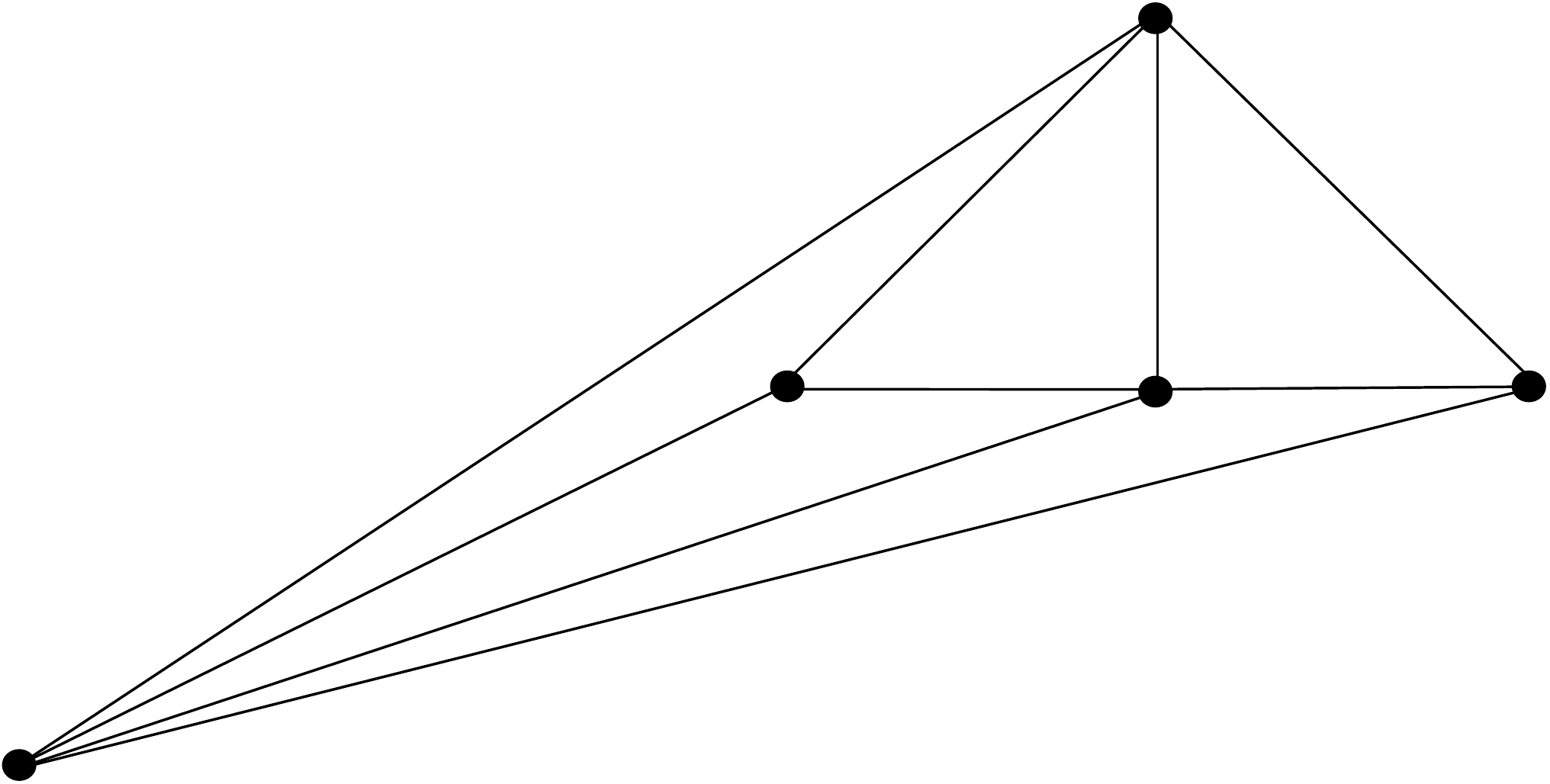}}
\caption{A height one slice of the vectors (\ref{fan-2}), providing the toric data for the resolved $\IC^3/\IZ_5$ orbifold.}
 \label{fanfan}
  \end{center}
\end{figure}  
\end{center}
\begin{example} \label{5-orbifold} {\it The resolved $\IC^3/\IZ_5$ orbifold}. Let us consider the 
CY given by the total resolution of the orbifold $\IC^3/\IZ_5$, where the action has weights $(3,1,1)$. 
This geometry has been studied in detail in various references, like for example \cite{xenia,mr,karp, coates}, 
and (refined) topological string amplitudes on this background have been 
recently calculated in \cite{kpsw}. The vectors of charges are given by 
\be
\label{5o-cv}
Q^1= (-3, 1,1,1,0), \quad Q^2= ( 1,0,0,-2,1). 
\ee
To parametrize the moduli space, we introduce five variables $x_0, \cdots, x_5$, as well as the combinations
\be
\label{z1-z2}
\ba
z_1&=\frac{x_1 x_2 x_3}{x_0^3},\\
z_2&=\frac{x_0 x_4}{x_3^2}.
\ea
\ee
A useful choice of vectors for this example is 
\be
\ba
\overline \nu^{(0)}&= (1,0,0), \\
\overline \nu^{(1)}&= (1,1,0), \\
\overline \nu^{(2)}&= (1,0,1), \\
\overline \nu^{(3)}&= (1,-1,-1), \\
\overline \nu^{(4)}&= (1,-2,-2), \\
\ea
\ee
and the equation for the Riemann surface reads, after setting $x_1=x_2=x_4=1$, 
\be
\label{mc-2}
\re^x+ \re^y + \re^{-2 x -2 y}+ x_3 \re^{-x-y}+ x_0=0. 
\ee
However, it is easy to see that one can also choose the vectors
\be
\label{fan-2}
\ba
\overline \nu^{(0)}&= (1,-1,0), \\
\overline \nu^{(1)}&= (1,0,1), \\
\overline \nu^{(2)}&= (1,-3,-1), \\
\overline \nu^{(3)}&= (1,0,0), \\
\overline \nu^{(4)}&= (1,1,0), \\
\ea
\ee
which leads to the equation
\be
\label{mc-1}
\re^x+ \re^y + \re^{-3 x -y}+ x_0 \re^{-x}+ x_3=0. 
\ee
In \figref{fanfan} we show the vectors $\nu_i$ for the system (\ref{fan-2}) (this is sometimes called a height one slice of the fan (\ref{fan-2})). 
Of course, although we have chosen the same notations, the variables $x$, $y$ appearing in (\ref{mc-1}) are not the same ones appearing in (\ref{mc-2}). Rather, 
they are related by a canonical transformation, 
\be
-x-y \rightarrow x, \qquad 2x+y\rightarrow y. 
\ee
In this case, the two canonical functions $\CO_1(x,y)$ and $\CO_2(x,y)$ are given by 
\be
\label{o12-ex}
\ba
\CO_1(x,y)&=\re^x+ \re^y + \re^{-2 x -2 y}+ x_3 \re^{-x-y}, \\
\CO_2(x,y)&= \re^x+ \re^y + \re^{-3 x -y}+ x_0 \re^{-x}, \\
\ea
\ee
and the moduli are 
\be
\kappa_1= x_0, \qquad \kappa_2=x_3. 
\ee
In the coordinates appropriate for $\CO_1(x,y)$, we have $\CP_{12}= \re^{-x-y}$, while in the coordinates appropriate for $\CO_2(x,y)$, we have $\CP_{21}=\re^{-x}$. 
In terms of the three-term operators introduced in \cite{kas-mar}, 
\be
\label{ttops}
\mO_{m,n}=\re^\mx +\re^{\my} + \re^{-m \mx - n \my},
\ee
the unperturbed operators are 
\be
\mO_1^{(0)}=\mO_{2,2}, \qquad  \mO_2^{(0)}=\mO_{3,1}. 
\ee
The theory of the operators (\ref{ttops}) has been developed in some detail in \cite{kas-mar}, and it will be quite useful to test some of our results later on. 
\qed
\end{example}

\begin{example} \label{a2-ex} {\it The resolved $\IC^3/\IZ_6$ orbifold, or $A_2$ geometry}. 
Let us now consider the total resolution of the orbifold $\IC^3/\IZ_6$, where the action has weights $(4,1,1)$. 
This is precisely the $A_2$ geometry studied in the first papers on local mirror symmetry \cite{kkv,ckyz}, 
which engineers geometrically $SU(3)$ Seiberg--Witten theory. It has also 
been studied in some detail in \cite{kpsw}. In this case, the charge vectors are 
\be
Q^1=(-2,1,0,0,1,0), \qquad Q^2= (1,-2,1,0,0,0), \qquad Q^3=(0,0,0,1,-2,1). 
\ee
Like before, we can parametrize the moduli space with six coordinates $x_i$, $i=0, \cdots, 5$, or in terms of 
\be
z_1={x_1 x_4 \over x_0^2}, \qquad z_2={x_0  x_2 \over x_1^2}, \qquad z_3= {x_3 x_5 \over x_4^2}. 
\ee
The coordinates $z_1$, $z_2$ are true moduli of the curve, while $z_3$ is rather a mass parameter \cite{kpsw}. A useful choice of vectors is, 
\be
\ba
\overline \nu^{(0)}&= (1,-1,0), \\
\overline \nu^{(1)}&= (1,0,0), \\
\overline \nu^{(2)}&= (1,1,0), \\
\overline \nu^{(3)}&= (1,0,1), \\
\overline \nu^{(4)}&= (1,-2,0), \\
\overline \nu^{(5)}&= (1,-4,-1), 
\ea
\ee
and after setting $x_2=x_3=x_5=1$, we find the curve
\be
\re^x+ \re^y + \re^{-4x-y} + x_4 \re^{-2x} + x_0\re^{-x} + x_1=0. 
\ee
It is easy to see that there is another realization of this curve as 
\be
\re^{2x}+\re^y + \re^{-y -2x} + x_4 \re^{-x} + x_1\re^x+ x_0=0. 
\ee
Here, we can regard $x_4$ as a parameter, and $x_0$, $x_1$ as the moduli. The canonical operators derived from this geometry are then given by 
\be
\label{ex2-ops}
\ba
\CO_1(x,y)&=\re^x+ \re^y + \re^{-4x-y} + x_4 \re^{-2x} + x_0\re^{-x}, \\
\CO_2(x,y)&=\re^{2x}+\re^y + \re^{-y -2x} + x_4 \re^{-x} + x_1\re^x.
\ea
\ee
They can be regarded as perturbations of $\mO_{4,1}$, and of $\mO_{1,1}$, respectively.  
\qed
\end{example}

It was noted in \cite{ghm}, in the genus one case, that the most interesting operator was not really $\mO_S$, but rather its inverse 
$\rho_S$. The reason is that $\rho_S$ is expected to be of trace class and 
positive-definite, therefore it has a discrete, positive spectrum, and its Fredholm (or spectral) determinant is well-defined. 
It was rigorously proved in \cite{kas-mar} that, in many cases, this is the case, provided the parameters appearing in the operators satisfy certain positivity 
conditions. In analogy with the genus one case, we expect the operators 
\be
\label{Xops}
\rho_i =\mO_i^{-1}, \qquad i=1, \cdots, g_\Sigma
\ee
to exist, be of trace class and positive-definite. In the concrete examples that we have considered, this actually follows from 
the results in \cite{kas-mar}. In that paper, it was shown that 
\be
\label{rhomn}
\rho_{m,n}=\mO_{m,n}^{-1}
\ee
 exists and are of trace class. It was also shown that the inverse of 
\be
\mO_{m,n} + \mV,
\ee
where $\mV$ is positive and self-adjoint, is also of trace class. Clearly, the operators obtained by Weyl quantization of (\ref{o12-ex}) and (\ref{ex2-ops}) are of this type. 

\subsection{The generalized spectral determinant}
\label{gsd-sec}
According to the conjecture of \cite{ghm}, when the mirror curve has genus one, many important 
aspects of the spectral theory of $\rho_X$ can be encoded in the topological string 
amplitudes on $X$. We would like to generalize this to mirror curves of higher genus. What are the natural 
questions that we would like to answer from the point of view of spectral theory? 
Clearly, we would like to know the spectrum of the operators $\mO_i$ in terms 
of enumerative data of $X$, and in addition, as in \cite{ghm}, we would like to have precise formulae for the spectral determinants of their inverses $\rho_i$. 
However, one should note that, due to (\ref{op-rel}), the operators $\rho_i$ are closely 
related, and their spectra and spectral determinants are not independent. 

In a more fundamental sense, we need an appropriate multivariable generalization of the spectral determinant. In the 
genus one case, when $X$ is a local del Pezzo of the form (\ref{delpezzo}), there is one single modulus $\kappa$, 
and the spectral determinant  
\be
\label{det-prod}
\Xi_S(\kappa, \hbar) = {\rm det}\left(1+ \kappa \rho_S\right)
\ee
can be defined in at least three equivalent 
ways (see \cite{simon,simon-paper} for a detailed 
discussion of this issue). The first one is as an infinite product, 
\be
\Xi_S(\kappa, \hbar) = \prod_{n\ge 0} \left(1+ \kappa \re^{-E_n} \right), 
\ee
where we denoted the eigenvalues of the positive definite, trace class operator $\rho_S$ by $\re^{-E_n}$, $n=0, 1, \cdots$. A 
more useful definition, advocated by Grothendieck \cite{gro} and Simon \cite{simon,simon-paper}, involves the 
{\it fermionic spectral traces} $Z_S(N, \hbar)$, defined as 
\be
Z_S(N, \hbar) = \tr \left(\Lambda^N(\rho_S)\right),  \qquad N=1,2, \cdots
\ee
In this expression, the operator $\Lambda^N(\rho_S)$ is defined by $\rho_S^{\otimes N}$ acting on $\Lambda^N\left(L^2(\IR)\right)$. A theorem of 
Fredholm \cite{fredholm} asserts that, if $\rho_S(x_i, x_j)$ is 
the kernel of $\rho_S$, the fermionic spectral trace can be computed as a multi-dimensional integral, 
\be
\label{fred-th}
Z_S(N, \hbar) = {1 \over N!}  \int {\rm det}\left( \rho_S (x_i, x_j) \right) \,   \rd ^N x.
\ee
The spectral determinant is then given by the convergent series, 
\be
\Xi_S(\kappa, \hbar)=1+\sum_{N=1}^\infty Z_S(N, \hbar) \kappa^N.  
\ee
Another definition of the Fredholm determinant is based on the Fredholm--Plemelj's formula,  
\be
\label{f-p}
 \Xi_S (\kappa, \hbar)= \exp \left\{ -\sum_{\ell=1}^\infty {(-\kappa)^\ell \over \ell} \tr \rho_S^\ell \right\}.  
\ee
In the higher genus case, there should exist a generalization of the 
spectral determinant (\ref{det-prod}), depending on all the moduli $\kappa_1, \cdots, \kappa_{g_\Sigma}$. We 
also expect to have spectral traces depending on various integers $N_i$, $i=1, \cdots, g_\Sigma$. One motivation for this comes from the connection between 
fermionic spectral traces and matrix models 
developed in \cite{mz,kmz}: in the higher genus case, we expect to have a multi-cut matrix model, 
and there should be as many cuts as true moduli in the model. 

In order to construct this generalization, we consider the following operators,  
\be
\label{ajl}
\mA_{jl}= \rho_j^{(0)} \mP_{jl}, \quad j, l=1, \cdots, g_{\Sigma}.  
\ee
The operators $\rho_j^{(0)}$ were defined in (\ref{unp-inv}), while the operators $\mP_{jl}$ are defined by (\ref{op-rel}). 
We will assume that the $\mA_{jl}$ are of trace class (this can be verified in concrete examples). 
We now define the {\it generalized spectral determinant} as
\be
\label{gsd}
\Xi_X ( {\boldsymbol \kappa}; \hbar)= {\rm det} \left( 1+\kappa_1 \mA_{j1} +\cdots+ \kappa_{g_\Sigma} \mA_{j g_\Sigma}  \right). 
\ee
This definition does not depend on the index $j$: from the relationships (\ref{ozero-rels}) and (\ref{p-rels}), we find
\be
\mA_{il}= \mP_{ij}^{-1/2} \mA_{jl} \mP_{ij}^{1/2}. 
\ee
Different choices of the index lead to operators related by a similarity transformation, and their determinants are equal. The generalized spectral determinant (\ref{gsd}) can 
be of course regarded as the conventional spectral determinant of the operator
\be
\kappa_1 \mA_{j1} +\cdots+ \kappa_{g_\Sigma} \mA_{j g_\Sigma}.
\ee
As shown in \cite{simon-paper}, if the operators $\mA_{jl}$ are of trace class, as we are assuming here, (\ref{gsd}) is 
{\it an entire function on the moduli space parametrized by $\kappa_1, \cdots, \kappa_{g_{\Sigma}}$}. This function can be expanded 
around the origin ${\boldsymbol \kappa}=0$, as follows, 
\be
\label{or-exp}
\Xi_X ({\boldsymbol \kappa};\hbar)= \sum_{N_1\ge 0} \cdots \sum_{N_{g_\Sigma}\ge 0} Z_X(\boldsymbol{N}, \hbar) \kappa_1^{N_1} \cdots \kappa_{g_\Sigma}^{N_{g_\Sigma}},  
\ee
with the convention that 
\be
\label{zzeros}
Z_X(0, \cdots, 0;\hbar)=1. 
\ee
This expansion defines the (generalized) fermionic spectral traces $Z_X(\boldsymbol{N}, \hbar)$, as promised. These are crucial in our construction, since 
they will provide a non-perturbative definition of the topological string partition function on $X$. Fredholm's formula (\ref{fred-th}) can be now used to give an explicit expression 
for these traces. Let us consider the kernels $A_{jl}(x_m, x_n)$ of the operators defined in (\ref{ajl}), and let us construct the 
following matrix:
\be
R_j (x_m, x_n) = A_{jl}(x_m , x_n) \,\, \,\, \,\, \,\,  \text{if} \, \,  \,\, \,\, \,\,\sum_{s=1}^{l-1} N_s< m \le \sum_{s=1}^l N_s. 
\ee
Then, we have that
\be
\label{gen-fred-th}
Z_X(\boldsymbol{N}; \hbar)= {1\over N_1! \cdots N_{g_\Sigma}!} \int  {\rm det}_{m,n} \left(R_j(x_m, x_n)\right)\rd^N x, 
\ee
where
\be
N=\sum_{s=1}^{g_\Sigma} N_s. 
\ee
As we showed above, the definition does not depend on the choice of $j=1, \cdots g_\Sigma$. Note that the expansion (\ref{or-exp}) has detailed information about the traces 
of all the operators $\mA_{ji}$ and their products.

Let us write some of the above formula in the case $g_\Sigma=2$, 
since we will use them later in the paper. In this case, the fermionic spectral traces can be written as
\be
\label{gtwo-matrix}
Z_X(N_1, N_2;\hbar)={1\over N_1! N_2!} \int  {\rm det} \begin{pmatrix} A_{j1}(x_1, x_1)&  \cdots  & A_{j1}(x_1, x_N) \\
\vdots &  & \vdots\\
A_{j1}(x_{N_1}, x_1) & \cdots & A_{j1}( x_{N_1}, x_N) \\
A_{j2} (x_{N_1+1}, x_1) & \cdots & A_{j2}( x_{N_1+1}, x_N) \\
\vdots &  & \vdots \\
A_{j2} (x_{N}, x_1) & \cdots &A_{j2}( x_{N}, x_N) \end{pmatrix}\rd x_1 \cdots \rd x_{N}. 
\ee
One finds, for example
\be
\ba
Z_X(1, 1; \hbar)&= \tr \, \mA_{j1} \,  \tr  \, \mA_{j2} - \tr \left( \mA_{j1} \mA_{j2}\right)\\
&= \int \rd x_1 \rd x_2 \left( A_{j1}(x_1, x_1) A_{j2}(x_2, x_2) -  A_{j1}(x_1, x_2) A_{j2}(x_2, x_1)\right), 
\ea
\ee
as well as 
\be
\ba
Z_X(2,1; \hbar)&=\tr \left(\mA_{j1}^{2} \mA_{j2}\right)-\frac{1}{2} \tr \left(\mA_{j1}^2\right) \tr \, \mA_{j2}+\frac{1}{2} \left(\tr\, \mA_{j1} \right)^2
  \tr\, \mA_{j2}-\tr \, \mA_{j1} \,  \tr  \left(\mA_{j1} \mA_{j2}\right), \\
 Z_X (1,2; \hbar)&=\tr \left(\mA_{j1}\mA_{j2}^{2} \right)-\frac{1}{2} \tr \, \mA_{j1}\,  \tr \left(\mA_{j2}^2\right) +\frac{1}{2}   \tr\, \mA_{j1}\left(\tr\, \mA_{j2} \right)^2
-\tr  \left(\mA_{j1} \mA_{j2}\right)\tr \, \mA_{j2}.
\ea 
  \ee
As we mentioned above, the integral (\ref{gen-fred-th}) should be regarded as a generalized multi-cut matrix model integral.

What is the motivation for the definition (\ref{gsd})? We should expect the generalized spectral determinant to contain information about the 
operators (\ref{Xops}). To see that this is the case, let us consider the spectral determinant
\be
\label{spec-one}
{\rm det}\left(1+ \kappa_1 \rho_1 \right). 
\ee
By using Fredholm--Plemelj's formula (\ref{f-p}), the log of this function can be computed as
\be
\label{sumj}
-\sum_{\ell_1=1}^\infty {(-\kappa_1)^{\ell_1} \over \ell_1} \tr \left( \mathsf{O}^{(0)}_{1} + \mP_1 \right) ^{-\ell_1}, 
\ee
where
\be
\mP_1= \sum_{j\not=1} \kappa_j \mP_{1j}. 
\ee
We first note that, 
\be
\tr \, \left( \mathsf{O}^{(0)}_{1} + \mP_1\right) ^{-\ell}= \tr \left[ \left( \rho_1^{(0)}  {1\over 1+ \rho_1^{(0)} \mP_1} \right)^\ell \right].  
\ee
By expanding each denominator in a geometric power series, we find that (\ref{sumj}) is given by 
\be
\label{above-ex}
-\sum_{\ell_1\ge 1} \sum_{\ell_2 \ge 0}  \cdots \sum_{\ell_{g_\Sigma} \ge 0} {1\over \ell_1+ \cdots + \ell_{g_\Sigma} } (-\kappa_1)^{\ell_1} \cdots \left(-\kappa_{g_\Sigma} \right)^{\ell_{g_\Sigma}} \sum_{W \in \CW_{\boldsymbol{\ell}} } \tr (W). 
\ee
In this equation, 
\be
\boldsymbol{\ell}=\left( \ell_1, \cdots, \ell_{g_\Sigma} \right)
\ee
and $\CW_{\boldsymbol{\ell}}$ is the set of all possible ``words" made of $\ell_i$ copies of the letters $\mA_{1i}$ defined in (\ref{ajl}). 
It is easy to see that (\ref{above-ex}) is almost identical to
\be
\ba
\CJ_X (\boldsymbol{\kappa}; \hbar)&= \log\, \Xi_X ( {\boldsymbol \kappa}; \hbar)=
 -\sum_{\ell \ge 1} {(-1)^\ell \over \ell} \tr \left( \kappa_1 \mA_{11} +\cdots+ \kappa_{g_\Sigma} \mA_{1 g_\Sigma}  \right)^\ell \\
 &=-\sum_{\ell_1\ge 0} \sum_{\ell_2 \ge 0}  \cdots \sum_{\ell_{g_\Sigma} \ge 0} {1\over \ell_1+ \cdots + \ell_{g_\Sigma} } (-\kappa_1)^{\ell_1} \cdots \left(-\kappa_{g_\Sigma} \right)^{\ell_{g_\Sigma}} \sum_{W \in \CW_{\boldsymbol{\ell}} } \tr (W), 
 \ea
 \ee
 except that all the terms have a strictly positive power of $\kappa_1$. It follows that 
\be
{\rm det}\left(1+ \kappa_1 \rho_1 \right)= {\Xi_X ( {\boldsymbol \kappa}; \hbar) \over \Xi_X ( (0, \kappa_2, \cdots, \kappa_{g_\Sigma}); \hbar)}. 
\ee
In addition, a simple inductive argument shows that
\be
\label{strata}
\Xi_X ( {\boldsymbol \kappa}; \hbar)= {\rm det}\left(1+ \kappa_1 \rho_1\right) {\rm det}\left(1+ \kappa_2 \rho_2\Big|_{\kappa_1=0} \right) 
 \cdots {\rm det}\left(1+ \kappa_{g_\Sigma} \rho_{g_\Sigma} \Big|_{\kappa_1=\cdots = \kappa_{g_\Sigma-1}=0} \right). 
\ee
In this derivation, we have taken as our starting point the operator $\rho_1$ and its spectral determinant, but it is clear that we could have used any other operator $\rho_i$, 
$i=1, \cdots, g_\Sigma$. In particular, we have 
\be
\label{quot-i}
 {\rm det}\left(1+ \kappa_i \rho_i\right) ={\Xi_X ( {\boldsymbol \kappa}; \hbar) \over \Xi_X \left(\kappa_1, \cdots, \kappa_{i-1},0, \kappa_{i+1}, \cdots, \kappa_{g_\Sigma} ; \hbar\right) }, \qquad 
i=1, \cdots, g_\Sigma. 
 \ee
If we set all moduli to zero in (\ref{quot-i}), except for $\kappa_i$, we find
\be
{\rm det}\left(1+ \kappa_i \rho_i^{(0)}\right)=\Xi_X(0,\cdots, 0, \kappa_i, 0, \cdots, 0; \hbar), \qquad i=1, \cdots, g_\Sigma. 
\ee
Therefore, the generalized spectral determinant specializes to the spectral determinant of the unperturbed operators appearing in the different canonical forms of the curve. 
We will see for example that the generalized spectral determinant 
associated to the resolved $\IC^3/\IZ_5$ geometry gives, after suitable specializations, the 
spectral determinants of the operators $\rho_{3,1}$ and $\rho_{2,2}$. 

The attentive reader has probably noticed that the operators $\mA_{jl}$ defined in (\ref{ajl}) are not Hermitian, in general. 
However, the generalized spectral traces defined by (\ref{or-exp}) are real (for real $\hbar$). This follows immediately from (\ref{strata}), which expresses 
(\ref{gsd}) as a product of spectral determinants of Hermitian operators.   
 
The generalized spectral determinant (\ref{gsd}) vanishes in a codimension one submanifold of the moduli space. This submanifold is a global analytic set, since it is determined 
by the vanishing of an entire function (see \cite{syz}). It contains all the required information about the spectrum of the operators $\rho_i$ appearing in the quantization of the mirror curve. For example, it follows from (\ref{quot-i}) that 
it gives the spectrum of eigenvalues $\re^{-E^{(i)}_n}$ of a given operator $\rho_i$, as a function of the other moduli $\kappa_j$, $j\not=i$. Since 
this holds for the different operators $\rho_k$, $k=1, \cdots, g_\Sigma$, it follows that their spectra are closely related. Heuristically, this can be already seen from (\ref{op-rel}). Let us suppose that
$|\psi^{(i)}_n \rangle$ is an eigenstate of $\mO_i$, with eigenvalue $\lambda_n^{(i)}$, and for given values of the $\kappa_l$, $l\not=i$. Then, 
\be
\label{ji-waves}
|\psi^{(j)}_n \rangle = \mP_{ij}^{1/2} |\psi^{(i)}_n \rangle, 
\ee
is an eigenstate of $\mO_j$ with eigenvalue $-\kappa_j$, where the parameters $\kappa_l$ appearing in $\mO_j$, $l \not=j$, are the same ones that appear in $\mO_i$, for $l \not=i$, 
while $\kappa_i= - \lambda_n^{(i)}$. Of course, since $\mP_{ij}^{1/2}$ is not bounded, the relation (\ref{ji-waves}) only holds if the square integrability of the wavefunction is not jeopardized. This is the case in the examples that we have looked at, like the resolved $\IC^3/\IZ_5$ orbifold. 

Interestingly, the codimension one submanifold determined by the vanishing of the generalized spectral determinant has been recently proposed as 
the natural definition of the joint spectrum of the $g_\Sigma $-uple of non-commuting operators $\mA_{j1}, \cdots, \mA_{jg_\Sigma}$ \cite{syz,csz}.   

\subsection{Comparison to quantum integrable systems}
\label{comparison}
In the theory that we have developed in the previous sections, the quantization process leads to $g_\Sigma$ different operators. However, 
these operators are related by reparametrizations of the coordinates and the relation (\ref{op-rel}). In particular, 
there is a {\it single} quantization condition for all of them, given by the vanishing of the generalized spectral determinant (\ref{gsd}), as in \cite{ghm}. This vanishing condition 
selects a discrete family of codimension one submanifolds in the moduli space parametrized by $\kappa_1, \cdots, \kappa_{g_\Sigma}$. 
We will determine this family in some detail in the case of the $\IC^3/\IZ_5$ orbifold. 

As we mentioned in the Introduction, our quantization scheme might be counter-intuitive for readers familiar with quantum integrable systems, in which 
the quantization of a genus $g_\Sigma$ spectral curve leads typically to $g_\Sigma$ quantization conditions. In order to appreciate the difference between the 
two quantization schemes, let us review in some 
detail the case of the periodic Toda chain of $N$ sites. This system is classically integrable, with 
$g_\Sigma=N-1$ Hamiltonians in involution (see \cite{bbt} for an excellent exposition of the classical chain). 
In the quantum theory, the Hamiltonians can be diagonalized simultaneously and one obtains in this way $g_\Sigma$ quantization conditions 
that determine their spectrum completely \cite{gutz}. An elegant way to obtain the spectrum is by using Baxter's equation \cite{sklyanin,gp}, which in the case of the Toda chain is given by, 
\be
\label{q-toda}
\im^{N} Q(x+ \im \hbar) + \im^{-N}  Q(x- \im \hbar)=\Lambda(x) Q(x), 
\ee
where 
\be
\Lambda(x)=x^N- \sum_{j=1}^{N-1} x^{N-1-j} \kappa_j. 
\ee
The $\kappa_j$ can be interpreted as the Hamiltonians of the Toda chain. 
It was shown in \cite{gp} that, by requiring $Q(x)$ to be an {\it entire} function which 
decays sufficiently fast at infinity, 
one recovers the quantization conditions of \cite{gutz}. 

Baxter's equation can be obtained by formally ``quantizing" the 
spectral curve of the Toda chain, which can be written as 
\be
\label{spec-curve}
2 \cosh y=\Lambda(x). 
\ee
The conserved Hamiltonians $\kappa_1, \cdots, \kappa_{g_\Sigma}$ are the moduli of the curve. The variables  
$y$ and $x$ can be regarded as canonically conjugate variables, in which $y$ plays the r\^ole of the momentum. 
In order to quantize (\ref{spec-curve}), we promote $x$, $y$ to Heisenberg 
operators. In the position representation we have
\be
y \rightarrow  -\ri \hbar {\rd \over \rd x}. 
\ee
If we now regard (\ref{spec-curve}) as an operator equation, acting on a wavefunction of the form 
\be
\label{wv}
\psi(x)= \exp\left({ N \pi \over 2 \hbar}\right) Q(x), 
\ee
we recover Baxter's equation (\ref{q-toda}). As already noted by Gutzwiller \cite{gutz}, this procedure is purely formal, since the spectral curve (\ref{spec-curve}) does not determine 
by itself the conditions that $Q(x)$ has to satisfy, and one needs additional information. A more detailed analysis \cite{gutz,kl,an} shows that this information is 
provided by the standard $L^2(\IR^{g_\Sigma})$ integrability of the original 
many-body problem, which forces $Q(x)$ to be entire and to decay at infinity in a prescribed way. 

\begin{figure}[h]
\center
\includegraphics[scale=0.4]{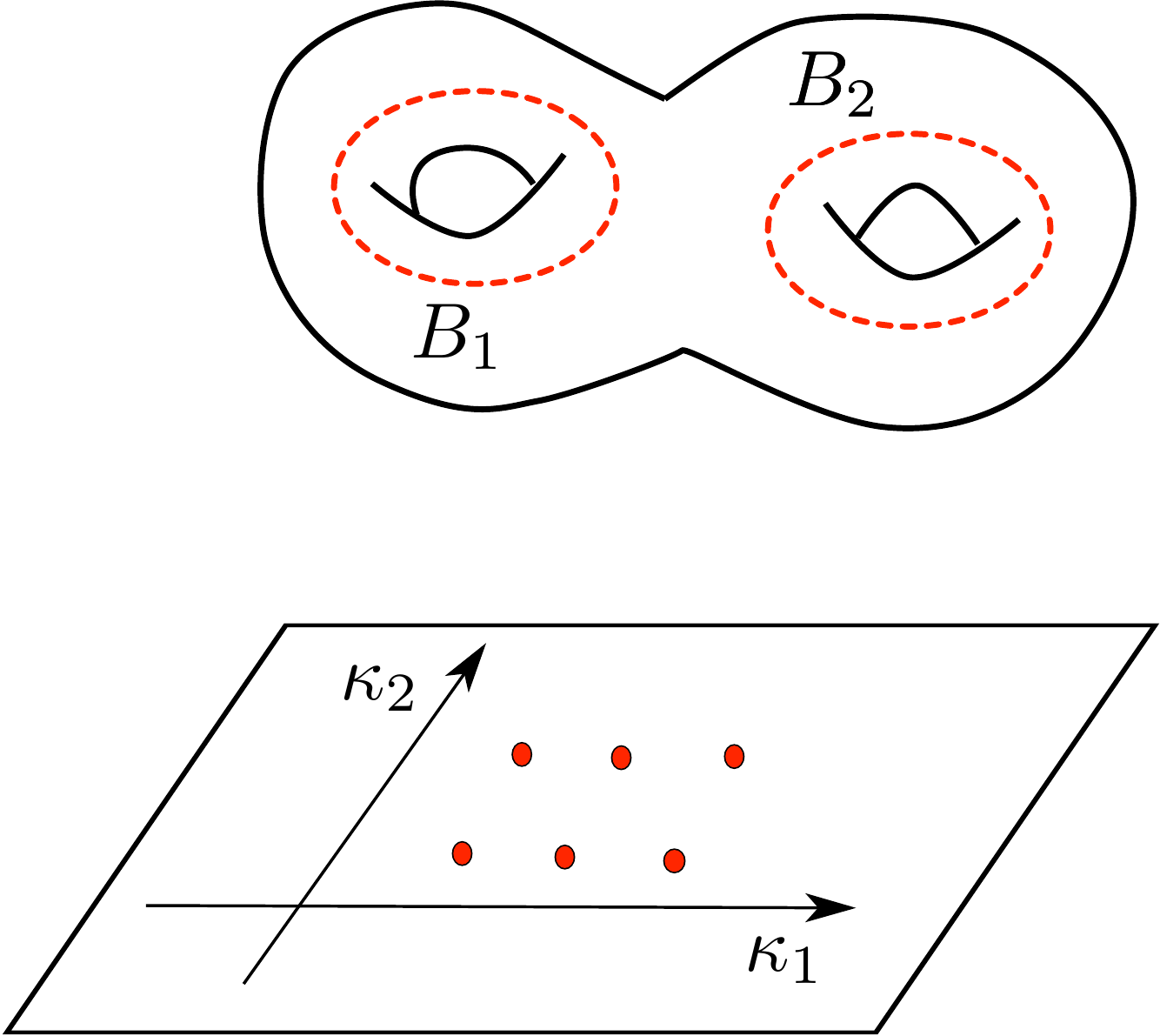}
\caption{In the quantum Toda lattice, the quantum spectral curve leads to $g_{\Sigma}=N-1$ quantization conditions, which select an infinite set of 
points in the moduli space parametrized by $\kappa_1, \cdots, \kappa_{g_{\Sigma}}$. We show a cartoon of this situation in the case $g_{\Sigma}=2$, which should be compared to the actual 
calculation in Figure 4 of \cite{matsuyama}.  
}
\label{discrete-ens}
\end{figure}

The resulting quantization conditions can be also analyzed in a WKB approximation \cite{gp}. If we use an ansatz for the wavefunction (\ref{wv}) of the form, 
\be
\psi(x) \sim \exp \left\{ -{\im \over \hbar} S(x; \hbar)\right\}, 
\ee
where
\be
\label{all-S}
S(x; \hbar)= \sum_{n \ge 0} \hbar^n S_n(x), 
\ee
the leading term is determined by $S'_0(x)= y(x)$, where $y(x)$ solves the equation for the spectral curve (\ref{spec-curve}), as expected. 
Based on the all-orders WKB solution (\ref{all-S}), we can define a ``quantum" differential as
\be
\lambda=\partial_x S(x; \hbar) \rd x.
\ee
Analyticity of $Q(x)$ leads to all-orders Bohr--Sommerfeld 
quantization conditions, 
\be
\label{toda-wkb}
 \oint_{B_i} \lambda=2 \pi \hbar n_i, \qquad i=1, \cdots, g_{\Sigma}, 
 \ee
 where $B_i$ are appropriate cycles on the curve (\ref{spec-curve}). 
 It was conjectured in \cite{ns} that these conditions can be derived from the Nekrasov--Shatashvili (NS) limit of the 
 instanton partition function of $SU(N)$, $\CN=2$ Yang--Mills theory. 
 This limit leads to a quantum-deformed prepotential $\CF^{\rm NS}(\boldsymbol{a}; \hbar)$, where 
 $\boldsymbol{a}=(a_1, \cdots, a_{g_\Sigma})$ are flat coordinates parametrizing the Coulomb branch and $g_{\Sigma}=N-1$ is the genus of the Seiberg--Witten curve. 
 The conjecture of \cite{ns} states that the periods appearing in (\ref{toda-wkb}) 
 are related to this prepotential by 
\be
\label{q-periods}
{\partial \CF^{\rm NS} \over \partial a_i}= \oint_{B_i} \lambda, \qquad  i=1, \cdots, g_{\Sigma}. 
\ee
In addition, the flat coordinates $a_i$ are related to the $\kappa_1, \cdots, \kappa_{g_{\Sigma}}$ through a ``quantum" mirror map, 
\be
a_i = \oint_{A_i} \lambda, \qquad i=1, \cdots, g_{\Sigma}, 
\ee
where $A_i$ are appropriate cycles on the spectral curve. This conjecture was verified, in the very first orders of the perturbative WKB expansion, in \cite{mirmor, mirmor2}. 
Additional evidence for this claim has been also provided in for example \cite{kt}. 

Therefore, in the case of quantum integrable systems of the Toda type, one has $g_{\Sigma}$ quantization 
conditions, which in the all-orders WKB quantization can be written as in (\ref{toda-wkb}). The solution to these conditions on the $g_{\Sigma}$-dimensional moduli 
space parametrized by the Hamiltonians $\kappa_1, \cdots, \kappa_{g_\Sigma}$ 
is a set of points, i.e. a submanifold of codimension $g_\Sigma$. In \figref{discrete-ens} we show a cartoon of how 
the quantization conditions, in the case of $g_{\Sigma}=2$, lead to such a discrete spectrum. This cartoon should be compared to Figure 4 of \cite{matsuyama}, which shows 
the result of the actual calculation. 

As we already noted, the quantum version of the Toda spectral curve does not lead by itself to a well-defined spectral problem: one needs 
additional conditions that follow from a detailed analysis of the 
original integrable system, which has $g_\Sigma$ Hamiltonians in involution and requires $g_\Sigma$ quantization conditions. However, this does not mean that 
a curve of genus $g_\Sigma$ should always lead, after quantization, to $g_\Sigma$ quantization conditions. The simplest example showing that this is not the case is 
a one-dimensional particle in an (even) confining potential, with a classical Hamiltonian
\be\label{h-v}
H(x,y)={y^2\over 2} + V(x), \qquad V(x)= x^{2N}+ c_{N-1} x^{2(N-2)}+\cdots+ c_0. 
\ee
The curve 
\be
\label{h-curve}
H(x,y)=E
\ee
has genus $g_{\Sigma}=N-1$. The ``quantization" of this curve leads to a standard eigenvalue problem for a Schr\"odinger equation with potential $V(x)$. For real $c_i$, $i=0, \cdots, N-1$, 
the spectrum is real and discrete, and there should be a single quantization condition, expressing the energy $E$ as a function of the parameter $c_i$ 
and the quantum number $n$. Semiclassically, and for $E$ sufficiently large, the quantization condition is simply 
given by the Bohr--Sommerfeld rule,  
\be
\label{bs}
\oint_B y(x) \rd x = 2 \pi \hbar n, \qquad n=0, 1, \cdots, 
\ee
where $B$ is the cycle associated to the turning points of the classical motion. Therefore, although the curve (\ref{h-v}) has genus $N-1$, when interpreted as describing a particle in an even, confining potential, 
its quantum version should lead to a single quantization condition, associated to the preferred cycle $B$. One could think that the other cycles of the higher genus curve do not play a r\^ole. However, this is not so. 
The reason is that, in the exact WKB method, one should consider complex trajectories around all possible cycles of 
the underlying curve, and these trajectories will lead to complex instanton corrections to (\ref{bs}), as first pointed out in the seminal paper \cite{bpv}.

\begin{figure}[h]
\center
\includegraphics[scale=0.25]{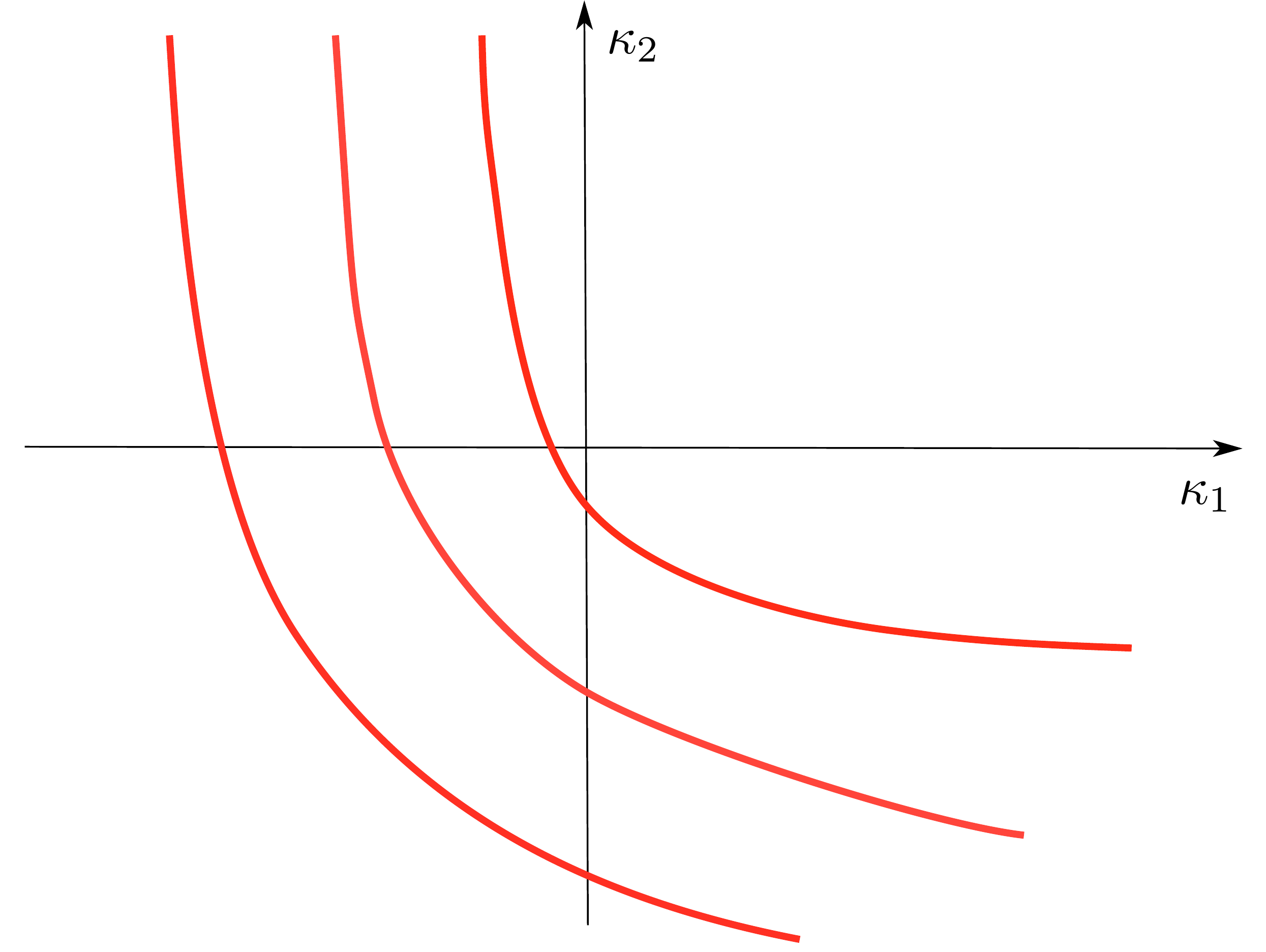}
\caption{The quantization of a higher genus mirror curve leads to a {\it single} quantization condition, which defines a discrete family of 
codimension one submanifolds in moduli space. Here, the $\kappa_i$ should be understood as $-\re^{E_i}$, where $E_i$ are the energies. This cartoon should 
be compared to the actual calculation for the resolved $\IC^3/\IZ_5$ geometry and $\hbar=2\pi$, in \figref{bluecurve}.
}
\label{codimg-f}
\end{figure}
The quantization of higher genus mirror curves considered in this paper is in fact very similar to the quantization of the curve (\ref{h-curve}): 
there is in principle no need to specify $g_\Sigma$ quantization conditions, 
since (at least in the cases we have considered) the relevant operators $\mO_i$ have a well-defined discrete, positive spectrum which is determined 
by a single quantization condition. This condition determines a discrete family of codimension one submanifolds in moduli space. A cartoon for what we expect 
when $g_{\Sigma}=2$ is shown in \figref{codimg-f}. At the same time, our quantization scheme leads to a genuine $g_\Sigma$-dimensional problem, as reflected in the fact that we 
have $g_\Sigma$ different operators 
and our generalized fermionic spectral traces depend on $g_\Sigma$ different integers. Our goal will be to determine the quantization 
condition, as well as the generalized spectral determinant (\ref{gsd}) and spectral traces, from the topological string amplitudes on $X$. The cartoon in \figref{codimg-f} can be compared to the actual calculation 
of such a family in the example of the resolved $\IC^3/\IZ_5$ geometry, and for $\hbar=2\pi$ in \figref{bluecurve}.

\sectiono{Spectral determinants and topological strings}

\subsection{A conjecture for the generalized spectral determinant}

We will now state our main conjecture, which generalizes \cite{ghm} to the higher genus case, and provides an 
explicit expression for the generalized spectral determinant (\ref{gsd}) in terms 
of topological string amplitudes. As in \cite{ghm}, the key object is the (modified) grand potential introduced in \cite{hmmo}. 

In order to state the conjecture, let us first recall some basic geometric ingredients. As we noted above, in the geometry there are $g_\Sigma$ moduli, $\kappa_i$, $i=1, \cdots, g_\Sigma$, and 
$r_\Sigma$ mass parameters, $\xi_j$, $j=1, \cdots, r_\Sigma$.  
The classical mirror map expresses the large radius, K\"ahler parameters $t_i$ of the CY 
in terms of ${\boldsymbol \kappa}$, $\xi_j$. We will also parametrize the moduli through the ``chemical potentials" $\mu_i$, defined by 
\be
\kappa_i =\re^{\mu_i}, \qquad i=1, \cdots, g_{\Sigma}. 
\ee
There are $ g_\Sigma+r_\Sigma$ K\"ahler parameters and their mirror map at large $\mu_i$ is of the form 
\be
\label{tmu} 
t_i \approx \sum_{j=1}^{g_\Sigma} C_{ij} \mu_j + \sum_{j=1}^{r_\Sigma} \alpha_{ij} t_{\xi_j}, \qquad i=1, \cdots, g_\Sigma+r_\Sigma, 
\ee
where $t_{\xi_j}$  are in general function of the mass parameters $\xi_j$ as discussed for instance in \cite{gkmr}.
%
In particular, this means that the complex moduli $z_i$ corresponding to the 
K\"ahler parameters $t_i$ are given by 
\be
\label{zmu}
-\log \, z_i= \sum_{j=1}^{g_\Sigma}C_{ij} \mu_j + \sum_{k=1}^{r_\Sigma} a_{ik}\log {\xi_k}, \qquad i=1, \cdots, g_\Sigma+r_\Sigma.
\ee
The coefficients $C_{ij}$ determine a $\left(g_\Sigma +r_\Sigma \right)  \times g_\Sigma$ matrix which can be read off from the 
intersection data of $X$. The truncated version 
\be \label{tmc} C_{ij}, \quad  i,j =1, \cdots, g_\Sigma \ee
is an invertible matrix  and it agrees (up to an overall sign) with the matrix $C_{ij}$ appearing in \cite{kpsw}. 
%
As first shown in \cite{acdkv}, the mirror map can be promoted to a {\it quantum} mirror map $t_i(\hbar)$ which depends now on $\hbar$. 
Explicit expressions for the quantum mirror map can be obtained in various ways, 
see for instance \cite{huang,hkrs} for examples. 
We note however that the algebraic mirror maps for $t_{\xi_j}$ remain undeformed \cite{hkrs}.  

In addition to the quantum mirror map, we need the following topological string theory ingredients. 
First of all, we have the conventional genus $g$ free energies $F_g({\bf t})$ of $X$, $g \ge 0$, in the so-called 
large radius frame. We have, at genus zero, 
\be
\label{gzp}
F_0({\bf t})={1\over 6} a_{ijk} t_i t_j t_k  + \sum_{{\bf d}} N_0^{ {\bf d}} \re^{-{\bf d} \cdot {\bf t}}.
\ee
At genus one, one has
\be
\label{gop}
F_1({\bf t})=b_i t_i + \sum_{{\bf d}} N_1^{ {\bf d}} \re^{-{\bf d} \cdot {\bf t}}, 
\ee
while at higher genus one finds 
\be
\label{genus-g}
F_g({\bf t})= C_g+\sum_{{\bf d}} N_g^{ {\bf d}} \re^{-{\bf d} \cdot {\bf t}}, \qquad g\ge 2, 
\ee
where $C_g$ is a constant, called the constant map contribution to the free energy \cite{bcov}. 
The total free energy of the topological string is formally defined as the sum, 
\be
\label{tfe}
F^{\rm WS}\left({\bf t}, g_s\right)= \sum_{g\ge 0} g_s^{2g-2} F_g({\bf t})=F^{({\rm p})}({\bf t}, g_s)+ \sum_{g\ge 0} \sum_{\bf d} N_g^{ {\bf d}} \re^{-{\bf d} \cdot {\bf t}} g_s^{2g-2},   
\ee
where
\be
F^{({\rm p})}({\bf t}, g_s)= {1\over 6 g_s^2} a_{ijk} t_i t_j t_k +b_i t_i + \sum_{g \ge 2}  C_g g_s^{2g-2}. 
\ee
As it is well-known \cite{gv}, the sum over Gromov--Witten invariants in (\ref{tfe}) 
can be resummed order by order in $\exp(-t_i)$, at all orders in $g_s$. This resummation involves a new 
set of enumerative invariants, the so-called Gopakumar--Vafa (GV)
invariants $n^{\bf d}_g$. Out of these invariants, one constructs the generating series 
\be
\label{GVgf}
F^{\rm GV}\left({\bf t}, g_s\right)=\sum_{g\ge 0} \sum_{\bf d} \sum_{w=1}^\infty {1\over w} n_g^{ {\bf d}} \left(2 \sin { w g_s \over 2} \right)^{2g-2} \re^{-w {\bf d} \cdot {\bf t}}, 
\ee
and one has, as an equality of formal series, 
\be
\label{gv-form}
F^{\rm WS}\left({\bf t}, g_s\right)=F^{({\rm p})}({\bf t}, g_s)+F^{\rm GV}\left({\bf t}, g_s\right). 
\ee
One can generalize the Gopakumar--Vafa invariants to defined the so-called 
{\it refined BPS invariants} \cite{ikv,ckk,no}. These invariants depend on the degrees ${\bf d}$ and on two non-negative 
half-integers, $j_L$, $j_R$. We will denote them by $N^{\bf d}_{j_L, j_R}$, and they are also integers. The 
Gopakumar--Vafa invariants are particular combinations 
of these refined BPS invariants, and one has the following relationship between generating functionals, 
\be
\label{ref-gv}
\sum_{j_L, j_R} \chi_{j_L}(q) (2j_R+1) N^{\bf d} _{j_L, j_R} = \sum_{g\ge 0} n_g^{\bf d} \left(q^{1/2}- q^{-1/2} \right)^{2g}, 
\ee
where $q$ is a formal variable and 
\be
\chi_{j}(q)= {q^{2j+1}- q^{-2j-1} \over q-q^{-1}}
\ee
is the $SU(2)$ character for the spin $j$. We note that the sums in (\ref{ref-gv}) are well-defined, since for given degrees ${\bf d}$ only a finite number of $j_L$, $j_R$, $g$ give a non-zero 
contribution. Out of these refined BPS invariants, one can define the so-called NS free energy, 
\be
\label{NS-j}
F^{\rm NS}({\bf t}, \hbar) ={1\over 6 \hbar} a_{ijk} t_i t_j t_k +b^{\rm NS}_i t_i \hbar +\sum_{j_L, j_R} \sum_{w, {\bf d} } 
N^{{\bf d}}_{j_L, j_R}  \frac{\sin\frac{\hbar w}{2}(2j_L+1)\sin\frac{\hbar w}{2}(2j_R+1)}{2 w^2 \sin^3\frac{\hbar w}{2}} \re^{-w {\bf d}\cdot{\bf  t}}. 
\ee
In this equation, the coefficients $a_{ijk}$ are the same ones that appear in (\ref{gzp}), while $b_i^{\rm NS}$ can be obtained by using mirror symmetry as in \cite{hk}. 
This generating functional involves a particular combination of the refined BPS invariants, which defines the NS limit of the refined topological string. The NS limit was first discussed in the 
context of gauge theory in \cite{ns}. By expanding (\ref{NS-j}) in powers of $\hbar$, we find the NS free energies at order $n$, 
\be
\label{ns-expansion}
F^{\rm NS}({\bf t}, \hbar)=\sum_{n=0}^\infty  F^{\rm NS}_n ({\bf t}) \hbar^{2n-1}, 
\ee
and the expression (\ref{NS-j}) can be regarded as a Gopakumar--Vafa-like resummation of the series in (\ref{ns-expansion}). We recall that the first term 
in this series, $F_0^{\rm NS}({\bf t})$, is equal to $F_0({\bf t})$, the standard genus zero free energy. Note that the term involving the coefficients $b_i^{\rm NS}$ contributes to 
$F_1^{\rm NS}({\bf t})$. 

With all these ingredients, we are ready to define, following \cite{hmmo}, the so-called {\it modified grand potential} of the CY $X$. It is the sum of two functions. The first one is
\be
\label{jm2}
\mathsf{J}^{\rm WKB}_X(\boldsymbol{\mu}, \boldsymbol{\xi}, \hbar)= {t_i(\hbar) \over 2 \pi}   {\partial F^{\rm NS}({\bf t}(\hbar), \hbar) \over \partial t_i} 
+{\hbar^2 \over 2 \pi} {\partial \over \partial \hbar} \left(  {F^{\rm NS}({\bf t}(\hbar), \hbar) \over \hbar} \right) + {2 \pi \over \hbar} b_i t_i(\hbar) + A({\boldsymbol \xi}, \hbar). 
\ee
Note that, in the second term, the derivative w.r.t. $\hbar$ does not 
act on the implicit dependence of $t_i (\hbar)$ (it is a true partial derivative).
The function $A({\boldsymbol \xi}, \hbar)$ is not known in closed form for arbitrary 
geometries, although detailed conjectures for its form exist in some cases. It is closely related to a resummed form of the constant map contribution appearing in (\ref{genus-g}). 
The function (\ref{jm2}) is perturbative in $\hbar$, and it can be in principle obtained by performing a resummation of the all-orders WKB expansion, 
hence its name. At leading order in $\hbar$, the quantum mirror map becomes the classical mirror map $t_i(\hbar) \approx t_i$, and 
\be
\mJ^{\rm WKB}_X ( \boldsymbol{\mu},  \boldsymbol{\xi}, \hbar) ={1\over \hbar} \mJ^X_0( \boldsymbol{\mu},  \boldsymbol{\xi}) +\cdots, 
\ee
where
\be
\label{semiJ}
\mJ^X_0( \boldsymbol{\mu},  \boldsymbol{\xi})= {1\over 2 \pi} \left(  t_i {\partial F_0\over \partial t_i} -2  F_0\right)+ 2 \pi b_i t_i
\ee
and $F_0$ is the genus zero free energy. 

The second function is 
the ``worldsheet" modified grand potential, which is obtained from the generating functional (\ref{GVgf}), 
\be
\label{jws}
\mathsf{J}^{\rm WS}_X(\boldsymbol{\mu}, \boldsymbol{\xi}, \hbar)=F^{\rm GV}\left( {2 \pi \over \hbar}{\bf t}(\hbar)+ \pi \ri {\bf B} , {4 \pi^2 \over \hbar} \right).
\ee
It involves a constant vector ${\bf B}$ (or ``B-field") which depends on the geometry under consideration. This vector should satisfy the following requirement: 
for all ${\bf d}$, $j_L$ and $j_R$ such that $N^{{\bf d}}_{j_L, j_R} $ is non-vanishing, we must have
\be
\label{B-prop}
(-1)^{2j_L + 2 j_R+1}= (-1)^{{\bf B} \cdot {\bf d}}. 
\ee
For local del Pezzo CY threefolds, the existence of such a vector was established in \cite{hmmo}. Note that the effect of this constant vector is to replace 
\be
\label{B-sign}
\re^{-{\bf t}} \rightarrow \re^{-{\bf t}- \pi \ri {\bf B}}
\ee
in the generating functional (\ref{GVgf}). Note as well that the string coupling constant $g_s$ is related to the Planck constant of the spectral problem by 
\be
g_s={4 \pi^2 \over \hbar}. 
\ee
Therefore, the strong string coupling regime corresponds to the semiclassical limit of the spectral problem, while the weakly coupled regime of the topological string 
corresponds to a highly quantum regime in the spectral problem. 

The {\it total, modified grand potential} is the sum of the above two functions, 
\be
\label{jtotal}
\mathsf{J}_{X}(\boldsymbol{\mu}, \boldsymbol{\xi},\hbar) = \mathsf{J}^{\rm WKB}_X (\boldsymbol{\mu}, \boldsymbol{\xi},\hbar)+ \mathsf{J}^{\rm WS}_X 
(\boldsymbol{\mu},  \boldsymbol{\xi} , \hbar), 
\ee
and it was first considered in \cite{hmmo}. When the mirror curve has genus one, it agrees with the modified grand potential of \cite{ghm}, although we have written 
it in a slightly different way. In particular, the modified grand potential of \cite{ghm} involves a perturbative part, a membrane part, 
and a worldsheet part. Here, we have put together 
the perturbative and the membrane part in the WKB grand potential. The quantity (\ref{jtotal}) has the following structure, 
\be
\label{j-larget}
\mathsf{J}_{X}(\boldsymbol{\mu}, \boldsymbol{\xi},\hbar)= {1\over 12 \pi \hbar} a_{ijk} t_i(\hbar) t_j(\hbar) t_k(\hbar) + \left( {2 \pi b_i \over \hbar} + {\hbar b_i^{\rm NS} \over 2 \pi} \right) t_i(\hbar) + 
\CO\left( \re^{-t_i(\hbar)}, \re^{-2 \pi t_i(\hbar)/\hbar} \right). 
\ee
The last term stands for a formal power series in $\re^{-t_i(\hbar)}$, $\re^{-2 \pi t_i (\hbar)/\hbar}$, whose coefficients depend explicitly 
on $\hbar$. However, this series is {\it a priori} ill-defined when $\hbar$ is a rational multiple of $\pi$. This is due to the double poles in the 
trigonometric functions appearing in (\ref{NS-j}) and (\ref{GVgf}). However, although the generating 
functionals (\ref{jm2}) and (\ref{jws}) diverge separately, the poles cancel 
in the sum \cite{hmmo}, as in the HMO cancellation mechanism discovered in \cite{hmo}. The presence of a $B$-field 
satisfying (\ref{B-prop}) is crucial for this cancellation. 
In the higher genus case, we have established 
the existence of such a ${\bf B}$ in the examples we have studied. Clearly, it would be important to determine ${\bf B}$ in full generality for any toric geometry. 
Note that, in addition, our expression for (\ref{jtotal}) is, as in \cite{ghm}, intrinsically a large radius expansion. We note that only at large radius we have geometric tools to 
sum up the $\hbar$ corrections at all orders.  

After introducing all of our ingredients, we are ready to state our main conjecture. We claim that the generalized 
spectral determinant (\ref{gsd}) is given by 
\be
\label{our-conj}
\Xi_X({\boldsymbol \kappa}; \hbar)= \sum_{ {\bf n} \in \IZ^{g_\Sigma}} \exp \left( \mathsf{J}_{X}(\boldsymbol{\mu}+2 \pi \ri  {\bf n}, \boldsymbol{\xi}, \hbar) \right). 
\ee
It is understood that the generalized spectral determinant also depends on the mass parameters $\boldsymbol{\xi}$, but we will not write this dependence explicitly. 
As in \cite{ghm}, the right hand side of (\ref{our-conj}) defines a quantum-deformed (or generalized) Riemann theta function by 
\be
\label{qtf}
\Xi_X({\boldsymbol \kappa}; \hbar)= \exp\left( \mathsf{J}_{X}(\boldsymbol{\mu}, \boldsymbol{\xi}, \hbar) \right) \Theta_X({\boldsymbol \kappa}; \hbar). 
\ee
We note that the function $A({\boldsymbol \xi}, \hbar)$ appearing in (\ref{jm2}) can be fixed by requiring that the expansion of the 
generalized spectral determinant around ${\boldsymbol \kappa}=0$ starts with $1$. The expression (\ref{our-conj}) looks rather formal, 
but in fact it can be computed systematically (for arbitrary $\hbar$) near the large radius point of moduli space, as shown in \cite{ghm} in the genus one case. Interestingly, if our conjecture is 
true, the resulting expression is in fact an {\it analytic} function on the CY moduli space. This is surprising, since the modified grand potential (\ref{jtotal}) is not analytic. However, the 
inclusion of the quantum theta function should cure the lack of analyticity. This is related to the observation in \cite{em} that including generalized theta functions in the total partition 
functions restores modular invariance. Note though that, in contrast to what happened in \cite{em}, the quantum theta function appearing in (\ref{qtf}) 
is well-defined, at least as an asymptotic expansion. In addition, and as we will see in the next section, when $\hbar=2 \pi$, the quantum theta function becomes a perfectly 
well-defined, ordinary theta function.  

\subsection{The maximally supersymmetric case} 

As in the genus one case, an important simplification in the above formulae occurs when $\hbar=2\pi$. 
In this case, the contribution to (\ref{GVgf}) involving invariants with $g\ge 2$ vanish. After carefully 
canceling the poles, one finds that (\ref{jtotal}) becomes
\be
\label{jx-ms}
\mJ_X(\boldsymbol{\mu}, \boldsymbol{\zeta}, \hbar)=  {1\over 4 \pi^2} \left\{ {1\over 2}  \sum_{i, j =1}^{g_\Sigma+r_\Sigma} t_i t_j {\partial^2 \widehat F_0  \over \partial t_i \partial t_j} 
-\sum_{i=1}^{g_\Sigma+r_\Sigma} t_i {\partial \widehat F_0 \over \partial t_i }+ \widehat F_0  \right\}  +\widehat F_1 + \widehat F_1^{\text{NS}}. 
\ee
In these formulae, the generating functionals $\widehat F_0$, $\widehat F_1$ and $\widehat F_1^{\text{NS}}$ are the same 
ones appearing in (\ref{gzp}), (\ref{gop}), (\ref{ns-expansion}), but where we perform the 
replacement (\ref{B-sign}) in the instanton expansion (i.e., we don't make such a replacement in the polynomial terms in ${\bf t}$.) In (\ref{jx-ms}), $t_i$ denotes 
the quantum mirror map evaluated at $\hbar=2 \pi$. It turns out that this equals the classical mirror map, up to a change of sign in the expansion in the moduli. This change of 
sign is precisely the one that would lead to  (\ref{B-sign}). 

As a consequence of this simplification, the quantum-deformed theta function becomes 
\be
\label{theta-hg}
\Theta_X({\boldsymbol \kappa}; 2 \pi)=\sum_{ {\bf n} \in \IZ^{g_\Sigma}} \exp\left[ \pi \ri ^t {\bf n} \tau {\bf n} + 2\pi \ri {\bf n} \cdot \boldsymbol {\upsilon} -{\ri \pi \over 3} a_{ijk} C_{il} C_{jm} C_{k p} n_l n_m n_p \right]. 
\ee
In this equation, $\tau$ is a $g_\Sigma\times g_\Sigma$ matrix given by 
\be
\label{tau-xi}
\tau_{l m}= -{1\over 2 \pi \ri} C_{jl} C_{k m} {\partial^2 \widehat F_0 \over \partial t_j\partial t_k}, \qquad l,m=1, \cdots, g_\Sigma,
\ee
where the sum over $k,j$ runs from 1 to $g_{\Sigma}+r_\Sigma$.
As explained in \cite{kpsw}, this is nothing but the $\tau$ matrix of the mirror curve. It is a symmetric matrix satisfying 
\be
{\rm Im}(\tau) >0. 
\ee
In addition, the vector $\boldsymbol {\upsilon}$ appearing in (\ref{theta-hg}) has components
\be 
\label{ups}
\upsilon_m= {C_{jm} \over 4 \pi^2} \left\{  {\partial^2 \widehat F_0 \over \partial t_j  \partial t_k} t_k  -  
{\partial \widehat F_0 \over \partial t_j } \right\} +  C_{jm} \left( b_j +  b_j^{\text{NS}}\right), \qquad m=1, \cdots, g_\Sigma, 
\ee
where the sum over $k,j$ runs over all the $g_{\Sigma}+r_\Sigma$ indices.
In all the examples we have considered, the cubic terms in (\ref{theta-hg}) can be traded by quadratic or linear terms. 
This adds constant, real shifts to $\tau$ and $\boldsymbol{\upsilon}$. The resulting 
matrix and vector will be denoted by $\widetilde \tau$ and $\widetilde {\boldsymbol {\upsilon}}$. In this way, (\ref{theta-hg}) becomes (up to an overall constant) a conventional higher genus Riemann--Siegel theta function on $\Sigma$, which we will write as 
\be
\label{til-te}
\Theta_X({\boldsymbol \kappa}; 2 \pi)=\sum_{ {\bf n} \in \IZ^g} \exp\left[ \pi \ri ^t {\bf n} \widetilde \tau {\bf n} + 2\pi \ri {\bf n} \cdot \widetilde {\boldsymbol {\upsilon}}\right]. 
\ee
Note that ${\rm Im}\left(\widetilde \tau\right)= {\rm Im}\left(\tau\right)$, therefore the theta function (\ref{til-te}) is still 
well defined. This result is a direct generalization of the genus one case considered in \cite{ghm}. 

As we have discussed, the quantization condition for the operators associated to $X$ is obtained by looking at the vanishing locus of the generalized spectral determinant. In the maximally 
supersymmetric case, this has a beautiful interpretation. The vanishing locus of (\ref{til-te}) on the Jacobi torus is by definition the {\it theta divisor} $D_\Theta$. The period 
$\widetilde {\boldsymbol {\upsilon}}$ can be regarded as a map 
from the moduli space $\CM$ parametrized by ${\boldsymbol \kappa}$, to the Jacobi torus, 
\be
\widetilde {\boldsymbol {\upsilon}}: {\cal M} \rightarrow \IT^{2g_\Sigma}. 
\ee
It follows that the vanishing locus giving the quantization condition can be geometrically interpreted as the 
inverse image of the theta divisor by the map $\widetilde {\boldsymbol {\upsilon}}$: 
\be
\widetilde {\boldsymbol {\upsilon}}^{-1} \left( D_\Theta \right) \subset \CM. 
\ee
Of course, the same interpretation can be made in the genus one case. In the generic case, one has to consider the quantum-deformed theta function, 
and its vanishing locus will be a quantum deformation of the locus 
above.

\subsection{Spectral traces at large $N$ and non-perturbative topological strings}
\label{NP}

One of the most surprising consequences of the correspondence between spectral theory and mirror symmetry is that the {\it conventional} 
topological string can be obtained from a 't Hooft-like limit 
of the fermionic spectral traces. In the case of genus one curves, this was explained in detail in \cite{mz}. First of all, note that these traces, 
which appear as coefficient in the expansion (\ref{or-exp}), can be written 
as 
\be
Z_X({\boldsymbol N}, \hbar)={1\over \left( 2 \pi \ri\right)^{g_\Sigma}} \oint_0 {\rd \kappa_1 \over \kappa_1^{N_1+1} } 
\cdots  \oint_0 {\rd \kappa_{g_\Sigma} \over \kappa_{g_\Sigma}^{N_{g_\Sigma} +1} } \, \Xi_X({\boldsymbol \kappa}; \hbar). 
\ee
We can now use the argument first presented in \cite{hmo}: the multi-contour integral can be written as an integral over 
$\mu_i$, from $-\ri \pi$ to $\ri \pi$. Since, according to our conjecture (\ref{our-conj}), the generalized 
spectral determinant can be obtained by summing over all displacements of the $\mu_i$ parameters in integer steps of $2 \pi \ri$, we can trade the sum over the $n_i$ by an integration 
along the whole imaginary axis, 
and we find 
\be
\label{multi-Airy}
Z_X({\boldsymbol N}, \hbar)={1\over \left( 2 \pi \ri\right)^{g_\Sigma}} \int_{-\ri \infty}^{\ri \infty} \rd \mu_1 \cdots \int_{-\ri \infty}^{\ri \infty} \rd \mu_{g_\Sigma} \,
 \exp \left\{ \mathsf{J}_X ({\boldsymbol \mu},  {\boldsymbol {\xi}}, \hbar) - \sum_{i=1}^{g_\Sigma} N_i \mu_i \right\}. 
\ee
As in \cite{mz}, we want to evaluate the asymptotic expansion of the fermionic spectral traces in the 't Hooft limit (\ref{thooft}). 
This can be done by evaluating the multi-integral in the saddle point approximation. 
We have to consider the limit in which 
\be
\label{thooft-mu}
 \hbar \rightarrow \infty, \qquad \mu_i \rightarrow \infty, \qquad {\mu_i \over \hbar} =\zeta_i \, \, \, \, \, \text{fixed}, \quad i=1, \cdots, g_\Sigma. 
 \ee
In this limit, the quantum mirror map becomes trivial, and the approximation (\ref{tmu}) is exact. 
We will also assume that the mass parameters $\boldsymbol{\xi}$ scale in such a way that 
\be
\label{mt}
m_j= \exp\left(-{2\pi \over \hbar} t_{\xi_j}\right), \qquad j=1, \cdots, r_\Sigma,
\ee
remain fixed in the 't Hooft limit (other scaling behaviors can be considered, as in \cite{mz}). In the study of the 't Hooft regime, we will denote $2 \pi {\bf t} /\hbar $ simply by $ {\bf t}$, in order
to avoid unnecessary additional notation. Note that, with this notation, we have from (\ref{tmu}) the relation 
\be
\label{tzeta}
t_i +\sum_{j=1}^{r_\Sigma}\alpha_{ij}\log m_j= 2 \pi \sum_{j=1}^{g_\Sigma} C_{ij} \zeta_j, \qquad i=1, \cdots, g_\Sigma+ r_\Sigma.
\ee
Then, in the limit (\ref{thooft-mu}), the modified grand potential has the genus expansion 
\be
\mJ^{\text{'t Hooft}}_X\left({\boldsymbol \zeta}, {\boldsymbol m}, \hbar \right)= \sum_{g=0}^\infty \mJ^{X}_g \left({\boldsymbol \zeta}, {\boldsymbol m}\right) \hbar^{2-2g}, 
\ee 
where
\be 
\label{gen-J-as}
\ba
 \mJ^{X}_0 \left( {\boldsymbol \zeta}, {\boldsymbol m}\right)&={1\over 16 \pi^4} \left(
 \widehat F_0 \left( {\bf t}\right) + 4 \pi^2 b_i^{\rm NS} t_i
  + 14 \pi^4 A_0 \left({\boldsymbol m}\right) \right), \\
 \mJ^{X}_1 \left( {\boldsymbol \zeta}, {\boldsymbol m}\right)&=  A_1\left({\boldsymbol m}\right)+  \widehat F_1 \left( {\bf t}\right), \\
  \mJ^{X}_g \left({\boldsymbol \zeta}, {\boldsymbol m}\right)&=  A_g \left({\boldsymbol m}\right)-C_g+ (4 \pi^2)^{2g-2}  \widehat F_g \left( {\bf t}\right), \qquad g\ge 2. 
  \ea
  \ee
The arguments ${\boldsymbol \zeta}$ and ${\boldsymbol m}$ of the modified grand potential are related to the K\"ahler parameters ${\bf t}$ by (\ref{mt}) and (\ref{tzeta}). 
We have assumed that the function $A\left({\boldsymbol{\xi}}, \hbar\right)$ has the expansion 
\be
A\left({\boldsymbol{\xi}}, \hbar\right)= \sum_{g=0}^\infty A_g({\boldsymbol m}) \hbar^{2-2g}. 
\ee
In (\ref{gen-J-as}), as in (\ref{jx-ms}), the $ \widehat F_g \left( {\bf t}\right)$ are the standard topological string free energies as a 
function of the K\"ahler parameters ${\bf t}$, after turning on the B-field. The saddle point of the integral (\ref{multi-Airy}) is given by 
\be
\label{saddle}
\lambda_i ={C_{ji} \over 8 \pi^3} \left( {\partial \widehat F_0 \over \partial t_j} + 4 \pi^2 b_j^{\rm NS} \right), \qquad i=1, \cdots, g_\Sigma.
\ee
One then finds that the fermionic spectral traces have an expansion of the form (\ref{logz-exp}). The leading function in this expansion is given by a Legendre transform, 
\be
\CF_0 ({\boldsymbol{\lambda}})=\mJ^X_0\left({\boldsymbol \zeta}, {\boldsymbol m} \right)- {\boldsymbol \lambda}\cdot  {\boldsymbol \zeta}.  
\ee
In particular, we find that
\be
\label{der-orbi}
{\partial \CF_0 \over \partial \lambda_i}= -\zeta_i =-\sum_{j=1}^{g_\Sigma}{C_{ij}^{-1} \over 2 \pi} \left( t_j  +\sum_{k=1}^{r_\Sigma}\alpha_{jk}\log m_k\right), \qquad i=1, \cdots, g_\Sigma,
\ee
where $C^{-1}$ denotes  the inverse of the truncated matrix \eqref{tmc}.
The higher genus corrections can be computed systematically. In view of \cite{abk}, their description is very simple. The integral (\ref{multi-Airy}) implements a 
symplectic transformation from the large radius frame, to a particular frame which we will call the {\it maximal conifold frame}. 
As in \cite{mz}, the 't Hooft coordinates $\lambda_i$ are flat coordinates in this frame, and the maximal conifold locus is defined by 
\be
\lambda_i=0, \qquad i=1, \cdots, g_\Sigma. 
\ee
This locus has dimension $r_\Sigma$, the number of mass parameters of the toric CY. In case there are no mass parameters, as in the example of the resolved $\IC^3/\IZ_5$ orbifold 
considered in this paper, the maximal conifold locus is in fact a point, and we will refer to it sometimes as the maximal conifold point. 
It follows that the functions $\CF_g({\boldsymbol{\lambda}})$ appearing in (\ref{logz-exp}) are the topological string genus $g$ free energies in the maximal conifold frame. 
Note that (\ref{saddle}) gives a prediction for the particular combination of periods which vanishes at the maximal conifold locus. As noted in \cite{kmz}, the coefficients 
of the constant trivial period are determined by the coefficients $b_i^{\rm NS}$, i.e. the coefficients of the linear terms in the next-to-leading NS free energy. As far 
as we know, this connection has not been noticed before and is a direct consequence of our conjecture (\ref{our-conj}). 

The main conclusion of this analysis is that, if (\ref{our-conj}) is correct, the fermionic spectral traces $Z_X(\boldsymbol{N}, \hbar)$ 
provide a non-perturbative definition of the genus expansion 
of the topological string (in the maximal conifold frame). This is of course the natural generalization of what was done 
in \cite{mz, kmz} in the case of genus one mirror curves.  We will provide some 
detailed verifications of this statement in the case of the resolved $\IC^3/\IZ_5$ geometry, in the next section.

Finally, let us note that the fermionic spectral traces can be also computed in the so-called {\it M-theory limit}, in which $N_i \gg 1$ but $\hbar$ is fixed. 
In this limit, $Z_X(\boldsymbol{N}, \hbar)$ is given, at leading order, by a multivariable generalization of the Airy function, 
extending in this way the results found in the genus one case in \cite{ghm}. In some cases, this generalization can be written as a product of conventional Airy functions. We will see a detailed example of this in section \ref{gsd-ex}.

\sectiono{Testing the conjecture}

In this section, we will perform a detailed test of the above conjectures in (arguably) the simplest toric geometry with a genus two mirror curve:  
the resolved $\IC^3/\IZ_5$ orbifold studied in the Example \ref{5-orbifold}. 

\subsection{The resolved $\IC^3/\IZ_5$ orbifold }

The toric description of the geometry is encoded in the charge vectors (\ref{5o-cv}). After setting $x_1=x_2=x_4=1$, we have 
\be
\label{z1z2}
z_1= {x_3 \over x_0^3}, \qquad z_2= {x_0 \over x_3^2}. 
\ee
Another useful set of parameters for the moduli space are, 
\be
u= z_1 z_2^3= x_3^{-5}, \quad v=z_1^2 z_2 =x_0^{-5}. 
\ee
This geometry has been discussed in detail in \cite{xenia,mr}, and it has a rich phase structure. The large radius point is, as usual, 
\be
z_1=z_2=0. 
\ee
In addition, there are two {\it half-orbifold points}. The first one is defined by
\be
\label{fop}
x_0=0, \quad u=0, 
\ee
while the second one is defined by 
\be
\label{sop}
x_3=0, \quad v=0. 
\ee
We note that these are the points which are suitable to study the operators $\mO_{3,1}$ and $\mO_{2,2}$, since in each case we are setting to zero 
the perturbation in (\ref{o12-ex}). The corresponding geometries are the canonical bundles over $\IP(1,3,1)$ and $\IP(1,2,2)$, respectively. 
The (full) orbifold point is simply 
\be
x_0=x_3=0. 
\ee
As in the genus one case considered in \cite{ghm}, studying the topological string around this point will make it possible to 
calculate the expansion (\ref{or-exp}) of the generalized spectral 
determinant.

\begin{figure}
\center
\includegraphics[height=5.3cm]{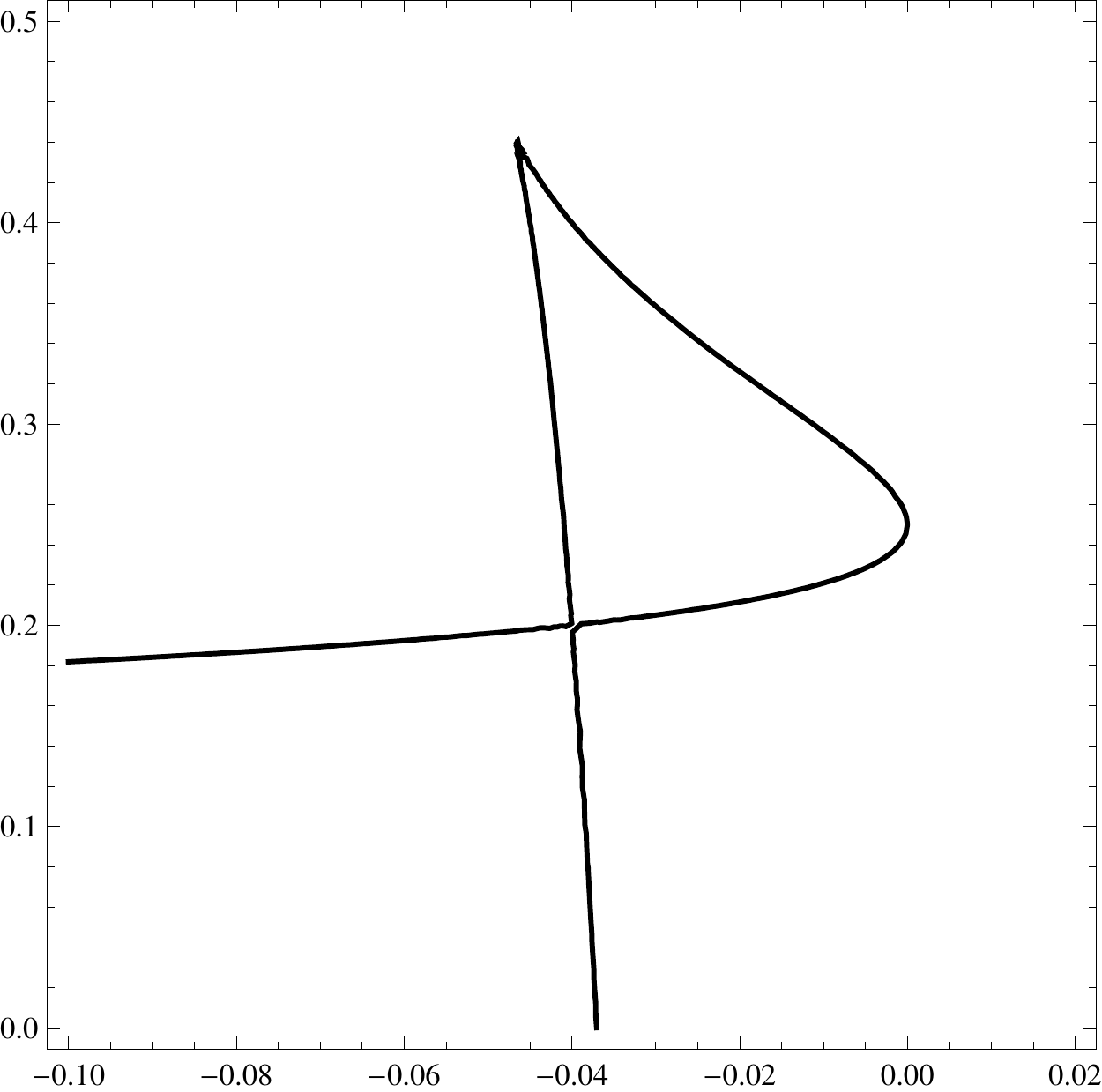} 
\caption{The conifold locus $\Delta(z_1, z_2)=0$ in the $(z_1, z_2)$ plane contains a point (\ref{mcp}) where two components cross transversally.}
\label{coni-locus}
\end{figure}

Another important region in the moduli space of the curve is the conifold locus, where the discriminant
\begin{equation}
\label{discri}\Delta(z_1,z_2)=3125 z_1^2 z_2^3+500 z_1 z_2^2+16 z_2^2-225 z_1 z_2-8 z_2+27 z_1+1,
\end{equation}
vanishes. The real part of this locus has various components, but there is a very special point at 
\be
\label{mcp}
z_1=-{1\over 25}, \qquad z_2={1\over 5}
\ee
where two components of the locus cross transversally (see \figref{coni-locus}). As we will see, this is the maximal conifold point at which the 't Hooft parameters $\lambda_1$, $\lambda_2$ 
vanish. This point controls the 't Hooft limit of the spectral traces, at weak 't Hooft coupling. 

The resolved $\IC^3/\IZ_5$ orbifold can be also realized as a perturbed local $\IP^2$ geometry. This can be easily seen by considering the equation 
(\ref{mc-2}) and performing the transformation $x+y \rightarrow -x$. After reinstating $x_4$ in the equation, and setting $x_1=x_2=x_3=1$, we find that (\ref{mc-2}) reads, 
 \be
 \label{pp2-curve}
 \re^x+ \re^y +\re^{-x-y}  + x_4 \re^{2x} + x_0=0, 
 \ee
 and we have
\be
z_1=x_0^{-3}, \qquad z_2= x_4 x_0. 
\ee
After Weyl quantization, we find the operator 
\be
\mO_{1,1}+ x_4 \re^{2\mx}, 
\ee
which is a perturbation of the operator $\mO_{1,1}$ obtained by quantizing the mirror curve of local $\IP^2$. 

Let us now review some of the topological string amplitudes on this geometry. Near the large radius point there are 
two flat coordinates, $t_1$, $t_2$. They can be expressed in terms of the moduli 
$z_1$, $z_2$ by the mirror map, 
\be
\ba
t_1&=-\Pi_{A_1} (z_1, z_2)=-\log(z_1)+ \CO(z_i), \\ 
t_2&=-\Pi_{A_2} (z_1, z_2)=-\log(z_2)+ \CO(z_i), 
\ea
\ee
where the periods $\Pi_{A_i}$, $\Pi_{B_i}$, $i=1,2$ are given in (\ref{genlr-pers}). By using (\ref{z1-z2}) and taking into account (\ref{tmu}), we conclude that the matrix $C_{ij}$ is given by 
\be
\label{C-mat}
C=\begin{pmatrix} 3 & -1\\ -1 &2 \end{pmatrix}. 
\ee
We will introduce, as usual, the exponentiated variables
\be
\label{QQ}
Q_1=\re^{-t_1}, \qquad Q_2=\re^{-t_2}. 
\ee
The large radius genus zero free energy is defined by the special geometry relations, 
\be
{\partial F_0 \over \partial t_1}= {1\over 10} \Pi_{B_1}, \qquad {\partial F_0 \over \partial t_2}= {1\over 10} \Pi_{B_2}, 
\ee
which leads to (see for example \cite{kpsw})
\be
\label{classf0}
F_0(t_1, t_2) = {1\over 15} t_1^3 +{1\over 10} t_1^2 t_2 +{3\over 10} t_1 t_2^2 +{3\over 10}t_2^3+F_0^{\rm inst}(t_1, t_2), 
\ee
where 
\be
\label{f0-ex}
F_0^{\rm inst}(t_1, t_2)=3 Q_1-2 Q_2-{45 \over 8} Q_1^2 +4 Q_1 Q_2 -
{Q_2^2 \over 4} + \cdots
\ee

The genus one free energies (both standard and refined) have been obtained in \cite{kpsw}. The (standard) genus one free energy is given by 
\be
\label{gone-exact}
F_1(t_1, t_2)=-{1\over 12} \log \left( \Delta z_1^{38/5} z_2^{39/5}\right)-{1\over 2} \log {\rm det}(J_{ij}), 
\ee
where
\be
J_{ij}= {\partial t_i \over \partial z_j}
\ee
is the Jacobian of the mirror map, and $\Delta$ is the discriminant (\ref{discri}). One finds, by explicit expansion, 
\be
\label{f1-ex}
F_1(t_1, t_2)={2 t_1 \over 15} +{3 t_2 \over 20} + {Q_1 \over 4} -{Q_2\over 6}-{3Q_1^2 \over 8} +{Q_1 Q_2 \over 3}-{Q_2^2 \over 12}+ \cdots
\ee
Similarly, one finds the NS refined free energy, 
\be
F_1^{\rm NS}(t_1, t_2)=- {1\over 24} \log \left( \Delta z_1^{-2} z_2^{-3}\right), 
\ee
which has the expansion
\be
\label{f1ns-ex}
F_1^{\rm NS}(t_1, t_2)= -{t_1 \over 12} - {t_2\over 8} -{7 Q_1 \over 8} +{Q_2 \over 6} + {129 Q_1^2 \over 16} -{5 Q_1 Q_2 \over 6}+{Q_2^2 \over 12}+ \cdots. 
\ee
Higher genus free energies, as well as higher $F^{\rm NS}_n(t_1, t_2)$, have been determined in \cite{kpsw} up to order $3$. We will however not use them in this paper.

\subsection{The generalized spectral determinant}
\label{gsd-ex}
The resolved $\IC^3/\IZ_5$ geometry involves two canonical operators, obtained by Weyl's quantization of (\ref{o12-ex}). They read, 
\be
\label{mos}
\ba
\mO_1&= \re^\mx + \re^\my+ \re^{-2 \mx -2 \my} + x_3 \re^{-\mx-\my}=\mO_{2,2}+ x_3 \re^{-\mx-\my}, \\
\mO_2&= \re^\mx + \re^\my+ \re^{-3 \mx - \my} + x_0 \re^{-\mx}=\mO_{3,1}+ x_0 \re^{-\mx}. 
\ea
\ee
As we have seen in section (\ref{gsd-sec}), the generalized spectral determinant can be expressed in many ways. A particularly useful representation in this geometry comes from (\ref{strata}), and we find
\be
\ba
\Xi(x_0,x_3; \hbar)&= {\rm det}\left (1+x_0 \left( \mO_{2,2}+ x_3 \mP_{12} \right)^{-1} \right) {\rm det}\left (1+x_3 \mO_{3,1}^{-1} \right) \\
&=  {\rm det}\left (1+ x_3 \left( \mO_{3,1}+ x_0 \mP_{21} \right)^{-1} \right) {\rm det} \left(1+ x_0 \mO_{2,2}^{-1} \right) . 
\ea
\ee
In particular, we have
\be
\ba
 {\rm det}\left (1+x_0 \left( \mO_{2,2}+ x_3 \mP_{12} \right)^{-1} \right)&= {\Xi (x_0, x_3; \hbar) \over \Xi(0,x_3; \hbar)}, \\
 {\rm det}\left (1+ x_3 \left( \mO_{3,1}+ x_0 \mP_{21} \right)^{-1} \right)&= {\Xi (x_0, x_3; \hbar) \over \Xi(x_0,0; \hbar)}. 
 \ea
 \ee
 The defining formula for the generalized spectral determinant is (\ref{gsd}). For convenience, we will choose as our reference operator $\mO_{3,1}$ (i.e. we will choose $j=2$ in (\ref{gsd})).
 The relevant operators are then, 
\be
\label{opes-ro}
\mA_{21}=\rho_{3,1} \mP_{21}, \qquad \mA_{22}= \rho_{3,1}, 
\ee
and we recall that 
\be
\mP_{21}=\re^{-\mx},
\ee
 where $\mx$ is the quantum Heisenberg operator appearing in $\mO_{3,1}$. 
 It is known from \cite{kas-mar} that $\rho_{3,1}$ is of trace class, and it can be easily checked that $\mA_{21}$ is of trace class as well.

We are now ready to write down the total grand potential, as it follows from our conjecture (\ref{our-conj}). The parameters entering the operators are written in terms 
of chemical potentials as, 
\be
x_0= \kappa_1= \re^{\mu_1}, \qquad x_3=\kappa_2=\re^{\mu_2}, 
\ee
and they are related to the complex moduli of the geometry by
\be
\label{zmu-ro}
\log z_1 =  - 3 \mu_1+\mu_2, \qquad 
			\log z_2 =\mu_1 - 2 \mu_2, 
			\ee
as it follows from (\ref{zmu}). The first thing we must know is the value of the appropriate B-field in (\ref{jws}). Since this geometry can be regarded 
as a perturbation of the local $\IP^2$ geometry when $z_2=0$, a natural guess is that 
\be
\label{b-ro}
{\bf B}= (1,0). 
\ee
It can be checked that, for this choice, (\ref{B-prop}) is satisfied\footnote{We would like to thank Albrecht Klemm for 
verifying explicitly that this is indeed the case for the refined BPS invariants of 
this geometry calculated in \cite{kpsw}.}. The insertion of this B-field in the worldsheet instanton piece is equivalent to 
changing the sign of $Q_1$ in the expansions at large radius (but not in the $\log$ terms). For example, for the very first terms, one finds, 
\be
\mJ^{\rm WS}(\mu_1, \mu_2; \hbar)= - \sum_{v=1}^\infty {1\over v} \left( 2 \sin {2 \pi^2 v\over \hbar} \right)^{-2} \left(3 \re^{-2 \pi v t_1/\hbar} +2 \re^{-2 \pi v t_2/\hbar}\right)+\cdots, 
\ee
and the sign in the first exponential (involving $t_1$) is the opposite one to what we had in (\ref{f0-ex}). 

The function $\mJ^{\rm WKB}(\mu_1, \mu_2; \hbar)$ can be computed in many different ways. The leading order terms 
at large $\mu_i$ can be read from (\ref{classf0}), (\ref{f1-ex}) and 
(\ref{f1ns-ex}). The semiclassical limit (\ref{semiJ}) can be checked as in \cite{mp}, by calculating semiclassical traces. 
This calculation is easy to do either when $x_3=0$, or when $x_0=0$. In these cases, the relevant 
operators are simply $\mO_{2,2}$, $\mO_{3,1}$, respectively, and the corresponding 
semiclassical grand potential is easy to calculate (see also \cite{hatsuda}). The classical spectral traces of these operators are 
\be
\ba
Z_\ell^{(0)}\left(\mO_{2,2} \right)&= \int {\rd x \rd y \over 2 \pi} {1\over \left( \re^x+ \re^y + \re^{-2x-2y}  \right)^{\ell}}={1\over 10 \pi} {\Gamma\left( \ell/5 \right) \Gamma\left(2 \ell/5\right)^2 \over \Gamma(\ell)},\\
Z_\ell^{(0)}\left(\mO_{3,1} \right)&= \int {\rd x \rd y \over 2 \pi} {1\over \left( \re^x+ \re^y + \re^{-3x-y}  \right)^{\ell}}={1\over 10 \pi} {\Gamma\left( \ell/5 \right)^2 \Gamma\left(3 \ell/5\right) \over \Gamma(\ell)}. 
\ea
\ee
We then find,  
\be
\ba
\mJ_0(\mu_1, 0)&= -\sum_{\ell=1}^\infty {(-\kappa_1)^\ell \over \ell} Z_\ell^{(0)}\left(\mO_{2,2} \right),\\ 
\mJ_0(0, \mu_2)&= -\sum_{\ell=1}^\infty {(-\kappa_2)^\ell \over \ell} Z_\ell^{(0)}\left(\mO_{3,1} \right).
\ea
\ee
From the point of view of the geometry, these are expansions near the orbifold point. It is easy to verify, 
by using the explicit formulae in section \ref{ap-orbis} of the Appendix, that these expansions are 
indeed reproduced by the r.h.s. of (\ref{semiJ}). Finally, the quantum mirror map entering in the expression (\ref{jtotal}) can be computed systematically, as shown in 
Appendix \ref{res-or-qmm}.  

As we explained above, in the maximally supersymmetric case $\hbar=2\pi$, the spectral determinant can be written down explicitly. The generalized theta function becomes in this 
case a standard Riemann theta function. Indeed, it can be easily checked that 
\be 
\label{otheta} 
\ba
&\Theta(\boldsymbol{\mu}; 2 \pi)\\
&=\sum_{n_1,n_2 \in \IZ}\exp \left[  \ri \pi \left( n_1^2 \tau_{11} +2 n_1 n_2\tau_{12} + n_2^2 \tau_{22} \right) + 2 \pi \ri \left( n_1 \upsilon_1+n_2 \upsilon_2\right) - \ri \pi \left( n_1+{8\over 3}n_2\right)\right], 
\ea
\ee
where the vector ${\boldsymbol{\upsilon}}$ is given in (\ref{ups}) (since we are considering a fixed CY example, we have removed 
the subscript $X$). Let us recall that the Riemann theta function with characteristics $\boldsymbol{\alpha}$, $\boldsymbol{\beta}$ is defined by 
\be \label{thetadef}  \vartheta \left[\begin{array}{cc} \boldsymbol{\alpha}\\ \boldsymbol{\beta}
 \end{array}\right] \left({\boldsymbol{z}}, \tau \right)= \sum_{ {\boldsymbol{n}} \in \IZ^2 }\exp \left[  \ri \pi ~^t ({\boldsymbol{n}}  + {\boldsymbol{\alpha}})  \tau 
({\boldsymbol{n}}  + {\boldsymbol{\alpha}}) +2 \pi \ri ({\boldsymbol{z}}  + {\boldsymbol{\beta}})  \cdot\left( {\boldsymbol{n}} + {\boldsymbol{\alpha}} \right) \right].
\ee
It follows that the generalized spectral determinant, for the maximally supersymmetric case, can be written as 
\be
\label{gsd-ro}
\Xi (\boldsymbol{\mu}; 2 \pi)=\exp\left( \mJ(\boldsymbol{\mu}; 2 \pi)\right)\vartheta \left[\begin{array}{cc} \boldsymbol{0}\\ \boldsymbol{\beta} \end{array}\right] ( {\boldsymbol{\upsilon}}, \tau). 
\ee
where 
\be
 \boldsymbol{\beta} =- \left({1\over 2},{4\over 3} \right) 
 \ee
 and $\mJ(\boldsymbol{\mu}; 2 \pi)$ is given by the specialization of (\ref{jx-ms}) to the resolved $\IC^3/\IZ_5$ orbifold. It is also easy to check from the 
 results in the Appendix \ref{res-or-qmm} that, for $\hbar=2 \pi$, the quantum mirror map becomes the classical mirror map, together with a change of sign 
 $z_1 \rightarrow -z_1$ in the polynomial part. 
 
The expression (\ref{gsd-ro}) embodies our conjecture for the case at hand. We can now test our conjecture by verifying that this formula indeed gives the right 
spectral properties and quantities (in the maximally supersymmetric case). In the rest of this section, we will check the predictions for the spectral traces.

The fermionic spectral traces $Z(N_1, N_2; \hbar)$ can be read off from the expansion of the 
spectral determinant around $\kappa_1=\kappa_2=0$, which in our case corresponds to $x_0=x_3=0$. This 
is the orbifold point. In the maximally supersymmetric case, this can be done by performing an analytic continuation of the various 
quantities involved in (\ref{otheta}) to the orbifold point. 
As in the case of local $\IP^2$ analyzed in \cite{ghm}, it is convenient to change the sign of $x_3$ and 
perform the expansion of a closely related theta function. Note that this 
leads to a change of sign $z_1 \rightarrow -z_1$. This has the effect of restoring the 
conventional sign of the standard topological string amplitudes (which 
we had to change, due to the B-field (\ref{b-ro})), but also changes the structure of the theta function, 
due to the shifts in the logarithms. After carefully keeping track of all these changes, we find,  
\be 
\label{SpDet}\Xi(x_0,-x_3; 2 \pi)= \re^{\mJ(x_0,x_3; 2 \pi)}  \re^{\ri \pi} \vartheta \left[\begin{array}{cc} \boldsymbol{\alpha} \\  \boldsymbol{\beta}   \end{array}\right] \left(  \boldsymbol{\upsilon},   \tau- S\right),
\ee
where
\be  
\label{sab-orbi} 
S=\left(\begin{array}{cc} 
{1/ 2}& {1/ 2}\\
{1/ 2}&0 \\
\end{array}\right),    \quad \boldsymbol{\alpha} =\left(0, {1\over 2}  \right), \quad   \boldsymbol{\beta} =-\left({3\over 8},{4\over 3} \right), 
\ee
and the quantities $\mJ(x_0,x_3; 2 \pi)$, $\tau$ and $\boldsymbol{\upsilon}$ are given by 
(\ref{jx-ms}), (\ref{tau-xi}) and (\ref{ups}) but they involve now the analytic continuation to the orbifold point of the 
{\it standard} genus zero and one free energies. After implementing this formula, one finds the expansion
\be
\label{xi-orbi}
\Xi (x_0, x_3; 2 \pi)=1+ Z(1, 0; 2\pi) x_0 + Z(0,1; 2\pi) x_3+ Z(1,1; 2\pi) x_0 x_3+ \cdots. 
\ee
The coefficients of this expansion involve derivatives of the Riemann theta function of genus two, but they can be evaluated numerically with high precision. We find, 
\be
\ba
\label{num-traces}
Z(1,0; 2 \pi)&=0.0552786404500042..., \\
Z(0,1; 2 \pi)&=0.0894427190999916...,\\
Z(1,1;2 \pi)&=0.0030770561988687...
\ea
\ee
As we explained in section \ref{gsd-sec}, these coefficients are defined as (generalized) fermionic spectral traces of the operators (\ref{opes-ro}). One has, for example, 
\be
\label{rho-trs}
\ba
Z(1,0; 2 \pi)&= \tr \left(\rho_{3,1} \mP_{21}\right)= \tr \rho_{2,2}, \\
Z(0,1;2 \pi)& = \tr \rho_{3,1}, \\
Z(1,1;2 \pi)&= \tr (\rho_{3,1} \mP_{21} \rho_{3,1}). 
\ea
\ee
 In \cite{kas-mar}, the integral kernels of the operators $\rho_{m,n}$ were obtained in closed form, in terms of the 
quantum dilogarithm. Therefore, the traces (\ref{rho-trs}) can be computed explicitly. Since these results will be 
also used in the analysis near the maximal conifold point, let us briefly summarize them. Let us denote by $ \fad(x)$ Faddeev's quantum dilogarithm \cite{faddeev, fk} 
(we follow the notations in \cite{kas-mar}). We define as well the functions (see also \cite{ak})
\be
\label{mypsi-def}
\mypsi{a}{c}(x)= \frac{\re^{2\pi ax}}{\fad(x-\im(a+c))}. 
\ee
Let $\mq$, $\map$ be operators satisfying the normalized 
Heisenberg commutation relation
\be
[\map, \mq]=(2 \pi \im)^{-1}. 
\ee
They are related to the Heisenberg operators $\mx$, $\my$ appearing in $\mO_{m,n}$ by the following linear canonical transformation: 
\begin{equation}
\mathsf{x}\equiv 2\pi\mathsf{b}\frac{(n+1)\mathsf{p}+n\mathsf{q}}{m+n+1},\quad \mathsf{y}\equiv -2\pi\mathsf{b}\frac{m\mathsf{p}+(m+1)\mathsf{q}}{m+n+1}, 
\end{equation}
so that $\hbar$ is related to $\mb$ by
\be
\label{b-hbar}
\hbar=\frac{2\pi\mathsf{b}^2}{m+n+1}. 
\ee
Then, in the momentum representation associated to $\map$, 
the operator $\rho_{m,n}$ has the integral kernel, 
\begin{equation}
\label{ex-k}
\rho_{m,n}(p,p')=\frac{\overline{\mypsi{a}{c}(p)}\mypsi{a}{c}(p')}{2\mathsf{b}\cosh\left(\pi\frac{p-p'+\im (a+c-nc)}{\mathsf{b}}\right)},
\end{equation}
where $a$, $c$ are given by 
\be
a =\frac{m \mb}{2(m+n+1)}, \qquad c=\frac{\mb}{2(m+n+1)}. 
\ee
By using these results, we can easily compute the kernel of $\mA_{21}$, and one finds
\be
 \langle p|  \rho_{3,1} \mP_{21} |p'\rangle=\re^{- {4 \pi \mb p' \over 5} } \re^{-{2 \pi \mb^2 \im \over 25}}  \rho_{3,1}\left(p, p'+ {\im \mb \over 5}\right).
 \ee
 Therefore, we find the following integral representation
 \be
 \label{cross-terms}
\tr (\rho_{3,1} \mP_{21} \rho_{3,1})=\re^{-{2 \pi \mb^2 \im \over 25}}  \int \re^{- {4 \pi \mb p' \over 5} }   \rho_{3,1}\left(p, p'+ {\im \mb \over 5}\right)\rho_{3,1}\left(p',p\right)\, \rd p\,  \rd p' .
 \ee

In the maximally supersymmetric case, $\hbar=2 \pi$, the spectral theory of these operators also simplifies, as noted already in \cite{km}, 
and one can use the results of \cite{garou-kas} to show that the 
integral kernels above become elementary functions. 
The trace of $\rho_{m,1}$ was computed in \cite{kas-mar} for any $m$ and $\hbar=2 \pi$, and one finds 
\be
\tr \, \rho_{3,1}={1\over 5 {\sqrt{5}}}. 
\ee
A similar computation shows that 
\be
\label{rho22trace}
\tr \, \rho_{2,2}= {1\over 50} \left( 5 - {\sqrt{5}}\right). 
\ee
These agree {\it precisely} with the predictions (\ref{num-traces}) of the spectral determinant (\ref{gsd-ro}). 
A numerical calculation of the double-integral (\ref{cross-terms}) makes it also possible to verify the prediction 
in (\ref{num-traces}) for $Z(1,1;2\pi)$. 

We should note that, although we evaluated the expansion (\ref{xi-orbi}) numerically, its coefficients can be computed analytically in terms 
of derivatives of the Riemann--Siegel theta function. For example, one finds
\be
\label{ft-siegel}
Z(1,0; 2 \pi)=\frac{ 2^{9/5}\ri \sqrt{\frac{\pi }{5}} \Gamma \left(\frac{9}{10}\right)}{\Gamma \left(\frac{1}{5}\right)^2} \left( \Theta_{11}(\boldsymbol{\upsilon}_0,\tau_0)+\frac{1}{2} \Theta_{12}(\boldsymbol{\upsilon}_0, \tau_0)-\Theta_{22}(\boldsymbol{\upsilon}_0,\tau_0)\right),
\ee
where 
\be \Theta_{ij} (\boldsymbol{\upsilon}_0,\tau_0)={\partial_{\tau_{ij}}\vartheta \left[\begin{array}{cc} \boldsymbol{\alpha} \\  \boldsymbol{\beta}   \end{array}\right](\boldsymbol{\upsilon}_0,\tau_0) \over \vartheta \left[\begin{array}{cc} \boldsymbol{\alpha} \\  \boldsymbol{\beta}   \end{array}\right](\boldsymbol{\upsilon}_0,\tau_0) } ,  \quad   \boldsymbol{\alpha} =\left(0, {1\over 2}  \right), \quad   \boldsymbol{\beta} =\left({3\over 8},{4\over 3} \right), 
\ee
and 
\be \boldsymbol{\upsilon}_0=\left(-{11 \over 40},-{1 \over 30}\right),  \quad \tau_0=  \left(
\begin{array}{cc}
 -\frac{1}{2}+\frac{1}{2} \ri \sqrt{\frac{1}{5} \left(5+2 \sqrt{5}\right)} & \quad -\frac{1}{2}+\ri \sqrt{\frac{1}{4}-\frac{1}{2 \sqrt{5}}} \\
 -\frac{1}{2}+\ri \sqrt{\frac{1}{4}-\frac{1}{2 \sqrt{5}}} & \quad  \ri \sqrt{\frac{1}{10} \left(5+\sqrt{5}\right)} \\
\end{array}
\right).\ee
Moreover, by requiring that $Z(0,0; 2\pi)=1$, we find the following identity:
\be \vartheta \left[\begin{array}{cc} \boldsymbol{\alpha} \\  \boldsymbol{\beta}   \end{array}\right](\boldsymbol{\upsilon}_0,\tau_0) =-\frac{5^{7/40} \Gamma \left(\frac{1}{5}\right)^{3/2}}{2^{2/5} \left(5+\sqrt{5}\right)^{3/5} \pi  \sqrt{\Gamma \left(\frac{3}{5}\right)}} ,\ee
which we checked numerically with high precision.
The fact that (\ref{ft-siegel}) agrees with (\ref{rho22trace}) is another manifestation of the highly non-trivial content of our conjecture (\ref{our-conj}). 

There is yet another method to evaluate the spectral traces, which can be also applied away from the maximally supersymmetric case. This method, which goes back to \cite{hmo}, 
is based on the integral formula (\ref{multi-Airy}), and in using directly the large radius expansion of the modified grand potential. In the genus one case, where there is one single integration, 
this leads to an expression for the fermionic spectral traces given by an infinite sum of Airy functions, in which each term is exponentially suppressed with respect to the preceding one. It turns 
out that this method can be generalized to the resolved $\IC^3/\IZ_5$, as follows. The modified grand potential is given by 
\be
\mJ (\mu_1, \mu_2; \hbar)=\mJ^{({\rm p})} (\mu_1, \mu_2; \hbar)+ \mJ^{({\rm np})}(\mu_1, \mu_2; \hbar). 
\ee
Here, the perturbative part is the cubic polynomial in the $\mu_i$s, 
\be
\label{jp2}
\ba
\mJ^{({\rm p})} (\mu_1, \mu_2; \hbar)&={1\over \pi \hbar} \left( \frac{3 \mu _1^3}{4}-\frac{3 \mu _2 \mu _1^2}{4}+\frac{\mu _2^2 \mu _1}{4}+\frac{2 \mu
			_2^3}{3} \right) \\
			& +{1\over 2} \left( {\pi \over \hbar} -{\hbar \over 8 \pi} \right) \mu_1 +{1\over 3} \left( {\pi \over \hbar} -{\hbar \over 4 \pi} \right) \mu_2+A(\hbar), 
			\ea
			\ee
while the non-perturbative part $\mJ^{({\rm np})}(\mu_1, \mu_2; \hbar)$ contains the exponentially small corrections appearing in the expression (\ref{jtotal}), and it is a power 
series in $z_1$, $z_2$. We recall that the complex moduli $z_{1,2}$ are related to the parameters $\mu_{1,2}$ by (\ref{zmu-ro}). 
We can now make a change of variables such that the cubic polynomial appearing in (\ref{jp2}) does not contain mixed terms, 
\be
\mu_1= \nu_1 + {\nu_2 \over 3},  \qquad \mu_2 =\nu_2. 
\ee
We find, 
\be
\mJ^{({\rm p})} (\nu_1, \nu_2; \hbar)=\sum_{i=1}^2 \left( {C_i(\hbar) \over 3} \nu_i^3 + B_i(\hbar) \nu_i \right) + A(\hbar), 
\ee
where
\be
C_1(\hbar) ={9 \over 4 \pi \hbar}, \qquad C_2(\hbar)= {25 \over 12 \pi \hbar}, 
\ee
and
\be
B_1(\hbar)= {1\over 2} \left( {\pi \over \hbar} -{\hbar \over 8 \pi} \right), \qquad B_2(\hbar)={1\over 2} \left( {\pi \over \hbar} -{5\hbar \over 24 \pi} \right). 
\ee
It follows from (\ref{multi-Airy}) that the fermionic spectral traces are given, at leading order, by 
\be
Z(N_1, N_2; \hbar)\approx Z^{({\rm p})}(N_1, N_2; \hbar), 
\ee
where
\be 
\label{zpns}
Z^{({\rm p})}(N_1, N_2; \hbar)=\re^{A(\hbar)} \prod_{i=1}^2 \left(C_i(\hbar)  \right)^{-1/3} {\rm Ai} \left[ \left(C_i(\hbar) \right)^{-1/3}  \left( M_i- B_i(\hbar) \right) \right], 
\ee
and the $M_i$ are defined by the condition, 
\be
\boldsymbol{\mu} \cdot \boldsymbol{N}=\boldsymbol{\nu} \cdot \boldsymbol{M}, 
\ee
so that, in this case,
\be
M_1= N_1, \qquad M_2={N_1 \over 3} + N_2. 
\ee
It is also clear how to incorporate the corrections due to $\mJ^{({\rm np})}(\mu_1, \mu_2; \hbar)$. We can write, 
\be
\re^{\mJ^{({\rm np})}(\mu_1, \mu_2; \hbar)} = \sum_{i,j \geq 0} P_{i,j}(\mu_1,\mu_2; \hbar) z_1^i z_2^j, 
\ee
where the $P_{i,j}\left(\mu_1,\mu_2; \hbar \right)$ are polynomials in $\mu_1,\mu_2$, and $P_{0,0} = 1$. Then, a simple computation shows that
\be
Z(N_1, N_2; \hbar)=\sum_{i,j\ge 0} P_{i,j}\left(-\partial_{N_1},-\partial_{N_2}; \hbar \right) Z^{({\rm p})}\left(N_1+3 i -j, N_2-i+ 2j; \hbar \right). 
\ee
The leading term in this expression is of course given by (\ref{zpns}), while the remaining series gives, for $N_i$ large, exponentially small corrections. As in the case of 
genus one mirror curves, this expansion seems to converge rapidly, and we have verified that, for $\hbar=2 \pi$, it reproduces the spectral traces computed above.  
In addition, we found the following educated guess for the value of $A(2 \pi)$, 
\be
A(2 \pi)= \frac{1}{10} \log \left(\frac{25}{2} \left(5+\sqrt{5}\right)\right)-\frac{3 \zeta (3)}{5 \pi ^2}. 
\ee

The formula (\ref{zpns}) generalizes the results involving Airy functions found in Chern--Simons--matter theories \cite{fhm,mp} and in the case of topological strings 
on local del Pezzo surfaces \cite{ghm}. It has been recently shown in \cite{ayz} that the Airy behavior of the topological string partition function is a universal feature. 
In \cite{ayz}, this behavior (involving a single Airy function) was obtained by considering a one-dimensional slice of the moduli space. It would be interesting 
to see if the argument of \cite{ayz} can be used to derive (\ref{zpns}). Note that, if $g_\Sigma$ is large enough, we cannot put to zero all the crossing terms in the cubic polynomial 
appearing in $\mJ^{({\rm p})} (\boldsymbol{\mu}; \hbar)$, and the leading behavior of the fermionic spectral traces will be given by a generalization of the Airy function which does not 
reduce to a product of elementary Airy functions. 

\subsection{Quantization conditions}

One of the most important results of \cite{ghm} is that, in the case of mirror curves of genus one, the quantization condition for the 
spectrum of the corresponding operator can be 
read from the vanishing of the (deformed) theta function entering in the spectral determinant. As it was already pointed out in \cite{ghm}, there is 
a natural generalization of this conjecture to the higher genus case, by considering 
the vanishing of the higher genus, deformed theta function in (\ref{qtf}). However, the higher genus case is richer (and slightly more 
complicated) due to the fact that there are many operators $\mO_i$, $i=1, \cdots, g_\Sigma$, which one can associate to the same geometry. Let us explain this in some more detail. 

The vanishing of the 
generalized spectral determinant gives a {\it global} quantization condition, which 
defines a discrete family of codimension one submanifolds in moduli space. In many cases, a given point in the vanishing locus solves 
the spectral problem for different (related) operators. For example, the resolved $\IC^3/\IZ_5$ orbifold leads to two different operators (\ref{mos}). 
A point $(x_0, x_3)$ in the vanishing locus of the spectral determinant, with $x_0<0$ and $x_3<0$,
 can be interpreted in two ways: either as an eigenvalue $-x_0$ of the operator $\mO_1$, which depends on $x_3$, or an 
eigenvalue $-x_3$ for the operator $\mO_2$, which depends on $x_0$. This follows from the discussion around (\ref{ji-waves}). However, if the 
point in the vanishing locus occurs at $x_0=0$, it can not be interpreted in terms of $\mO_1$, since this operator is positive-definite and all its eigenvalues 
are strictly positive.

 \begin{center}
 \begin{figure} \begin{center}
 {\includegraphics[scale=0.48]{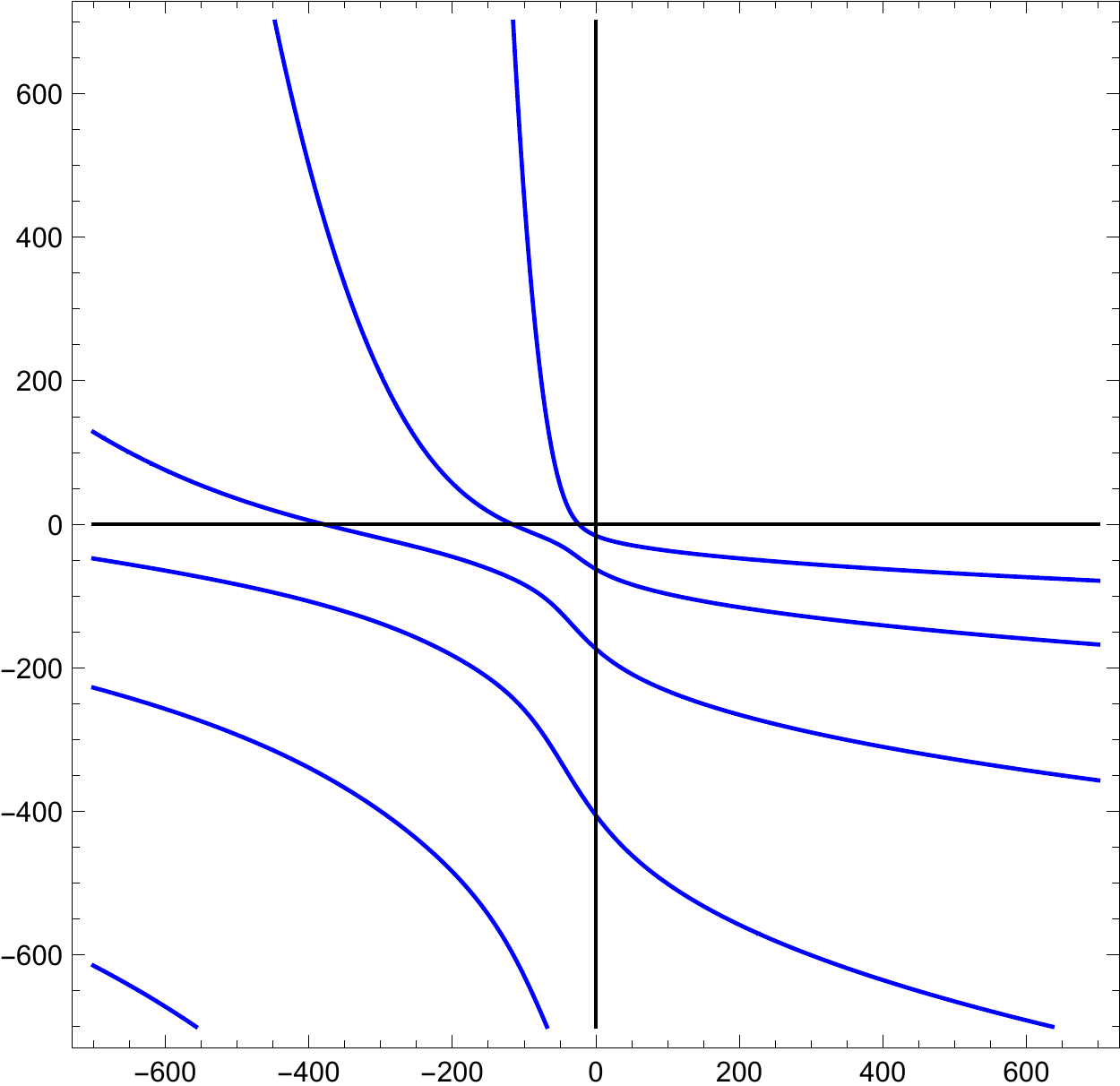}}
\caption{The curves represent the locus in the $(x_0,x_3)$ plane in which the generalized spectral determinant \eqref{gsd-ro} vanishes. They can be labelled by the quantum number $n=0, 1, \cdots$ appearing 
in the generalized Bohr--Sommerfeld quantization condition. The uppermost curve corresponds to $n=0$. }
 \label{bluecurve}
  \end{center}
\end{figure}  
\end{center}

In this section we will obtain the quantization condition for $\IC^3/\IZ_5$, in the maximally supersymmetric case, and 
verify explicitly that it solves many different spectral problems. In particular, we will be able 
to write exact quantization conditions for the unperturbed operators $\mO_{3,1}$ and $\mO_{2,2}$. 
In order to have a first view of the vanishing locus of the generalized spectral determinant \eqref{gsd-ro} in the moduli space 
parametrized by $(x_0, x_3)$,  we can simply plot it by using the expansion (\ref{xi-orbi}) (we assume that $x_0$ and $x_3$ are real). The result is shown in \figref{bluecurve}. 
It consists of a discrete family of curves, and 
each curve crosses both the negative $x_3$ and $x_0$ axis. Note that there are no solutions in which both $x_0$ and $x_3$ are positive. 
The vanishing locus obtained in this way has all the expected properties: the intersection with the axis $x_3=0$ and $x_0=0$ gives the spectrum of the operators $\rho_{2,2}$ and 
$\rho_{3,1}$. The discrete family of curves correspond to the quantum numbers $n=0, 1, \cdots$ of the generalized Bohr--Sommerfeld quantization condition.

  \begin{table}[t] 
\centering
   \begin{tabular}{l l }
  \\
Order& $E_0 $  \\
\hline
 1 &                    2.8953686937107540094\\
 5 &   \underline{3.1640650}172200080194\\
 8&     \underline{3.164065078132119}2069\\
 10 &  \underline{3.1640650781321190565}\\
 \hline
Numerical value &
                         $3.1640650781321190565$   \\
\end{tabular}
\caption{ The ground state energy $E_0$ for the operator $\rho_{2,2}$, as obtained from the vanishing locus of the spectral determinant $\Xi(x_0, 0; 2 \pi)$. This is a power 
series around $x_0=0$, and to obtain the energy we truncate it at a given order in $x_0$. 
As we keep more and more terms in the series, we quickly approach the ground state energy obtained 
by numerical methods.}
 \label{Tabo22}
\end{table}

As a first test that this vanishing locus produces the actual spectrum, we can compute the ground 
state energy for the operators $\rho_{2,2}$ and $\rho_{1,1}$, by using the diagonalization 
method of \cite{hw}, and compare it with the zeros of the spectral determinant $\Xi(x_0, 0; 2 \pi)$ and $\Xi(0, x_3; 2 \pi)$, 
as we keep more and more terms in their polynomial expansion. We recall that, if these 
functions vanish at $(x_0, 0)$ and $(0,x_3)$, respectively, the energies are given by
\be
E=\log\left(-x_0\right), \qquad E=\log\left(-x_3\right). 
\ee
As we see in the tables \ref{Tabo22} 
and \ref{Tabo31}, the answer obtained from the spectral determinant converges rapidly to the correct value.

  \begin{table}[t] 
\centering
   \begin{tabular}{l l }
  \\
Order& $E_0 $  \\
\hline
 1   & \underline{2}.4141568686511505619\\
 5 &   \underline{2.77000}28996745256210\\
 8&    \underline{2.7700040488404}954468\\
 10 & \underline{2.7700040488404460337}\\
 \hline
Numerical value &$2.7700040488404460337$   \\
\end{tabular}
\caption{ The ground state energy $E_0$ for the operator $\rho_{3,1}$, as obtained from the vanishing locus 
of the spectral determinant $\Xi(0, x_3; 2 \pi)$. We follow the same procedure as 
in Table \ref{Tabo22}.}
 \label{Tabo31}
\end{table}

The expansion (\ref{xi-orbi}) around the orbifold point is very convenient for small energies, but it does not make contact 
with the WKB expansion for the operators $\rho_{2,2}$ and $\rho_{3,1}$. 
We can however obtain alternative formulations of the exact quantization condition for these operators by using expansions 
appropriate for the half-orbifold points. Let us first consider 
the operator $\rho_{2,2}$. In principle, the zeroes of the spectral determinant occur at negative values of $x_0$ and $x_3$, 
but it is convenient to change their signs so that they occur along the positive real axis. 
In the case of $\rho_{2,2}$, we change the sign of $x_0$, which involves changing the sign of both $z_1$ and $z_2$. 
The quantization condition is given by the vanishing of the theta function 
\be 
\label{theta22}
\Theta_{2,2}(E)= \vartheta \left[\begin{array}{cc} \boldsymbol{\alpha} \\  \boldsymbol{\beta}   \end{array}\right] \left(  \boldsymbol{\upsilon},   \tau-S\right), 
\ee
where
\be \boldsymbol{\alpha} =\left({1 \over 2},  0 \right), \quad   \boldsymbol{\beta} =\left(-{3 \over 8}, {7 \over 24} \right), \qquad S=\begin{pmatrix} 1/2& 1/2\\
1/2 & 1/2 \end{pmatrix}. 
\ee
In this theta function, $\boldsymbol{\upsilon}$, $\tau$ are computed by using the analytic continuations (\ref{ABhor2}), we set $x_3=0$, and $E=-\log(Y)$. 

In the case of $\rho_{3,1}$, we change the sign of $x_3$, which involves changing the sign of $z_1$. We already did this 
in the calculation near the full orbifold point, 
and we find that the quantization condition is given by the vanishing of the theta function, 
\be
\label{theta31}
\Theta_{3,1}(E)= \vartheta \left[\begin{array}{cc} \boldsymbol{\alpha} \\  \boldsymbol{\beta}   \end{array}\right] \left( \boldsymbol{\upsilon},   \tau-S \right), 
\ee
where $\boldsymbol{\alpha} $, $\boldsymbol{\beta}$ and $S$ are given in (\ref{sab-orbi}),  $\boldsymbol{\upsilon}$, $\tau$ are 
computed by using the analytic continuations (\ref{Ahor}), (\ref{Bhor}), 
we set $x_0=0$, and $E=-\log(X)$. 

It is interesting to see in some detail how the above quantization conditions agree, in the limit of large energies, with the semiclassical result. Let us consider for example 
the operator $\rho_{3,1}$. The semiclassical quantization condition can be obtained by using for example Fermi gas technology. 
The grand potential for the operator $\rho_{3,1}$ at large $\mu$ is given by 
\be
\CJ(\mu, \hbar)\approx  {25 \over 36 \pi \hbar} \mu^3 + \left( {\pi \over 2 \hbar} -{5 \hbar \over 48 \pi}\right) \mu, \qquad \mu \gg 1. 
\ee
This follows from formulae (5.8) and (B.2) of \cite{hatsuda}. Using the general results of \cite{mp}, one finds that the quantization condition at large $E$ is given by 
\be
\label{bohr-s}
{\rm vol}(E) = 2\pi \hbar\left( n+{1\over 2}\right), 
\ee
where
\be
\label{ap-vol}
{\rm vol}(E) \approx {25 \over 6} E^2-{7 \pi^2 \over 18} -{5 \hbar^2 \over 24 \pi},  \qquad E \gg 1.
\ee
This includes the first order correction in $\hbar^2$. How can this be obtained from $\Theta_{3,1}(E)$? First, we have to 
understand the structure of the various functions involved in the higher 
genus theta function. One finds, from the formulae in Appendix \ref{ap-orbis}, 
\be
\ba
\tau_{11}&= \frac{\ri \sqrt{3}}{2}+\mathcal{O}(X^{10/3}),\\
  \tau_{12}&= -\frac{\ri}{2 \sqrt{3}}+\mathcal{O}(X^{5/3}), \\
    \tau_{22}&=-{25 \ri \over 6 \pi} \log(X)+ {\ri \over 6 {\sqrt{3}}}+\mathcal{O}(X^{5/3}),
  \ea\ee
 as well as
  \be \ba 
\upsilon_1&= \frac{5}{24}+\mathcal{O}(X^{5/3}),\\
 \upsilon_2&=\frac{25 \log ^2(X)}{24 \pi ^2}-\frac{1}{36} +\mathcal{O}(X^{5/3}).
 \ea \ee
The terms involving positive powers of $X$ are exponentially small corrections. We would like now to obtain the 
vanishing condition for $\Theta_{3,1}(E)$ at leading order, neglecting these small 
corrections. It is easy to see, from the above behaviors, that the leading contribution comes from the terms with $n_2=0, -1$ 
in the theta function. More precisely, one finds the vanishing condition 
\be
\label{cosv2}
\cos\left( \pi \left(- \upsilon_2+1/3\right)+ \phi  \right)=0. 
\ee
Here, $\phi$ is the argument of the (genus one) Jacobi theta function 
\be
\label{gone-theta}
\vartheta \left[\begin{array}{cc} 0 \\  0  \end{array}\right] \left( {5 \over 12}- {\ri \over 4 {\sqrt{3}}},   {1\over 2}+ {\ri {\sqrt{3}} \over 2} \right). 
\ee
Numerically, we have verified that 
\be
\phi=-{\pi \over 18}. 
\ee
Therefore, we find that the quantization condition, at leading order, is 
\be
\cos \left( \pi \left( {11\over 36} -{25  E^2 \over 24 \pi^2} \right) \right)=0, 
\ee
which is precisely what one obtains from (\ref{ap-vol}) and (\ref{bohr-s}) when $\hbar=2 \pi$. We find it remarkable that the 
argument of the theta function (\ref{gone-theta}) is rational and has the right value 
to reproduce the next-to-leading WKB quantization condition. Of course, one can check explicitly that the zeroes of 
the theta function (\ref{theta31}) give the spectrum of $\rho_{3,1}$ for $\hbar=2 \pi$ with very high precision.

 \begin{center}
 \begin{figure} \begin{center}
 {\includegraphics[scale=0.4]{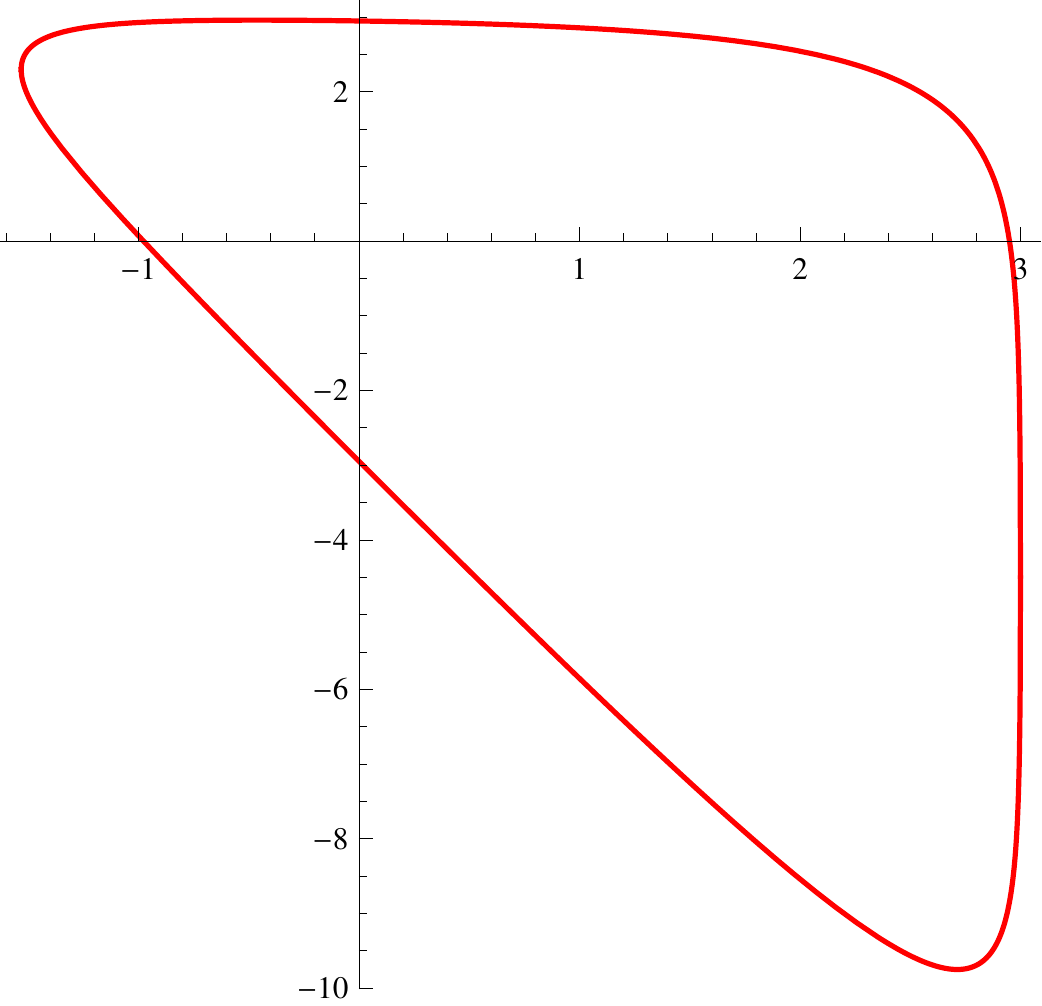}} \qquad \qquad  {\includegraphics[scale=0.3]{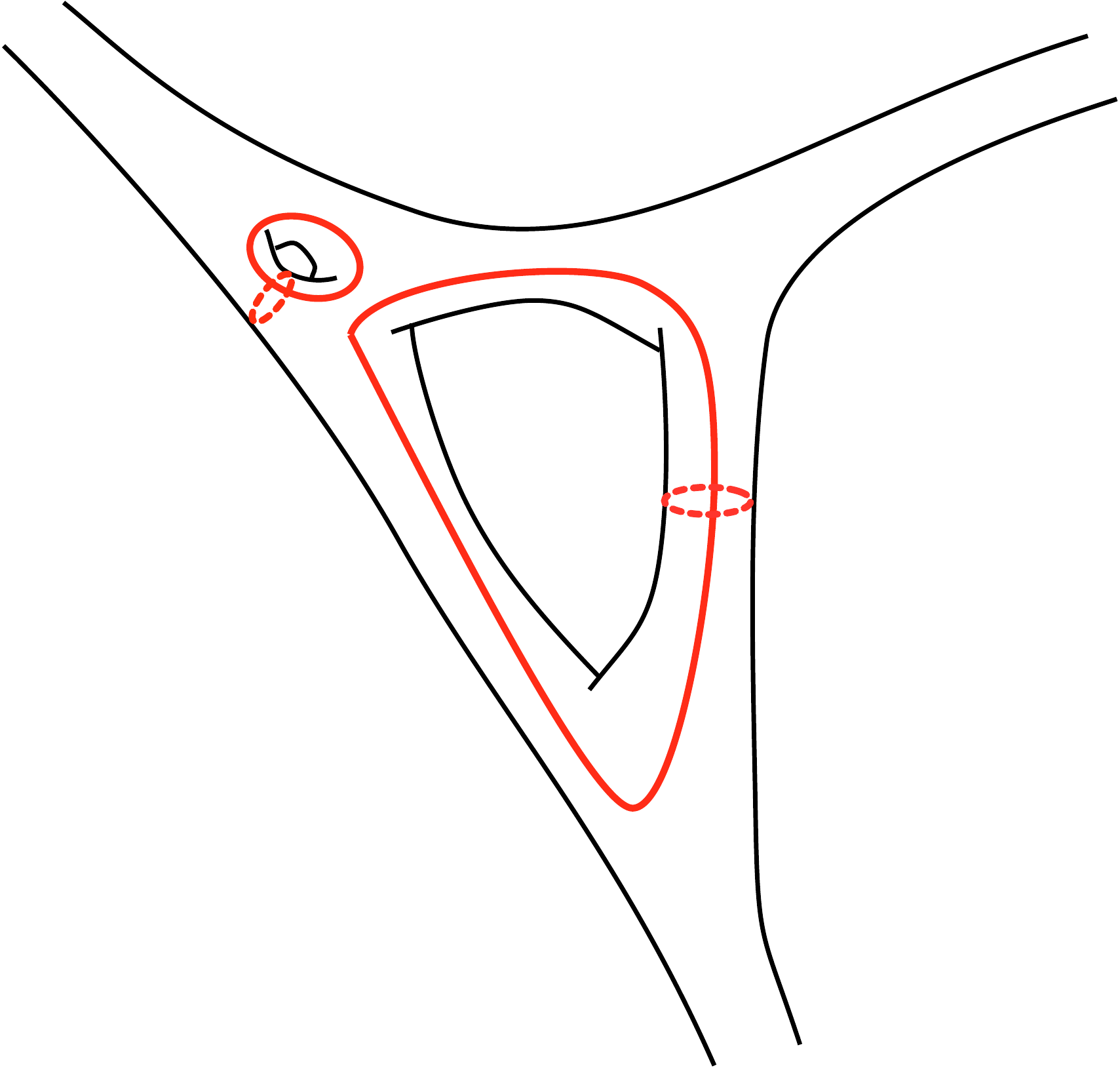}}
\caption{On the right hand side, we show the boundary of the region (\ref{cre}). This is the B-cycle which leads to the perturbative WKB quantization condition. However, 
when we consider the full complexified genus two curve, we find other cycles which correspond to complex instantons and also contribute to the quantization condition.}
 \label{hg-curve}
  \end{center}
\end{figure}  
\end{center}

The quantization condition encoded in the theta functions (\ref{theta22}) and (\ref{theta31}) has a nice interpretation in terms of complex instantons. As we have seen in the example 
of (\ref{theta31}), which corresponds to the operator $\rho_{3,1}$ this condition is given, at leading order, by (\ref{cosv2}). The perturbative part of this 
quantization condition involves the combination of B-periods appearing in $\upsilon_2$,
\be
\label{fcb}
- \Pi_{B_1} + 2 \Pi_{B_2}.  
\ee
This follows from (\ref{ups}) and the matrix (\ref{C-mat}). The B-cycle (\ref{fcb}) has a very concrete incarnation as the boundary of the region
\be
\label{cre}
\CR(E)=\{ (x,y) \in\IR^2: \CO_{3,1}(x,y) \le \re^E \}, 
\ee
which is shown in the left hand side of \figref{hg-curve} (for $E=3$). As in \cite{mp,km}, $\upsilon_2$ 
also involves corrections coming from complex instantons associated to the dual A-period. However, there are further subleading corrections involving the other handle 
of the Riemann surface. These are due to complex instantons associated to other combinations of A and B periods. The effects of these instantons are 
encoded in the genus two theta function, 
through the dependence in for example $\upsilon_1$, which involves 
\be
3\Pi_{B_1} - \Pi_{B_2}. 
\ee
This is in agreement with the principle put forward in \cite{bpv}: in a exact WKB analysis, all periods appearing in the complexified Hamiltonian contribute to the 
quantization condition. This is illustrated in the right hand side \figref{hg-curve}, which shows the underlying genus two curve and its ``hidden" cycles. One remarkable 
implication of our conjecture (\ref{our-conj}) is that all these complex instanton effects are encoded in the higher genus theta function (or a deformation thereof, for 
general values of $\hbar$).

The vanishing locus of the spectral determinant contains as well information about the perturbed operators $\mO_1$, $\mO_2$ which are obtained by 
quantizing the functions (\ref{o12-ex}). Let us consider for example the operator $\mO_2$, which is a perturbation of the operator $\mO_{3,1}$. 
Given a value of the perturbation, $x_0\geq 0$, the conjecture predicts the spectrum as follows. We look at the values of $x_3$ such that $\Xi\left(x_0,x_3^{(n)};2 \pi\right)$ 
vanishes. The spectrum of $\mO_2$, for the given value of $x_0$, is then
\be 
\label{te31}
\left\{-\re^{E_n} \right\}_{n=0, 1, \cdots}=\left\{ x_3^{(n)}: \, \Xi\left(x_0,x_3^{(n)};2 \pi\right)=0 \right\}.
\ee
Graphically, these values are obtained by taking the intesection of the curves in \figref{bluecurve} with the vertical line $x_0=$ constant. The predictions can be compared by 
the spectrum obtained by numerical diagonalization. We find an excellent agreement, as we show for $x_0=20$ in Table \ref{Tabo31new}. 
Of course, completely similar considerations apply to the perturbed operator $\mO_1$. 

 \begin{table}[t] 
\centering
   \begin{tabular}{l l l}
  \\
Order& $E_0 $  & $E_1$\\
\hline
 4 &   \underline{3.122}3827669081676 & \underline{4.2}33804854297745\\ 
 6 &   \underline{3.122038}8008498759 & \underline{4.286}273969753037\\ 
 9&    \underline{3.12203875419326}48& \underline{4.286366387}547196\\ 
 12 & \underline{3.1220387541932659} & \underline{4.286366387477153}\\
 \hline
Numerical value &
                         $3.1220387541932659$  & $4.286366387477153 $ \\
\end{tabular}
\\
\caption{ The ground state and first excited energies, $E_0$ and $E_1$, for the perturbed $\rho_{3,1}$ operator, 
as obtained from the vanishing locus of the spectral determinant $\Xi(20, x_3; 2 \pi)$. This is a power 
series around $x_3=0$, and to obtain the energies we truncate it at a given order in $x_3$. As we keep more and more terms in the series, 
we quickly approach the  energy obtained 
by numerical methods.}
 \label{Tabo31new}
\end{table}

 \begin{center}
 \begin{figure} 
 \begin{center}
 {\includegraphics[scale=0.31]{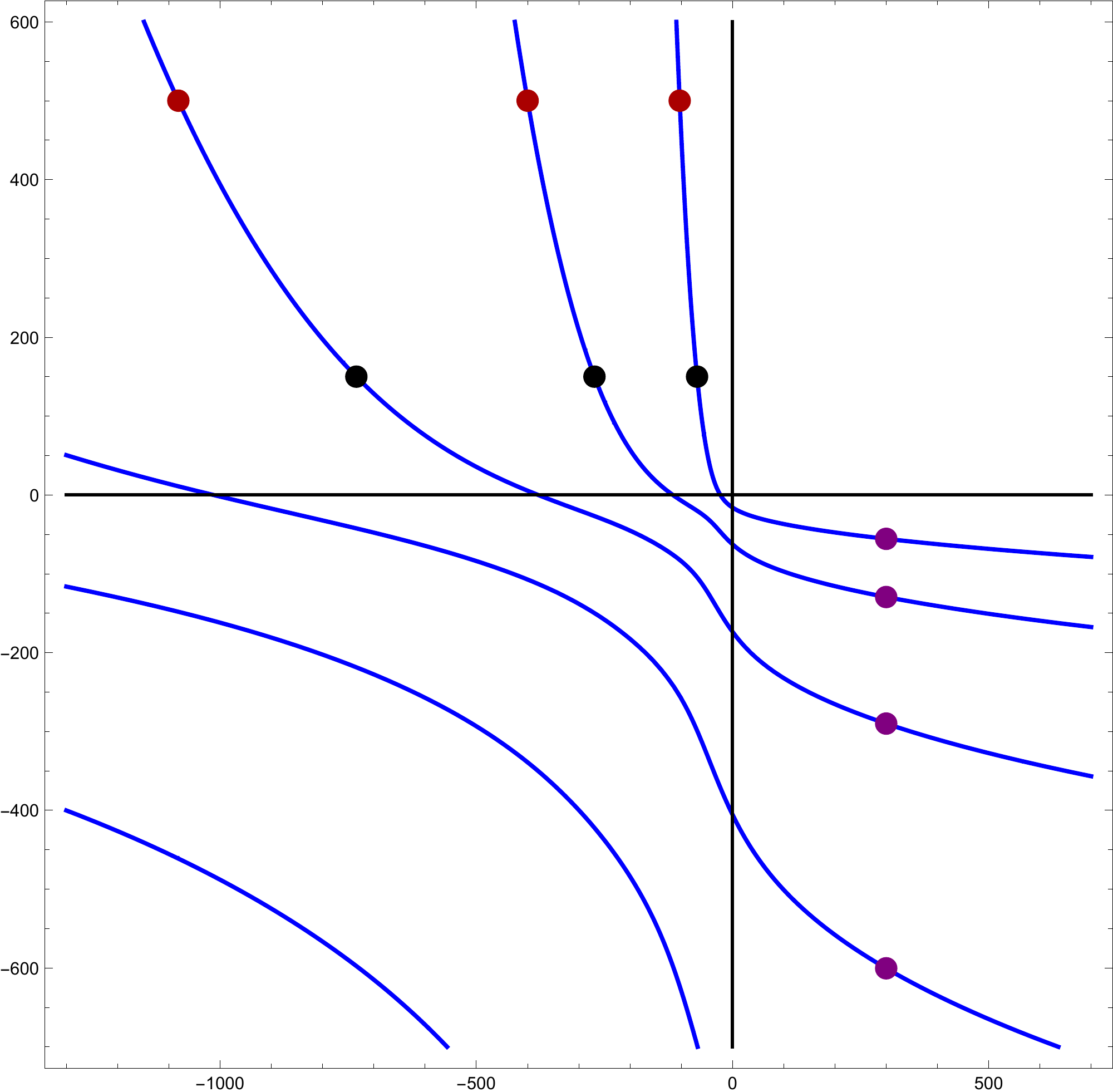}}
\caption{The blue lines represent the locus in the $(x_0,x_3)$ plane where the spectral determinant \eqref{SpDet} vanishes. The horizontal group of dots in red in the 
second quadrant represents the points $(x^{(n)}_0,x_3)=(-\re^{E_n},500)$, $n=0,1,2$, which give the spectrum of the operator $\mO_{1}$ for the value $x_3=500$. 
The vertical group of dots in purple, in the fourth quadrant, represents the points 
  $(x_0,x_3^{(n)})=(300,-\re^{E_n})$, $n=0,1,2,3$, giving the spectrum of the operator $\mO_2$ with $x_0=300$. Finally, the horizontal group of dots in black, in the second 
  quadrant, represents the points
  $(x^{(n)}_0,x_3)=(-\re^{E_n} 6,6^3)$, which encode the spectrum of the perturbed 
$\IP^2$ operator \eqref{pp2-operator} with $x_4=6^{-5}$. }
 \label{zerosb}
  \end{center}
\end{figure}  \end{center}

The vanishing locus of the spectral determinant determines also the spectrum of the operator  
\be
\label{pp2-operator}
\mO_{1,1}+ x_4 \re^{2 \mx}, 
\ee
which is obtained by quantization of the mirror curve in the form (\ref{pp2-curve}). This is a perturbation of the operator $\mO_{1,1}$, which is obtained by quantizing 
the mirror curve to local $\IP^2$. To determine the spectrum, we proceed as follows. Given a value of the perturbation $x_4$, we have a corresponding value of $x_3$ given by
\be  
x_3=x_4^{-3/5}.
 \ee 
This automatically determines an infinite, discrete series of (negative) values of $x_0$, $x_0^{(n)}$, $n=0, 1, \cdots$, in the vanishing locus of the spectral determinant. 
Then the energy levels of the operator (\ref{pp2-operator}) are determined by
\be 
\label{tep2}
-\re^{E_n}=  x_4^{1/5} x^{(n)}_0, \qquad n=0, 1, \cdots
\ee
We find again an excellent agreement between the numerical spectrum, as obtained by diagonalization of (\ref{pp2-operator}), and the one predicted by \eqref{tep2}.

In \figref{zerosb} we illustrate these considerations for different cases. The dots indicate the spectrum as computed numerically, 
by diagonalization of the operators. The vertical group of dots in the fourth quadrant 
correspond to a perturbed operator $\mO_2$ with $x_0=300$. The horizontal group of dots at the top of the second quadrant 
corresponds to a perturbation of the operator $\mO_1$ with $x_3=500$. Finally, the 
horizontal group of dots at the bottom of the second quadrant corresponds to the perturbed operator $\mO_{1,1}$ with $x_4=6^{-5}$. In all cases, 
we find perfect agreement between the numerical results 
and the prediction from the vanishing locus. 

\begin{center}
 \begin{figure} \begin{center}
 {\includegraphics[scale=0.28]{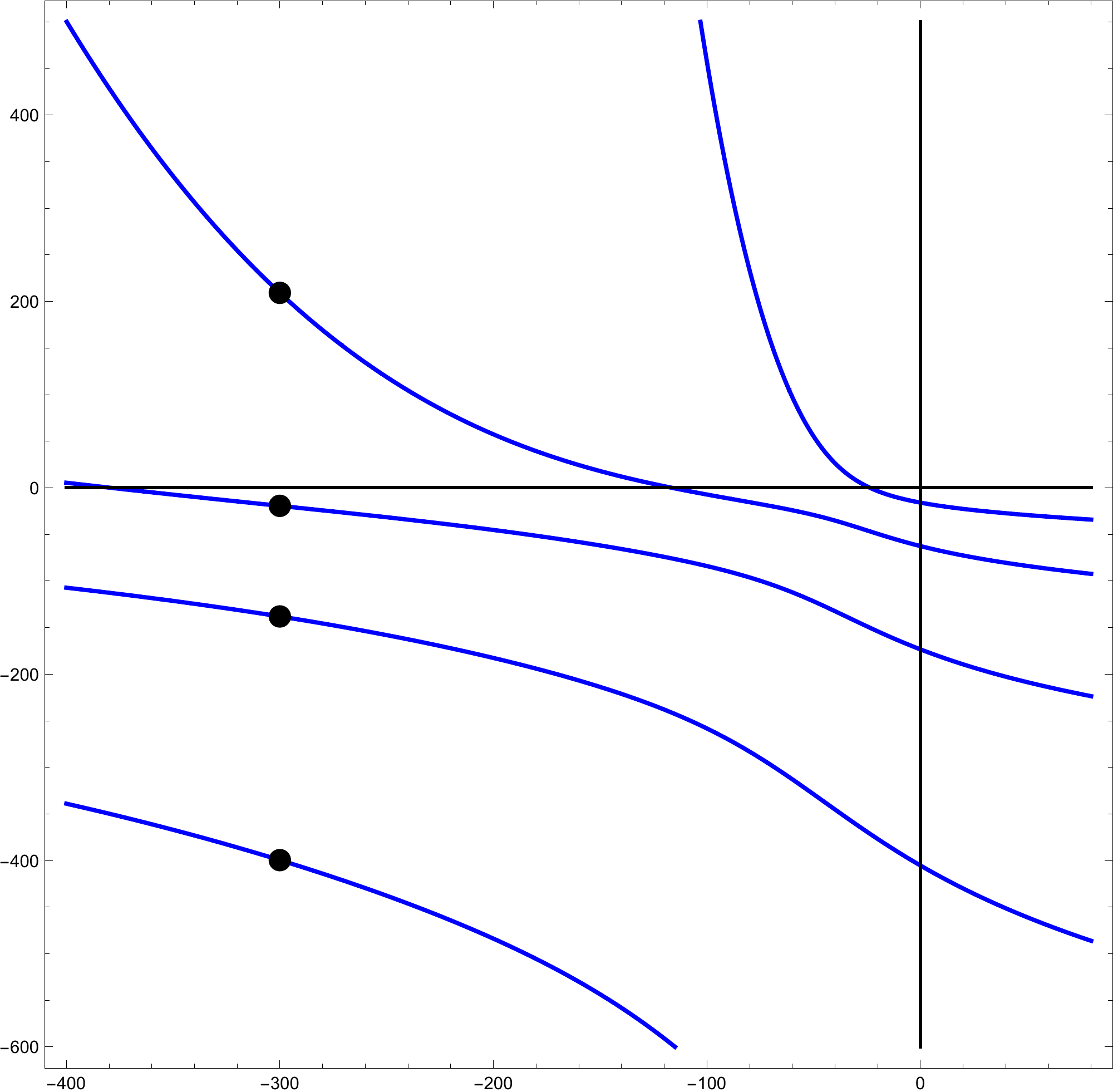}}
\caption{ The dots represent the points $(x_0,x^{(n)}_3)=(-300,-\re^{E_n})$, $n=1, 2, 3, 4$, giving the spectrum of the operator $\mO_2$ with a negative value of $x_0=-300$.}
 \label{xonegativo}
  \end{center}
\end{figure}  \end{center}

It turns out that one can also consider negative values of the perturbations. For example, one can consider $x_0<0$ for the operator $\mO_2$ in (\ref{mos}). The 
generalized spectral determinant predicts that, in this case, 
the values of $x_3^{(n)}=-\re^{E_n}$ for the 
first eigenstates will be positive, while the remaining values will be negative. 
This is easy to understand from the explicit expression in (\ref{mos}): 
the operator $\mO_{3,1}$ gives positive contributions to the exponentiated energy, while the perturbation gives a negative contribution. 
For the low-lying eigenstates, the perturbation takes over, while for the higher, excited states, the operator $\mO_{3,1}$ takes over. 
An example of such a situation is shown in \figref{xonegativo}, for $x_0=-300$. 
Again, the predictions are in perfect agreement with the numerical results. 

\subsection{The large $N$ limit of spectral traces}

As we explained in section \ref{NP}, the generalized spectral determinant provides a non-perturbative completion 
of the conventional topological string free energy. The genus expansion of the 
topological string, in the frame associated to the maximal conifold locus, appears as an asymptotic expansion of the 
fermionic spectral traces $Z_X (\boldsymbol{N}, \hbar)$. This is however a non-trivial 
statement, since it is based on the conjecture that the non-perturbative corrections to the spectral problem are encoded in 
the conventional topological string. It was pointed out in \cite{mz,kmz} that this 
statement can be however checked if one can expand the spectral traces in the strong coupling limit $\hbar \rightarrow \infty$. 
One can then compare this expansion with the predictions of the topological 
string. We will now perform such a comparison. 

Let us first calculate the asymptotic expansion (\ref{logz-exp}) directly on the operator side. In our case, this reads
\be
\log Z(N_1, N_2; \hbar)= \hbar^2 \CF_0(\lambda_1, \lambda_2) + \CF_1(\lambda_1, \lambda_2)+ \cdots
\ee
We note that, when $N_2=0$ or $N_1=0$, the l.h.s. reduces to the fermionic spectral trace of the operators $\rho_{2,2}$ or $\rho_{3,1}$, respectively. It follows that
\be
\label{fg-both}
\CF_g(\lambda_1, \lambda_2)= \CF_g^{(2,2)}(\lambda_1) + \CF_g^{(3,1)}(\lambda_2)+ \CO\left( \lambda_1 \lambda_2 \right).  
\ee
The expansions of $\CF_g^{(2,2)}(\lambda)$ and $\CF_g^{(3,1)}(\lambda)$ near $\lambda=0$ were worked out in \cite{mz}, 
for small $g$, and directly from the spectral theory. One finds, for the leading terms, 
\be
\label{g-zero}
\ba
\CF_0^{(2,2)}(\lambda)&= {\lambda^2 \over 2} \left( \log\left( \lambda \sigma_{1} \right) -{3\over 2} \right) - c_1 \lambda-\frac{1}{75} \sqrt{65-22 \sqrt{5}} \pi ^2 \lambda ^3-\frac{\left(174 \sqrt{5}-425\right) \pi ^4 \lambda ^4}{11250}\\
&+\frac{4 \left(145-59 \sqrt{5}\right) \sqrt{5-2 \sqrt{5}} \pi ^6 \lambda^5}{46875}+ \CO(\lambda^6), \\
\CF_0^{(3,1)}(\lambda)&= {\lambda^2 \over 2} \left( \log\left( \lambda \sigma_{2} \right) -{3\over 2} \right) - c_2 \lambda-\frac{1}{75}\sqrt{65+22 \sqrt{5}} \pi ^2 \lambda ^3+\frac{\left(425+174 \sqrt{5}\right) \pi ^4 \lambda ^4}{11250}\\
   &-\frac{4 \sqrt{14530+\frac{32482}{\sqrt{5}}} \pi ^6 \lambda ^5}{9375}+ \CO(\lambda^6). 
\ea
\ee
In these equations, 
\be
\sigma_1= \frac{2}{25} \sqrt{10-2 \sqrt{5}} \pi ^2, \qquad \sigma_2=\frac{2}{25} \sqrt{10+2 \sqrt{5}} \pi ^2. 
\ee
The coefficients $c_{1,2}$ can be expressed in terms of the Bloch--Wigner function, 
\be
D_2(z)= {\rm Im}\left( {\rm Li}_2 (z) \right) + \log |z| {\rm arg} (1-z), 
\ee
where arg denotes the branch of the argument between $-\pi$ and $\pi$. We have, 
\be
\label{c-dilogs}
c_1= {5 \over 2 \pi^2} D_2\left( \re^{2 \pi \ri \over 5} {1+ {\sqrt{5}} \over 2} \right), \qquad
c_2={5 \over 2 \pi^2} D_2\left( \re^{ \pi \ri \over 5} {1+ {\sqrt{5}} \over 2} \right). 
\ee
For the next-to-leading function, one finds, 
\be
\label{g-one}
\ba
\CF_1^{(2,2)}(\lambda)&=-{1\over 12} \log \left(\lambda \hbar\right)+\zeta'(-1)+ \frac{\sqrt{725-178 \sqrt{5}} \pi ^2}{150} \; \lambda+\frac{\left(174 \sqrt{5}-425\right) \pi ^4}{11250} \; \lambda^2
			\\
			&-\frac{4\sqrt{112450-\frac{249538}{\sqrt{5}}} \pi ^6}{28125} \; \lambda^3+ \CO(\lambda^4) , \\
\CF_1^{(3,1)}(\lambda)&= -{1\over 12} \log \left(\lambda \hbar\right)+\zeta'(-1)+\frac{\sqrt{725+178 \sqrt{5}} \pi ^2}{150} \; \lambda-\frac{\left(425+174 \sqrt{5}\right) \pi ^4 }{11250} \; \lambda ^2
		\\
		&+\frac{4\sqrt{112450+\frac{249538}{\sqrt{5}}} \pi ^6}{28125} \; \lambda ^3+ \CO(\lambda^4). 
\ea
\ee

The results above do not determine the crossing terms. To obtain these, we have to calculate fermionic traces with both $N_1$, $N_2$ different from zero, and 
expand them at large $\hbar$. These expansions can be obtained, as in \cite{mz,kmz}, by writing 
integral expressions for the traces and expanding them around the Gaussian point. For example, for the calculation of $Z(1,1; \hbar)$ we need the 
integral expression (\ref{cross-terms}), and we obtain
\be
\label{first-crossed}
\ba
\log Z(1,1;\hbar)&={1\over 2} \log \left[ \frac{2 \left(7 \sqrt{5}-15\right) \pi ^2}{625 \hbar^2} \right] -c_1- c_2 + 
\frac{\sqrt{\frac{1}{2} \left(1205+31 \sqrt{5}\right)} \pi ^2}{75 \hbar}\\
&-\frac{8   \left(\left(5+11 \sqrt{5}\right) \pi ^4\right)}{625 \hbar^2}+\cdots. 
\ea
\ee
We have examined the very first terms in the large $\hbar$ expansion of $Z(1,1; \hbar)$, $Z(2,1; \hbar)$ and $Z(1,2; \hbar)$, which allows us to  
determine the coefficients of the cross-terms $\lambda_1 \lambda_2$, $\lambda_1^2 \lambda_2$, $\lambda_1 \lambda_2^2$ in $\CF_0(\lambda_1, \lambda_2)$. In this way we find, 
 \be
 \ba
 \CF_0(\lambda_1, \lambda_2)&=\CF^{(2,2)}_0(\lambda_1) + \CF^{(3,1)}_0(\lambda_2)+ \alpha _{12} \lambda _1 \lambda _2\\
 & +\frac{4}{25}
		\sqrt{5+2 \sqrt{5}} \pi ^2 \lambda _1 \lambda _2^2+\frac{4}{25} \sqrt{5-2 \sqrt{5}} \pi ^2
		\lambda _1^2 \lambda _2+\cdots
		\ea
		\ee
		where
\be
\label{al-12}
\alpha_{12}= -\log\left[ {3+ {\sqrt{5}} \over 2} \right].
\ee
Note that, in comparing an expansion at small $N_1$, $N_2$ like (\ref{first-crossed}) to (\ref{g-zero}), (\ref{g-one}), we cannot use the 
asymptotic expansion of the Barnes functions $G_2(N_1+1)$, $G_2(N_2+1)$, which give the very first terms in (\ref{g-zero}), (\ref{g-one}). 
Rather, we have to subtract these terms from the asymptotic expansion, and replace them by the exact values of the Barnes functions, 
similarly to what was done in \cite{msw} in a related context. 
 
 We now want to compare these results with the predictions of (\ref{multi-Airy}). According to (\ref{saddle}), the 't Hooft parameters are given by 
 \be
 \label{lam-van}
 \ba
 \lambda_1&={1\over 8 \pi^3} \left( 3 {\partial  \widehat F_0 \over \partial t_1}-{\partial  \widehat F_0 \over \partial t_2} -{\pi^2 \over 2} \right), \\
  \lambda_2&={1\over 8 \pi^3} \left( - {\partial  \widehat F_0 \over \partial t_1}+2 {\partial  \widehat F_0 \over \partial t_2} -{2\pi^2 \over 3} \right). 
  \ea
  \ee
  We recall that the prepotential $\widehat F_0(t_1, t_2)$ appearing here is the standard large radius prepotential of this geometry, but after turning on the B-field (\ref{b-ro}). 
  As in the genus one case, we expect the $\lambda_i$ to be vanishing flat coordinates around the point in the conifold locus 
characterized by two vanishing periods. The natural candidate is 
the maximal conifold point (\ref{mcp}). In Appendix \ref{mcp-periods} we have found flat coordinates $t_{1,2}^c$ around this point by solving the Picard--Fuchs equations. 
Note however that we have to turn on a B-field, which is equivalent to changing $z_1 \rightarrow -z_1$ in the results of that Appendix. We will keep the same notation for the resulting flat coordinates after this change of sign. A detailed numerical 
analysis shows that indeed
\be
t_i^c= r_i \lambda_i,  \qquad i=1, 2, 
\ee
where
\be
r_1=4 \pi^2 {\sqrt{ 1- {2\over {\sqrt{5}}}}}, \qquad r_2=4 \pi^2 {\sqrt{ 1+{2\over {\sqrt{5}}}}}. 
\ee
As we noted in section \ref{NP}, the constants in (\ref{lam-van}) are determined by the coefficients $b_i^{\rm NS}$ 
and the matrix (\ref{C-mat}). It was observed in the Appendix to \cite{xenia} that the combinations 
appearing in (\ref{lam-van}) are precisely vanishing flat coordinates along the two different branches of the 
conifold locus which intersect at the maximal conifold point. Interestingly, we can predict 
these combinations from our main conjecture (\ref{our-conj}), as it has been already noted in \cite{kmz}.

According to (\ref{der-orbi}), the leading term in the expansion (\ref{logz-exp}) is determined by the equations, 
\be
{\partial \CF_0 \over \partial \lambda_1}= {1\over 10 \pi} \Pi^{-}_v, \qquad 
{\partial \CF_0 \over \partial \lambda_2}= {1\over 10 \pi} \Pi^{-}_u,
\ee
where $\Pi^{-}_{v,u}$ are the combinations of A-periods written down in (\ref{piuv}), (\ref{Apers}), but after changing $z_1 \rightarrow -z_1$ in the power series expansion. 
In order to integrate these equations and expand them around $\lambda_1=\lambda_2=0$, so as to make contact with the expansions of the spectral traces, 
we have to consider the analytic continuation of the periods $\Pi^{-}_{v,u}$ 
around the maximal conifold point, i.e. we have to express them as a linear combination of the flat coordinates $\lambda_{1,2}$ and the logarithmic periods $S_{1,2}$ in (\ref{S-pers}). 
This seems to be difficult, analytically. However, our conjecture {\it predicts} that this combination should be 
\be
\label{an-pi}
\ba
{1\over 10 \pi} \Pi_v^{-}&= {1\over r_1} S_1 + \left( \log {\sigma_1 \over r_1} -1\right) \lambda_1 -c_1+ \alpha_{12} \lambda_2, \\
{1\over 10 \pi} \Pi_u^{-}&= {1\over r_2} S_2 + \left( \log {\sigma_2 \over r_2} -1\right) \lambda_2 -c_2+ \alpha_{12} \lambda_1, 
\ea
\ee
where $\alpha_{12}$, given in (\ref{al-12}), is the coefficient of $\lambda_1 \lambda_2$ in $\CF_0(\lambda_1, \lambda_2)$. We have verified (\ref{an-pi}) numerically. In particular, 
a remarkable consequence of (\ref{an-pi}) is the following. Let us write (\ref{Apers}) as 
\be
\ba
 \Pi_{u}(z_1, z_2) &= \log (z_1z_2^3) + \widetilde \Pi_u(z_1, z_2), \\
 \Pi_{v}(z_1, z_2)&=\log (z_1^2 z_2) +\widetilde \Pi_v(z_1, z_2).
\ea
\ee
Then, if we denote the coordinates of the maximal conifold point (\ref{mcp}) as $z_{1,2}^c$, we find, by evaluating (\ref{an-pi}) at $(-z_1^c, z_2^c)$, that
\be
\label{idents}
\ba
-{1\over 25} \left( \log\left| z_1^c (z_2^c)^3\right| + \widetilde \Pi_{u}(z^c_1, z^c_2)  \right)&={1\over \pi} D_2\left( \re^{ \pi \ri \over 5} {1+ {\sqrt{5}} \over 2} \right),\\
-{1\over 25} \left( \log\left| (z_1^c)^2 z_2^c\right| + \widetilde \Pi_{v}(z^c_1, z^c_2) \right)&={1\over \pi} D_2\left( \re^{ 2 \pi \ri \over 5} {1+ {\sqrt{5}} \over 2} \right),  
\ea
\ee
where we took into account the expressions (\ref{c-dilogs}). 
Similar identities, evaluating the A-periods at the conifold point in terms of the dilogarithm function, 
were already predicted by the conjecture of \cite{ghm} in the genus one case, as explained in \cite{mz,kmz}. For elliptic mirror curves, some of these identities 
have been known in the mathematics and physics literature \cite{rv,yang,dk}. In our approach, these identities follow from the presence of 
the {\it quantum} dilogarithm in the integral kernel of the corresponding operators \cite{kas-mar}. As already emphasized in \cite{mz,kmz}, the fact that these identities are true 
is a highly non-trivial test of the spectral theory/mirror symmetry correspondence of \cite{ghm} that we are developing in this paper for the higher genus case. 
In particular, the identities (\ref{idents}), 
which we have verified with high numerical precision, do not seem to be known in the mathematics 
literature\footnote{After the first version of this paper appeared, 
Charles Doran and Matt Kerr proved these identities by using the techniques of \cite{dk}.}. 

Once (\ref{an-pi}) has been established, we can integrate it to obtain, up to a constant, 
\begin{align}
\label{cf0}
		\begin{split}
			\CF_0(\lambda_1, \lambda_2)&={\lambda_1^2 \over 2} \left( \log\left( \lambda_1 \sigma_{1} \right) -{3\over 2} \right) - c_1 \lambda_1+{\lambda_2^2 \over 2} \left( \log\left( \lambda_2 \sigma_{2} \right) -{3\over 2} \right) - c_2 \lambda_2+\alpha_{12} \lambda_1 \lambda_2
			\\
			&-\frac{1}{75} \sqrt{65-22 \sqrt{5}} \pi ^2 \lambda _1^3+\frac{4}{25} \sqrt{5+2
				\sqrt{5}} \pi ^2 \lambda _2^2 \lambda _1+\frac{4}{25} \sqrt{5-2 \sqrt{5}} \pi ^2 \lambda _2
			\lambda _1^2
			\\
			&-\frac{1}{75} \sqrt{65+22 \sqrt{5}} \pi ^2 \lambda _2^3+\frac{\left(425-174
				\sqrt{5}\right) \pi ^4 \lambda _1^4}{11250}-\frac{32 \left(15+7 \sqrt{5}\right) \pi ^4
				\lambda _2^3 \lambda _1}{1875}-
			\\
			&-\frac{8}{125} \pi ^4 \lambda _1^2 \lambda _2^2-\frac{32
				\left(-15+7 \sqrt{5}\right) \pi ^4 \lambda _2 \lambda _1^3}{1875}+\frac{\left(425+174
				\sqrt{5}\right) \pi ^4 \lambda _2^4}{11250}+ \CO(\lambda^5).
				\end{split}
				\end{align}
As in the case of mirror curves of genus one, (\ref{multi-Airy}) predicts that $\CF_1(\lambda_1, \lambda_2)$ is given, up to an additive constant, by the genus one free energy in the maximal conifold 
frame. We can now use (\ref{gone-exact}) and the mirror map near the maximal conifold point to obtain, up to an additive constant, 
\begin{align}
\label{cf1}
\begin{split}
\CF_1(\lambda_1, \lambda_2) &= -\frac{1}{12} \log \left(\lambda _1 \lambda
				_2\right)+\frac{1}{150} \sqrt{725-178 \sqrt{5}} \pi ^2 \lambda _1+\frac{1}{150}
				\sqrt{725+178 \sqrt{5}} \pi ^2 \lambda _2
				\\
				&-\frac{\left(425-174 \sqrt{5}\right) \pi ^4 \lambda
					_1^2}{11250}+\frac{184 \pi ^4 \lambda _1 \lambda _2}{375 \sqrt{5}}-\frac{\left(425+174
					\sqrt{5}\right) \pi ^4 \lambda _2^2}{11250}
				\\
				&-\frac{4 \sqrt{112450-\frac{249538}{\sqrt{5}}}
					\pi ^6 \lambda _1^3}{28125}-\frac{16 \sqrt{35050+\frac{73142}{\sqrt{5}}} \pi ^6 \lambda _2^2
					\lambda _1}{9375}
				\\
				&+\frac{16 \sqrt{35050-\frac{73142}{\sqrt{5}}} \pi ^6 \lambda _2 \lambda
					_1^2}{9375}+\frac{4 \sqrt{112450+\frac{249538}{\sqrt{5}}} \pi ^6 \lambda
					_2^3}{28125}+\CO(\lambda^4). 
					\end{split}
					\end{align}
If we compare the expressions (\ref{cf0}), (\ref{cf1}) with the results obtained from spectral theory, we find complete agreement. 
In particular, the free energies (\ref{g-zero}), (\ref{g-one}) for the matrix models 
associated to the operators $\rho_{3,1}$ and $\rho_{2,2}$ are recovered as topological string free energies 
in the maximal conifold frame, restricted to the two branches $\lambda_1=0$, $\lambda_2=0$ 
of the conifold locus, respectively. In addition, one can check that the cross-terms (\ref{cf0}), (\ref{cf1}) 
reproduce the expansions of the spectral traces with both $N_1$ and $N_2$ different from zero. For example, 
one finds that the term of order $\hbar^{-2}$ in $\log Z(1,1; \hbar)$, which is the term written down in the second line of (\ref{first-crossed}), 
precisely equals the sum of the coefficients of the quartic terms in $\CF_0(\lambda_1, \lambda_2)$, 
plus the sum of the coefficients of the quadratic terms in $\CF_1 (\lambda_1, \lambda_2)$. 

The conclusion of this rather lengthy and detailed analysis is that the spectral theory associated to the 
mirror curve of the resolved $\IC^3/\IZ_5$ orbifold provides a non-perturbative description of 
topological strings on this toric CY threefold. More precisely, the fermionic spectral traces $Z(N_1, N_2; \hbar)$, which 
are perfectly well-defined, can be expanded in a 't Hooft limit 
which reproduces the genus expansion of the topological 
string free energy, as we have verified in detail. 

It is also possible to write the expression (\ref{gtwo-matrix}) 
in the form of a two-cut matrix model. To do this, one has to use the explicit expression for the kernels (\ref{ex-k}) as well as 
the Cauchy identity, similar to what was done in \cite{mz,kmz} for genus one mirror curves. A straightforward calculation shows that the matrix model calculating the 
spectral trace $Z(N_1, N_2; \hbar)$ is given by 
\be
 \ba  Z(N_1,N_2; \hbar)&={1\over N_1! N_2!} \int{ {\rm d} ^N u \over (2 \pi)^N} \prod_{i=1}^{N_1} \left| \Psi_{{3 \mb\over 10},{\mb \over10}}\left( {\mb u_i \over 2 \pi} \right) \right|^2
 \prod_{j=1+N_1}^{N_2}\re^{-{\mb^2 u_j \over 5}}\left| \Psi_{{2\mb\over 10},{\mb \over10}}\left( {\mb u_j \over 2 \pi} \right) \right|^2\\
 & \times {\prod_{i<j}2\sinh\left( {u_i-u_j \over 2}+\ri \pi \Delta_{i,j} \right)2\sinh\left({u_i-u_j \over 2}-\ri \pi \Delta_{i,j} \right)\over \prod_{i,j}2\cosh\left({u_i-u_j \over 2} + \pi \ri c_{i,j} \right)}, 
\ea
\ee
where 
\be  
\Delta_{i,j} = \begin{cases}
0  \qquad \text{if $ i,j \leq N_1$ or $i,j > N_1$}, 
\\
 {1/10} \qquad \text{if $i\leq N_1$ and $ j> N_1$, }\\ 
 -{1/10} \qquad \text{otherwise,} \end{cases}
 \ee
 and
\be  c_{i,j} =  \begin{cases}
{3 / 10}   \qquad \text{if $ i,j \leq N_1$, }   
\\
 {1/ 10}  \qquad \text{if $ i,j \geq N_1 $, }  \\ 
 {2/10} \qquad \text{otherwise} 
 \end{cases} 
 \ee

In principle, 
our conjecture provides such a matrix-model-like description of the topological string for {\it all} toric CY threefolds, and 
it would be very interesting to test it in more higher genus examples.

\sectiono{Conclusions and future prospects}

In this paper we have extended the correspondence of \cite{ghm} to mirror curves of higher genus. 
This generalization requires 
many new ingredients: on the spectral theory side, we need 
a generalized spectral determinant which gives an entire function on the moduli space. This leads to a single quantization condition, 
in contrast to what happens in many quantum integrable systems. We have seen 
that this quantization condition captures in detail the spectrum of the operators appearing in the quantization of the curve. In addition, the 
fermionic spectral traces, which are obtained by expanding the generalized spectral determinant, provide a non-perturbative 
definition of the all-genus topological string in a certain conifold frame. 
All these considerations have been analyzed in detail in the example of the resolved $\IC^3/\IZ_5$ 
orbifold. 

The results presented in this paper open different avenues for future research. The general theory presented here grew out 
of a detailed analysis of the resolved $\IC^3/\IZ_5$ orbifold, and it would be very important to consider other higher genus examples in order to test it more carefully. 
It would be also important to extend our checks (which were mostly done in the maximally supersymmetric case) to arbitrary values of $\hbar$. 
We should note however that this seems to require a deeper understanding of the all-genus topological string away from the 
large radius point. Already in the genus one case, it was noted in \cite{ghm} that for example the expansion of the spectral determinant near orbifold points is 
only feasible in the maximally supersymmetric case, since we do not have systematic resummations of the topological string amplitudes at those points. In the case of higher genus 
curves, there are even more limitations of this type. For example, the operators $\rho_{3,1}$ and $\rho_{2,2}$ correspond 
to half-orbifold points of the geometry, and it seems difficult to write down an explicit quantization condition for these operators in terms of half-orbifold quantities for general $\hbar$. 
Clearly, more work is needed along this direction. 

Another related question is the following. In the case of genus one curves and for general $\hbar$, it has been shown in \cite{wzh} that 
the condition for the vanishing of the spectral determinant (i.e. the quantization condition) 
can be written in a closed form, in terms of the NS free energy (\ref{NS-j}). It would be interesting to see if a similar simple form can be found 
in the higher genus case. This might be however more difficult than for mirror curves of genus one. Recall that, in the maximally supersymmetric case, 
the quantization condition involves  the vanishing of the usual Riemann theta function. When $g_\Sigma \ge 2$, however, 
the theta divisor has a more complicated parametrization than in genus one, involving in particular the Abel map, and it is not clear that one can write a simple quantization condition 
even when $\hbar=2\pi$. 

As we have emphasized, the quantization scheme for mirror curves of higher genus that we are proposing in this paper is 
different from the more conventional procedure based on 
an underlying quantum integrable system. On the other hand, a construction by Goncharov and Kenyon associates an 
integrable system to any toric CY manifold \cite{gk} (see also \cite{fm,franco}). The quantization of this system 
leads to $g_\Sigma$ quantization conditions for the 
moduli of the curve. It would be very interesting to understand the precise relation between the quantization of the Goncharov--Kenyon system 
and the quantization procedure developed here. 

The results obtained in this paper might have implications for the study of non-perturbative aspects of Chern--Simons--matter theories. Some of the 
models studied in \cite{mori,hhm} in the Fermi gas approach involve operators which are obtained from the quantization of higher genus curves. 
The methods and ideas developed in this paper should be 
useful in their study. 

Of course, there remain deep conceptual questions concerning the origin of the correspondence between 
spectral theory and topological strings. From a mathematical point of view, it would be important 
to develop a version of the complex WKB method which makes it possible to understand the structure of non-perturbative corrections postulated in the 
conjecture of \cite{ghm} and the extension 
studied here. From a physical point of view, it would be important to know whether there is a full-fledged field theory behind the operators obtained by quantization. 
As pointed out in \cite{ghm}, the behavior of the 
fermionic spectral traces at large $N$ suggests that it could be a theory of M2 branes. Finding and describing in 
detail such a theory would lead to a much deeper understanding of topological string theory in the toric case.

\section*{Acknowledgements}
First of all, we would like to thank Jie Gu, Albrecht Klemm and Jonas Reuter for initial collaboration in this project and for sharing with us some of their 
results on the resolved $\IC^3/\IZ_5$ orbifold. We are grateful to Andrea Brini, Charles Doran, Sebasti\'an Franco, 
Krzysztof Gawedzki, Gian Michele Graf, Yasuyuki Hatsuda, Rinat Kashaev, Matt Kerr and Szabolcs Zakany for useful discussions and correspondence. 
This work is supported in part by the Fonds National Suisse, 
subsidies 200021-156995 and 200020-141329, and by the NCCR 51NF40-141869 ``The Mathematics of Physics" (SwissMAP).

\appendix 

\sectiono{Special geometry of the resolved $\IC^3/\IZ_5$ orbifold}

In this Appendix we collect necessary information on the resolved $\IC^3/\IZ_5$ orbifold, in particular its periods. Many of these results have appeared 
before, in \cite{mr,xenia, kpsw}. 

\subsection{Periods at large radius}
\label{ap-lr}
The moduli space of complex structures of this CY is parametrized by the complex variables $z_1, z_2$ introduced in 
(\ref{z1-z2}). The periods of this geometry are solutions to the Picard--Fuchs equations determined by the following  operators \cite{kpsw}, 
		\begin{align}
			\label{PFsystem}
			\begin{split}
				\mathcal{L}_1=&-2 \Theta _{2,1}+\Theta _{3,0}+z_1 \left(-2 \Theta _{0,1}+3 \Theta _{0,2}-\Theta _{0,3}+6 \Theta _{1,0}-\right.
				\\
				&\left.-18 \Theta
				_{1,1}+9 \Theta _{1,2}+27 \Theta _{2,0}-27 \Theta _{2,1}+27 \Theta _{3,0}\right),
			\end{split}
			\\
			\mathcal{L}_2=\;&\Theta _{0,2}-3 \Theta _{1,1}+z_2 \left(-2 \Theta _{0,1}-4 \Theta _{0,2}+\Theta _{1,0}+4 \Theta _{1,1}-\Theta
			_{2,0}\right),
			\\
			\mathcal{L}_3=\;&\Theta _{2,1}+z_1 z_2 \left(-2 \Theta _{0,2}+2 \Theta _{0,3}+7 \Theta _{1,1}-13 \Theta _{1,2}-3 \Theta _{2,0}+24
			\Theta _{2,1}-9 \Theta _{3,0}\right),
		\end{align}
	%
where $\Theta _{i,j}$ stands for the logarithmic derivative of order $i$ w.r.t. $z_1$ and of order $j$ w.r.t. $z_2$.
The standard way to solve these equations in the large radius point (see for example \cite{hkt}) is to consider the fundamental period $\varpi_0(\rho_1,\rho_2)$, given by
\be
\label{fplr} \varpi_0(\rho_1,\rho_2)=\sum_{\ell,n \geq 0}\frac{\Gamma (\rho_1+1)^2 
\Gamma (\rho_2+1) \Gamma (\rho_1-2 \rho_2+1) \Gamma (-3 \rho_1+\rho_2+1)z_1^{\ell+\rho_1}z_2^{k+\rho_2} }
{\Gamma (\ell+\rho_1+1)^2 \Gamma (k+\rho_2+1) \Gamma (\ell-2 k+\rho_1-2 \rho_2+1) \Gamma (-3 \ell+k-3 \rho_1+\rho_2+1)}, 
\ee
and take derivatives of this quantity w.r.t. $\rho_i$, $i=1,2$. We will then define, 
\be
\label{genlr-pers}
\ba & {\Pi}_{A_i}={\partial \varpi_0 (\rho_1,\rho_2) \over \partial \rho_i}\big|_{\rho_1=\rho_2=0}, \qquad i=1,2, \\
& {\Pi}_{B_1}=\left(2\partial_{\rho_1}^2+2\partial_{\rho_1} \partial_{\rho_2}+3 \partial_{\rho_2}^2\right)\varpi_0 (\rho_1,\rho_2)\big|_{\rho_1=\rho_2=0},\\
& {\Pi}_{B_2}=\left(\partial_{\rho_1}^2+6 \partial_{\rho_1} \partial_{\rho_2}+9 \partial_{\rho_2}^2\right)\varpi_0 (\rho_1,\rho_2)\big|_{\rho_1=\rho_2=0}.
\ea\ee 
We will also denote
\be
\label{oij}
  \varpi_{ij} = {\partial^2  \varpi_0\over \partial \rho_i \partial \rho_j}\bigg|_{\rho_1=\rho_2=0}. 
  \ee
One has, explicitly, 
\be
\ba
\Pi_{A_1}&= \log(z_1)-6 z_1 -z_2 +45 z_1^2 -{3z_2^2 \over 2}+\cdots,\\
\Pi_{A_2}&=\log(z_2)+2z_1 + 2 z_2-15 z_1^2 +3 z_2^2 +\cdots, \\
\Pi_{B_1}&= 2 \log^2(z_1) +2 \log(z_1) \log(z_2) + 3 \log^2(z_2) +  \cdots, \\
\Pi_{B_2}&= \log^2(z_1) + 6 \log(z_1) \log(z_2) + 9 \log^2(z_2) + \cdots .
\ea
\ee
In terms of the variables (\ref{QQ}), we have
\be
\ba 
z_1&=Q_1+ 6 Q_1^2 + Q_1 Q_2 + 9 Q_1^3 + 10 Q_1^2 Q_2 + 56 Q_1^4 + 26 Q_1^3 Q_2 + 4 Q_1^2 Q_2^2+ \cdots, \\
z_2&=Q_2- 2 Q_1 Q_2- 2Q_2^2 + 6 Q_1 Q_2^2 + 5 Q_1^2 Q_2 -3 Q_2^3 -32 Q_1^3 Q_2 - 10 Q_1 Q_2^3 - 4 Q_2^4 +\cdots
\ea
\ee
There are two combinations of the A-periods which play an important r\^ole, since they can be regarded as the flat coordinates corresponding to the moduli $x_3$, $x_0$. 
They are given by, 
\be
\label{piuv}
\Pi_u= \Pi_{A_1} + 3 \Pi_{A_2}, \qquad \Pi_v=2  \Pi_{A_1} +  \Pi_{A_2}. 
\ee
As already noted in \cite{mr}, their expansions can be written in closed form:
\be
\label{Apers}
\ba
\Pi_{u}&= \log u + 5 \sum_{(m,r)'}^\infty  {\Gamma(5m+2r) \over \Gamma(1+r) \Gamma(1+3m+r) \Gamma(1+m)^2} (-u)^m z_2^r, \\
\Pi_{v}&=\log v + 5 \sum_{(n,r)'}^\infty  {\Gamma(5n + 3r) \over \Gamma(1+r) \Gamma(1+2n+r)^2 \Gamma(1+n)} (-v)^n (-z_1)^r.
\ea
\ee
Here, $(m,r)'$ and $(n,r)'$ means that the corresponding pairs run over 
non-negative pairs of integers, except $(0,0)$. In the same way, one finds the explicit 
expression 
\be
\ba
\Pi_{B_2}
&=-\log^2(u)+2 \Pi_{A_1} \log(u)\\
& +10 \sum_{m,r} {5 \psi(5m+2r)-2 \psi(1+m) -3 \psi(1+3m+r) \over\Gamma(1+r) \Gamma(1+m)^2\Gamma(1+3m+r)} \Gamma(5m+2r) (-u)^m z_2^r. 
\ea
\ee

\subsection{Periods at the (half)-orbifold points}
\label{ap-orbis}
We also need the analytic continuation of these periods to the other significant points in moduli space. 
Let us first consider the half-orbifold point (\ref{fop}). Near this point, $z_1$ is large but $z_2$ is small. To perform the analytic continuation, it 
is is convenient to write the fundamental period in the Mellin--Barnes form. One has 
\be\label{fp} \ba \varpi_0(\rho_1,\rho_2)=&
 \sum_{k\geq 0}\Gamma (\rho_1+1)^2 \Gamma (\rho_2+1)(-1)^{k+1} \sin (\pi  (3 \rho_1-\rho_2)) \Gamma (\rho_1-2 \rho_2+1) \Gamma (-3 \rho_1+\rho_2+1) \\
&\times \int_{\mathcal{C}}\rd t\frac{ \Gamma (-t) \Gamma (t+1) z_2^{k+\rho_2}z_1^{\rho_1+t} \Gamma (-k+3 t+3 \rho_1-\rho_2)}{\pi  \Gamma (k+\rho_2+1) \Gamma (t+\rho_1+1)^2 \Gamma (-2 k+t+\rho_1-2 \rho_2+1)}.\ea\ee
Here, $\mathcal{C}$ is a contour running parallel to the imaginary axis. By closing the contour on the r.h.s. and picking up the residues at
\be t=\ell, \quad \ell \geq 0\ee
we obtain \eqref{fplr}. But we can close the contour on the l.h.s. and pick up the residues at
\be t=\frac{1}{3} (k-n-3 \rho_1+\rho_2),\quad  n\geq 0,  \ee
and we have
\be \label{fpor} \ba \varpi_0 (\rho_1,\rho_2)& =\sin (3 \pi  \rho_1-\pi  \rho_2) \sum_{k ,n \geq 0}(-1)^{k+n+1}x_0^nX^{\frac{1}{3} (5 k+n+5 \rho_2)} \csc \left(\frac{1}{3} \pi  (k-n-3 \rho_1+\rho_2)\right) \\
& \times \frac{ \Gamma (\rho_1+1)^2 \Gamma (\rho_2+1) \Gamma (\rho_1-2 \rho_2+1) \Gamma (-3 \rho_1+\rho_2+1) }{3 n! \Gamma (k+\rho_2+1) \Gamma \left(\frac{1}{3} (-5 k-n-5 \rho_2+3)\right) \Gamma \left(\frac{1}{3} (k-n+\rho_2+3)\right)^2}, \ea\ee
where we introduced the variable
\be
X={1\over x_3} 
\ee
and we used
\be z_2= x_0 X^2, \quad z_1={1 \over X x_0^3}.\ee
The analytic continuation of the large radius periods is simply obtained by taking the derivatives of  $\varpi_0(\rho_1,\rho_2)$ in the form (\ref{fpor}) 
as in (\ref{genlr-pers}). For the A-periods, it is 
easier to consider the combinations $\Pi_u$, $\Pi_v$ in (\ref{piuv}). Note that the analytic continuation of the period $\Pi_u$ is straightforward, and we simply obtain 
\be
\Pi_u(x_0,X)=5\log X+5\sum_{(m,r)^{'}} { \Gamma (5 m+2 r)   \over \Gamma (m+1)^2  \Gamma (r+1)  \Gamma (3 m+r+1)} (-1)^m X^{5m+2r} x_0^r. 
\ee
The analytic continuation of $\Pi_v$ gives, 
\be
 \Pi_v(x_0,X)={5\over 3}\log X+{5\over 3}\sum_{(p,t)^{'}}  \frac{(-1)^t  \Gamma \left(\frac{1}{3} (p+5 t)\right)}{\Gamma (p+1) \Gamma (t+1) \Gamma \left(\frac{t-p}{3}+1\right)^2}(-x_0)^p X^{\frac{1}{3} (p+5 t)}. 
 \ee
 We have, 
\be\label{Ahor}\ba \Pi_{A_1}(x_0,X)&={3\over 5}\Pi_v(x_0,X)-{1\over 5}\Pi_u(x_0,X),\\
    \Pi_{A_2}(x_0,X)&={2\over 5}\Pi_u(x_0,X)-{1\over 5}\Pi_v(x_0,X).
   \ea\ee
 It is also useful to consider the following periods, 
 \be \ba \pi_{13}(x_0,X)&=-\frac{\Gamma \left(\frac{2}{3}\right)^2}{\Gamma \left(\frac{1}{3}\right)}\sum_{(p,t)^{''}} \frac{(-1)^t (-x_0)^p X^{\frac{1}{3} (p+5 t)} 
 \Gamma \left(\frac{1}{3} (p+5 t)\right)}{\Gamma (p+1) \Gamma (t+1) \Gamma \left(\frac{t-p}{3}+1\right)^2}, \\
  \pi_{23}(x_0,X)&=\frac{2 \Gamma \left(\frac{1}{3}\right)^2}{\Gamma \left(\frac{2}{3}\right)} \sum_{(p,t)^{'''}} \frac{(-1)^t (-x_0)^p X^{\frac{1}{3} (p+5 t)}
   \Gamma \left(\frac{1}{3} (p+5 t)\right)}{\Gamma (p+1) \Gamma (t+1) \Gamma \left(\frac{t-p}{3}+1\right)^2},
 \ea\ee
  where  the sum
   \be \sum_{(p,t)^{''}} 
   \ee 
   runs over all non-negative integers $p,t$ such that $(p,t)\neq (0,0)$ and $\frac{1}{3} (p+5 t)-\frac{1}{3}\in \mathbb{N}$, while
  \be \sum_{(p,t)^{'''}} \ee
   runs over all non-negative integers $p,t$ such that $(p,t)\neq (0,0)$ and $\frac{1}{3} (p+5 t)-\frac{2}{3}\in \mathbb{N}$.  For the B-periods, one similarly finds that $\Pi_{B_2}$ has a straightforward 
   analytic continuation, while for $\Pi_{B_1}$ we find
   \be
   \label{Bhor}
  \Pi_{B_1}(x_0,X)=\frac{10 \pi \Gamma \left(\frac{1}{3}\right)}{3 \sqrt{3} \Gamma \left(\frac{2}{3}\right)^2} \pi_{13}(x_0,X)+\frac{5 \pi \Gamma \left(\frac{2}{3}\right)}{3 \sqrt{3} \Gamma \left(\frac{1}{3}\right)^2} \pi_{23}(x_0,X)+{1\over 3} \Pi_{B_2}(x_0,X)-\frac{10 \pi ^2}{9}, 
  \ee
see \cite{coates} for a similar derivation. 

A similar calculation can be done for the other half-orbifold point, at $x_3=0$. Let us denote 
\be
Y={1\over x_0}.
\ee
It is convenient to consider the large radius period, but after changing the sign of $z_2$. 
For this half-orbifold point, the particular combination of A-periods which has an easy analytic continuation is $\Pi_v$, and it reads, 
\be
\Pi_v=5 \log Y + 5 \sum_{(n,r)'}^\infty  {\Gamma(5n+3r) \over \Gamma(1+r) \Gamma(1+2n+r)^2 \Gamma(1+n)} Y^{5n+3r} (-x_3)^r. 
\ee
The rest of the periods at large radius can be written as linear combinations of the following series, 
\be
\ba
\pi_{1/2}&=-Y^{1/2} \sum_{l,s\ge 0} {\Gamma\left( {1 \over 2} +2l+s \right) \over \Gamma (2+2s-l) \Gamma(1+l)^2 \Gamma\left( {1\over 2} +l-s\right) }Y^{2l+s}(-x_3)^{2s+1-l}  \\
\pi_{3/2}&=-Y^{3/2}\sum_{s \ge 0} \sum_{l=0}^{2s+1}{3  \Gamma\left( {1 \over 2} +2l+s \right) \over \Gamma (2+2s-l) \Gamma(1+l)^2 \Gamma\left( {1\over 2} +l-s\right) } \\
&\qquad  \times 
\left(4 \psi(1+l)+ \psi(1/2+l-s)- 5 \psi(1/2-2l-s) -4  \log(4) \right)   Y^{2l+s-1} (-x_3)^{2s+1-l}, \\
\pi_1&=-\sum_{s\ge 1} \sum_{l =0}^{s-1} {2 (s-1-l)!  (2l+s-1)! \over (2s-l)! \ell!^2 }  (-1)^{s+l} Y^{2l+s}(-x_3)^{2s-l}, \\
\widetilde \Pi_B&=5 \sum_{(m,r)'}  {\Gamma(5m+3r) \over \Gamma(1+r) \Gamma(1+2m+r)^2 \Gamma(1+m)} \\
& \qquad \times  \left(4 \psi(1+2m+r)+ \psi(1+m)- 5 \psi(5m+3r) \right) Y^{5m+3r} (-x_3)^r . 
\ea
\ee
One finds, 
\be
\ba
\varpi_{11}&= -{25 \over 4} \log^2(Y)+ {5\over 2} \log(Y) \left( \pi_{1/2}+ \Pi_v\right) -2\log(4) \pi_{1/2} -{1\over 6} \pi_{3/2}-{1\over 2}\widetilde \Pi_B +{5\over 4} \pi_1+{\pi^2\over 3}, \\
\varpi_{12}&=- {5\over 2} \log(Y)  \pi_{1/2} +\log(16) \pi_{1/2}+{1\over 6} \pi_{3/2}-{\pi^2\over 3}, \\
\varpi_{22}&=0, 
\ea
\ee
and we finally obtain the following analytic continuation formulae, 
\be \label{ABhor2}
\ba
\Pi_{A_1}(Y,x_3)&={1\over 2}\left( \Pi_v +\pi_{12}\right), \\
\Pi_{A_2}(Y,x_3)&= -\pi_{12},\\
\Pi_{B_1}(Y,x_3)&= 2\varpi_{11}+2\varpi_{12}, \\
\Pi_{B_2}(Y,x_3)&= \varpi_{11}+6\varpi_{12} .
\ea
\ee

Let us finally consider the analytic continuation near the full orbifold point, $x_3=x_0=0$. After some computations, one finds that the fundamental period becomes
  \be 
  \label{fp-orb}\ba \varpi_0(\rho_1,\rho_2)=&\sum_{k,n}\frac{(-1)^{k+n} x_3^{k}x_0^n \Gamma (\rho_1+1) \Gamma (\rho_2+1)^2 \Gamma (\rho_1-3 \rho_2+1) \Gamma (-2 \rho_1+\rho_2+1)}{15 k! n! \Gamma \left(\frac{1}{5} (-3 k-n+5)\right) \Gamma \left(-\frac{k}{5}-\frac{2 n}{5}+1\right)^2}
  \\&\times \sin (\pi  (\rho_1-3 \rho_2))g(\rho_1,\rho_2,k,n),\ea\ee
  where
\be\ba &  g(\rho_1,\rho_2,k,n)= -\sin \left(\frac{1}{3} \pi  (n+5 \rho_1)\right) \csc \left(\frac{1}{15} \pi  (3 k+n+5 \rho_1)\right) \csc \left(\frac{1}{3} \pi  (n-\rho_1+3 \rho_2)\right)\\
&-\sin \left(\frac{1}{3} \pi  (n+5 \rho_1-1)\right) \sec \left(\frac{1}{30} \pi  (6 k+2 n+10 \rho_1-5)\right) \csc \left(\frac{1}{3} \pi  (-n+\rho_1-3 \rho_2+1)\right)
\\ &-\sin \left(\frac{1}{3} \pi  (n+5 \rho_1+1)\right) \sec \left(\frac{1}{30} \pi  (6 k+2 n+10 \rho_1+5)\right) \csc \left(\frac{1}{3} \pi  (n-\rho_1+3 \rho_2+1)\right).\ea\ee
Then one can check that
\be \label{Aorbi} \ba \Pi_{A_1}(x_0,x_3)&=-\sum_{m,r}\frac{\pi  (-1)^{m+r} x_0^m x_3^{r} \left((-1)^m \csc \left(\frac{1}{5} \pi  (2 m+r)\right)-3 \csc \left(\frac{1}{5} \pi  (m+3 r)\right)\right)}{5 m! r! \Gamma \left(-\frac{m}{5}-\frac{3 r}{5}+1\right) \Gamma \left(-\frac{2 m}{5}-\frac{r}{5}+1\right)^2} ,\\
 \Pi_{A_2}(x_0,x_3)&=\sum_{m,r} \frac{\pi  (-1)^{m+r} x_0^m x_3^{r} \left(2 (-1)^m \csc \left(\frac{1}{5} \pi  (2 m+r)\right)-\csc \left(\frac{1}{5} \pi  (m+3 r)\right)\right)}{5 m! r! \Gamma \left(-\frac{m}{5}-\frac{3 r}{5}+1\right) \Gamma \left(-\frac{2 m}{5}-\frac{r}{5}+1\right)^2}.
\ea\ee
This result  can also be obtained with the results of \cite{mr}. The B-periods can then be obtained by computing derivatives of (\ref{fp-orb}). One finds the closed form expression, 
  \be \ba \Pi_{B_1}(x_0,x_3)&=-\frac{8 \pi ^2}{3}+\sum_{(m,r)'}  \frac{ x_3^{m} x_0^r \Gamma \left(\frac{1}{5} (m+2 r)\right)^2}{m! \Gamma (r+1) \Gamma \left(-\frac{3 m}{5}-\frac{r}{5}+1\right)}, \\
  \Pi_{B_2}(x_0,x_3)&=-2\pi^2+{1\over 2}\sum_{(m,r)'}\frac{\sec \left(\frac{1}{5} \pi  (4 r-3 m)\right) (-x_3)^m x_0^r \Gamma \left(\frac{1}{5} (m+2 r)\right)^2}{m! \Gamma (r+1) \Gamma \left(-\frac{3 m}{5}-\frac{r}{5}+1\right)}.
 \ea \ee

\subsection{Periods at the maximal conifold point}

\label{mcp-periods}
The maximal conifold point is defined by (\ref{mcp}). Near this point, we have the vanishing coordinates $\rho_i$, $i=1,2$, defined by 
\begin{equation}
		\label{zetaVars}
		z_1=-\frac{1}{25}+\rho_1, \qquad z_2=\frac{1}{5}+\rho_2.
	\end{equation}
We expect to have two flat coordinates $t_1^c$, $t_2^c$ vanishing at the maximal conifold point, and two periods vanishing like $t_i^c \log t_i^c$. In addition, 
near the maximal conifold point, we expect the discriminant (\ref{discri}) to vanish like 
\be
\Delta\sim t_1^c t_2^c. 
\ee
One can indeed find two solutions of the Picard--Fuchs equations with the required behavior. In terms of these flat coordinates, the local coordinates $\rho_i$, $i=1,2$ are given by the 
expansions, 
	\begin{align}
		\label{mirrorMap}
		\begin{split}
			\rho_1 = &\;\frac{\left(25-11 \sqrt{5}\right) t_2^c+\left(25+11 \sqrt{5}\right) t_1^c}{250}
			\\
			&+\frac{\left(337 \sqrt{5}-755\right) (t_2^c)^2-\left(755+337 \sqrt{5}\right) (t_1^c)^2}{2500}+\mathcal{O}\left(t^c\right)^3,
		\end{split}
		\\
		\begin{split}
			\rho_2 = &\;\frac{t_1^c-t_2^c}{5 \sqrt{5}}+\frac{\left(7 \sqrt{5}-10\right) (t_2^c)^2+10 t_1^c t_2^c-\left(10+7 \sqrt{5}\right) (t_1^c)^2}{250}+\mathcal{O}\left(t^c\right)^3.
		\end{split}
	\end{align}
Similarly, one finds the following solutions to the Picard--Fuchs equations, 
\be
\label{S-pers}
\ba
S_1 &= t^c_1 \log \left(t^c_1\right) -\frac{1}{100} \left(10+\sqrt{5}\right)( t^c_1)^2+\frac{2}{25} \left(5-2 \sqrt{5}\right) t^c_1
			t^c_2+ \frac{1}{25} \left(5-2 \sqrt{5}\right) (t^c_2)^2+\cdots, \\
S_2 &= t^c_2 \log \left(t^c_2\right) +\frac{1}{25} \left(5+2 \sqrt{5}\right) (t^c_1)^2+\frac{2}{25} \left(5+2 \sqrt{5}\right) t^c_1
			t^c_2+ \frac{1}{100} \left(-10+\sqrt{5}\right)( t^c_2)^2+\cdots
						\ea
			\ee

\subsection{Quantum mirror map}
\label{res-or-qmm}
As in \cite{acdkv,hkrs}, the quantum mirror map of the $\IC^3/\IZ_5$ geometry can be computed directly by quantizing the mirror curve with 
Weyl's prescription, and solving the resulting difference equation. Let us consider the second function in (\ref{o12-ex}). After appropriate rescalings of $x$, $y$, we find 
that the equation $\CO_2(x,y)+x_3=0$ can be written as 
\be
\re^x + \re^y +z_1 z_2^3 \re^{-3x -y} + z_2  \re^{-x}+1=0.
\ee
Let us introduce the function
\be
V(x)={\psi(x- \ri \hbar) \over \psi (x)}, 
\ee
where $\psi(x)$ is a wavefunction in the $x$-representation. 
Therefore, the equation 
\be
\left(\re^\mx + \re^\my +z_1 z_2^3 \re^{-3\mx -\my} + z_2  \re^{-\mx}+1\right) |\psi \rangle=0
\ee
becomes
\be
X+ z_2 X^{-1} +1 + V(X)+ {z_1 z_2^3 q^{-3} X^{-3} \over V(q^2 X)}=0,
\ee
where
\be
q=\re^{\ri \hbar/2}, \qquad X=\re^x. 
\ee
We can now solve systematically for $V(X)$ as a power series in $z_1$, $z_2$. The quantum A-period is then given by 
\be
\Pi_u(z_1, z_2; \hbar)=\log u +\widetilde \Pi_u(z_1, z_2; \hbar), 
\ee
where
\be
\widetilde \Pi_u(z_1, z_2;\hbar)=-5 \, {\rm Res}_{X=0}\left[{1\over X} \log \left(V(X)\right) \right]. 
\ee
We find, at the very first orders, 
\be
\ba
\widetilde \Pi_u(z_1, z_2;\hbar)&= 5 z_2 + {15  z_2^2 \over 2} + {50 z_2^3\over 3}-\frac{5 z_2^3 \left(4 \left(q^6+q^4+q^2+1\right) z_1-35 q^3 z_2\right)}{4 q^3}\\
&-\frac{ z_2^4
   \left(5 \left(q^{10}+7 q^8+7 q^6+7 q^4+7 q^2+1\right) z_1-126 q^5 z_2\right)}{ q^5}+\CO(z_i^6). 
   \ea
   \ee
It is easy to check that, when $\hbar\rightarrow0$, we recover the classical $\Pi_u$ period in (\ref{Apers}).

We can compute another, independent quantum period by using the representation (\ref{pp2-curve}) of the geometry. After appropriate rescalings, we can write it as
\be
\re^x+\re^y + z_1 \re^{-x-y} + z_2 \re^{2x}+1=0, 
\ee
and following the same procedure we used above, we obtain the equation
\be
X+ z_2 X^{2} +1 + V(X)+ {z_1  q^{-1} X^{-1} \over V(q^2 X)}=0. 
\ee
Solving this, we can obtain the quantum A-period corresponding to $\Pi_{A_2}$. Namely, we find
\be
\Pi_{A_2}(z_1, z_2; \hbar)=\log u +\widetilde \Pi_{A_2}(z_1, z_2; \hbar), 
\ee
\be
\ba
\widetilde \Pi_{A_2} (z_1, z_2;\hbar)&= {\rm Res}_{X=0} \left[ {1\over X} \left( \log(V(X))-2  \log(V(X^{-1}))\right) \right]\\
&= (q+q^{-1}) z_1+2 z_2+ \frac{6 q^4z_2^2+\left(-2 q^8-7 q^6-12 q^4-7 q^2-2\right) z_1^2}{2 q^4}+\CO(z_i^3).
   \ea
   \ee


\begin{thebibliography}{99}
\bibliographystyle{plain}


   
\bibitem{ghm}
 A.~Grassi, Y.~Hatsuda and M.~Mari\~no, ``Topological Strings from Quantum Mechanics,''
  arXiv:1410.3382 [hep-th].

\bibitem{adkmv}
   M.~Aganagic, R.~Dijkgraaf, A.~Klemm, M.~Mari\~no and C.~Vafa, ``Topological strings and integrable hierarchies,''
  Commun.\ Math.\ Phys.\  {\bf 261}, 451 (2006)
  [hep-th/0312085].
    
\bibitem{acdkv}
M.~Aganagic, M.~C.~N.~Cheng, R.~Dijkgraaf, D.~Krefl and C.~Vafa, ``Quantum Geometry of Refined Topological Strings,''
  JHEP {\bf 1211}, 019 (2012)
  [arXiv:1105.0630 [hep-th]].
  
       \bibitem{ns}
   N.~A.~Nekrasov and S.~L.~Shatashvili, ``Quantization of Integrable Systems and Four Dimensional Gauge Theories,''
  arXiv:0908.4052 [hep-th].
    \bibitem{dmp}
   N.~Drukker, M.~Mari\~no and P.~Putrov, ``From weak to strong coupling in ABJM theory,''
  Commun.\ Math.\ Phys.\  {\bf 306}, 511 (2011)
  [arXiv:1007.3837 [hep-th]].
  \bibitem{mp}
   M.~Mari\~no and P.~Putrov, ``ABJM theory as a Fermi gas,''
  J.\ Stat.\ Mech.\  {\bf 1203}, P03001 (2012)
  [arXiv:1110.4066 [hep-th]].
  
    \bibitem{hmo}
  Y.~Hatsuda, S.~Moriyama and K.~Okuyama, ``Instanton Effects in ABJM Theory from Fermi Gas Approach,''
  JHEP {\bf 1301}, 158 (2013)
  [arXiv:1211.1251 [hep-th]].
  
    \bibitem{hmo2}
  Y.~Hatsuda, S.~Moriyama and K.~Okuyama, ``Instanton Bound States in ABJM Theory,''
  JHEP {\bf 1305}, 054 (2013)
  [arXiv:1301.5184 [hep-th]].
  \bibitem{hmmo}
Y.~Hatsuda, M.~Mari\~no, S.~Moriyama and K.~Okuyama, ``Non-perturbative effects and the refined topological string,''
  JHEP {\bf 1409}, 168 (2014)
  [arXiv:1306.1734 [hep-th]].
    \bibitem{km}
 J.~Kallen and M.~Mari\~no, ``Instanton effects and quantum spectral curves,''
  arXiv:1308.6485 [hep-th].
   \bibitem{kas-mar}
  R.~Kashaev and M.~Mari\~no, ``Operators from mirror curves and the quantum dilogarithm,''
  arXiv:1501.01014 [hep-th].
  \bibitem{mz}
M.~Mari\~no and S.~Zakany, ``Matrix models from operators and topological strings,''
  arXiv:1502.02958 [hep-th].
  \bibitem{kmz}
  R.~Kashaev, M.~Mari\~no and S.~Zakany, 
  ``Matrix models from operators and topological strings, 2,''
  arXiv:1505.02243 [hep-th].
  
  \bibitem{gkmr}
  J. Gu, A. Klemm, M. Mari\~no and J. Reuter, 
  ``Exact solutions to quantum spectral curves by topological string theory,''
  JHEP {\bf 1510}, 025 (2015)
  [arXiv:1506.09176 [hep-th]].
  
   \bibitem{rv}
 F. Rodriguez Villegas, ``Modular Mahler measures, I", in {\it Topics in number theory}, Kluwer Acad. Publ., Dordrecht, 1999, p. 17.
 
  \bibitem{dk}
 C. Doran and M. Kerr, ``Algebraic K-theory of toric hypersurfaces,"  Commun. Number Theory Phys. {\bf 5}, 397 (2011) [arXiv:0809.4669 [math.AG]].

  \bibitem{kkv}  
   S.~H.~Katz, A.~Klemm and C.~Vafa, ``Geometric engineering of quantum field theories,''
  Nucl.\ Phys.\ B {\bf 497}, 173 (1997)
  [hep-th/9609239].
  
   \bibitem{ckyz}
  T.~M.~Chiang, A.~Klemm, S.~T.~Yau and E.~Zaslow, ``Local mirror symmetry: Calculations and interpretations,''
  Adv.\ Theor.\ Math.\ Phys.\  {\bf 3}, 495 (1999)
  [hep-th/9903053].
  
  \bibitem{hv}
  K.~Hori and C.~Vafa,
  ``Mirror symmetry,''
  arXiv:hep-th/0002222.
  
  
\bibitem{Bat} V.V. Batyrev, ``Dual polyhedra and mirror symmetry for Calabi--Yau hypersurfaces in toric varieties,''
 J. Alg. Geom. {\bf 3} (1994) 493  [arXiv:alg-geom/9310003].
 
\bibitem{witten-phases}
   E.~Witten, ``Phases of N=2 theories in two-dimensions,''
  Nucl.\ Phys.\ B {\bf 403}, 159 (1993)
  [hep-th/9301042].
  
 \bibitem{akv}
 M.~Aganagic, A.~Klemm and C.~Vafa, ``Disk instantons, mirror symmetry and the duality web,''
  Z.\ Naturforsch.\ A {\bf 57}, 1 (2002)
  [hep-th/0105045].
      
 
\bibitem{mmopen}
 M.~Mari\~no, ``Open string amplitudes and large order behavior in topological string theory,''
  JHEP {\bf 0803}, 060 (2008)
  [hep-th/0612127].

\bibitem{bkmp}
V.~Bouchard, A.~Klemm, M.~Mari\~no and S.~Pasquetti, ``Remodeling the B-model,''
  Commun.\ Math.\ Phys.\  {\bf 287}, 117 (2009)
  [arXiv:0709.1453 [hep-th]].

 \bibitem{eo-proof}
B.~Eynard and N.~Orantin, ``Computation of Open Gromov--Witten Invariants for 
Toric Calabi--Yau 3-Folds by Topological Recursion, a Proof of the BKMP Conjecture,''
  Commun.\ Math.\ Phys.\  {\bf 337}, no. 2, 483 (2015)
  [arXiv:1205.1103 [math-ph]].
  
     \bibitem{hkp}
 M.~X.~Huang, A.~Klemm and M.~Poretschkin, ``Refined stable pair invariants for E-, M- and $[p, q]$-strings,''
  JHEP {\bf 1311}, 112 (2013)
  [arXiv:1308.0619 [hep-th]].
  
   \bibitem{kpsw}
 A.~Klemm, M.~Poretschkin, T.~Schimannek and M.~Westerholt-Raum, ``Direct Integration for Mirror Curves of Genus Two and an Almost Meromorphic Siegel Modular Form,''
  arXiv:1502.00557 [hep-th].
  
  \bibitem{xenia}
 X.~De la Ossa, B.~Florea and H.~Skarke, ``D-branes on noncompact Calabi-Yau manifolds: K theory and monodromy,''
  Nucl.\ Phys.\ B {\bf 644}, 170 (2002)
  [hep-th/0104254].
  
\bibitem{mr}
 S.~Mukhopadhyay and K.~Ray, ``Fractional branes on a noncompact orbifold,''
  JHEP {\bf 0107}, 007 (2001)
  [hep-th/0102146].
  
 \bibitem{karp}
 R.~L.~Karp, ``On the $\IC^n/\IZ_m$ fractional branes,''
  J.\ Math.\ Phys.\  {\bf 50}, 022304 (2009)
  [hep-th/0602165].
  
  \bibitem{coates}
T. Coates, ``Wall-crossings in toric Gromov--Witten theory, II: local examples," arXiv:0804.2592 [math.AG]. 
 \bibitem{simon} 
 B. Simon, {\it Trace ideals and their applications}, second edition, American Mathematical Society, Providence, 2000. 
 
 \bibitem{simon-paper}
 B. Simon, ``Notes on infinite determinants of Hilbert space operators," Advances in Mathematics {\bf 24}, 244 (1977).
 
 \bibitem{gro}
 A. Grothendieck, ``La th\'eorie de Fredholm," Bulletin de la Soci\'et\'e Math\'ematique de France {\bf 84}, 319 (1956). 
     

 \bibitem{fredholm}
I. Fredholm, ``Sur une classe d'\'equations fonctionnelles," Acta Mathematica {\bf 27}, 365 (1903).


\bibitem{syz} 
M. Stessin, R. Yang  and K. Zhu, ``Analyticity of a joint spectrum and a multivariable analytic Fredhom theorem," New York J. Math {\bf 17} (2011), 39.

\bibitem{csz}
I. Chagouel, M. Stessin and K. Zhu, ``Geometric spectral theory for compact operators," arXiv:1309.4375.

  \bibitem{bbt}
  O. Babelon, D. Bernard and M. Talon, {\it An introduction to classical integrable systems}, Cambridge University Press, 2003. 
   
 \bibitem{gutz}
   M.~C.~Gutzwiller, ``The Quantum Mechanical Toda Lattice,''
  Annals Phys. {\bf 124}, 347 (1980);  ``The Quantum Mechanical Toda lattice, II," Annals Phys. {\bf 133}, 304 (1981).
  
  \bibitem{sklyanin}
   E.~K.~Sklyanin, ``The Quantum Toda Chain,''
  Lect.\ Notes Phys.\  {\bf 226}, 196 (1985).
  
\bibitem{gp}
 M.~Gaudin and V.~Pasquier, ``The periodic Toda chain and a matrix generalization of the Bessel function's recursion relations,''
  J.\ Phys.\ A {\bf 25}, 5243 (1992).

\bibitem{kl}
 S.~Kharchev and D.~Lebedev, ``Integral representation for the eigenfunctions of quantum periodic Toda chain,''
  Lett.\ Math.\ Phys.\  {\bf 50}, 53 (1999)
  [hep-th/9910265].
  
  \bibitem{an}
  D. An, ``Complete set of Eigenfunctions of the quantum Toda chain,"  Lett.\ Math.\ Phys.\  {\bf 87}, 209 (2009).
  
 \bibitem{mirmor}
   A.~Mironov and A.~Morozov, ``Nekrasov Functions and Exact Bohr-Zommerfeld Integrals,''
  JHEP {\bf 1004}, 040 (2010)
  [arXiv:0910.5670 [hep-th]].
  
  \bibitem{mirmor2}
   A.~Mironov and A.~Morozov, ``Nekrasov Functions from Exact BS Periods: The Case of SU(N),''
  J.\ Phys.\ A {\bf 43}, 195401 (2010)
  [arXiv:0911.2396 [hep-th]].
    
    \bibitem{kt}
   K.~K.~Kozlowski and J.~Teschner, ``TBA for the Toda chain,''
  arXiv:1006.2906 [math-ph].

  
\bibitem{matsuyama}
A. Matsuyama, ``Periodic Toda lattice in Quantum Mechanics," Annals Phys. {\bf 222}, 300 (1992). 

  
\bibitem{bpv}  
 R.~Balian, G.~Parisi and A.~Voros, ``Discrepancies from asymptotic series and their relation to complex classical trajectories," Phys. Rev. Lett. {\bf 41}, 1141 (1978); 
 ``Quartic Oscillator,'' in {\it Feynman Path Integrals}, Lecture Notes in Physics {\bf 106}, 337 (1979).

  \bibitem{huang}
 M.~x.~Huang, ``On Gauge Theory and Topological String in Nekrasov-Shatashvili Limit,''
  JHEP {\bf 1206}, 152 (2012)
  [arXiv:1205.3652 [hep-th]].
  
         \bibitem{hkrs}
  M.~x.~Huang, A.~Klemm, J.~Reuter and M.~Schiereck, ``Quantum geometry of del Pezzo surfaces in the Nekrasov-Shatashvili limit,''
  JHEP {\bf 1502}, 031 (2015)
  [arXiv:1401.4723 [hep-th]].
  
  \bibitem{bcov}
 M.~Bershadsky, S.~Cecotti, H.~Ooguri and C.~Vafa, ``Kodaira-Spencer theory of gravity and exact results for quantum string amplitudes,''
  Commun.\ Math.\ Phys.\  {\bf 165}, 311 (1994)
  [hep-th/9309140].
   
    \bibitem{gv}
 R.~Gopakumar and C.~Vafa, ``M theory and topological strings. 2,''
  hep-th/9812127.
        
\bibitem{ikv}
A.~Iqbal, C.~Kozcaz and C.~Vafa, ``The Refined topological vertex,''
  JHEP {\bf 0910}, 069 (2009)
  [hep-th/0701156].
  
\bibitem{ckk}  J.~Choi, S.~Katz and A.~Klemm, 
``The refined BPS index from stable pair invariants,''
  Commun.\ Math.\ Phys.\  {\bf 328}, 903 (2014)
  [arXiv:1210.4403 [hep-th]].
  
   \bibitem{no}
  N.~Nekrasov and A.~Okounkov, ``Membranes and Sheaves,''
  arXiv:1404.2323 [math.AG].
  
  

  \bibitem{hk}
M.~x.~Huang and A.~Klemm, ``Direct integration for general $\Omega$ backgrounds,''
  Adv.\ Theor.\ Math.\ Phys.\  {\bf 16}, no. 3, 805 (2012)
  [arXiv:1009.1126 [hep-th]].
  
   \bibitem{em}
 B.~Eynard and M.~Mari\~no, ``A holomorphic and background independent partition function for matrix models and topological strings,''
  J.\ Geom.\ Phys.\  {\bf 61}, 1181 (2011)
  [arXiv:0810.4273 [hep-th]].
  
 \bibitem{abk}
   M.~Aganagic, V.~Bouchard and A.~Klemm, ``Topological Strings and (Almost) Modular Forms,''
  Commun.\ Math.\ Phys.\  {\bf 277}, 771 (2008)
  [hep-th/0607100].
  
  \bibitem{hatsuda}  
  Y.~Hatsuda, ``Spectral zeta function and non-perturbative effects in ABJM Fermi-gas,''
  arXiv:1503.07883 [hep-th].
  
  \bibitem{fk}
L.~D.~Faddeev and R.~M.~Kashaev, ``Quantum Dilogarithm,''
  Mod.\ Phys.\ Lett.\ A {\bf 9}, 427 (1994)
  [hep-th/9310070].
  
 \bibitem{garou-kas}
  S.~Garoufalidis and R.~Kashaev, ``Evaluation of state integrals at rational points,''
  arXiv:1411.6062 [math.GT].
    
  \bibitem{faddeev}
 L.~D.~Faddeev, ``Discrete Heisenberg-Weyl group and modular group,''
  Lett.\ Math.\ Phys.\  {\bf 34}, 249 (1995)
  [hep-th/9504111].
  
 \bibitem{fhm}
  H.~Fuji, S.~Hirano and S.~Moriyama, ``Summing Up All Genus Free Energy of ABJM Matrix Model,''
  JHEP {\bf 1108}, 001 (2011)
  [arXiv:1106.4631 [hep-th]].
  
  \bibitem{ayz}
  M.~Alim, S.~T.~Yau and J.~Zhou, ``Airy Equation for the Topological String Partition Function in a Scaling Limit,''
  arXiv:1506.01375 [hep-th].
 

\bibitem{hw}
 M.~x.~Huang and X.~f.~Wang, ``Topological Strings and Quantum Spectral Problems,''
  JHEP {\bf 1409}, 150 (2014)
  [arXiv:1406.6178 [hep-th]].




  
  \bibitem{ak}
J.~Ellegaard Andersen and R.~Kashaev, ``A TQFT from Quantum Teichm\"uller Theory,''
  Commun.\ Math.\ Phys.\  {\bf 330}, 887 (2014)
  [arXiv:a [math.QA]].
 
 
  \bibitem{msw}
 M.~Mari\~no, R.~Schiappa and M.~Weiss, ``Multi-Instantons and Multi-Cuts,''
  J.\ Math.\ Phys.\  {\bf 50}, 052301 (2009)
  [arXiv:0809.2619 [hep-th]].


\bibitem{yang}
 K.~Mohri, Y.~Onjo and S.~K.~Yang, ``Closed submonodromy problems, local mirror symmetry and branes on orbifolds,''
  Rev.\ Math.\ Phys.\  {\bf 13}, 675 (2001)
  [hep-th/0009072].
  
\bibitem{gk} 
 A.~B.~Goncharov and R.~Kenyon, ``Dimers and cluster integrable systems,''
  arXiv:1107.5588 [math.AG].

\bibitem{fm} 
V.~V.~Fock and A.~Marshakov, ``Loop groups, Clusters, Dimers and Integrable systems,''
  arXiv:1401.1606 [math.AG].

\bibitem{franco}
  R.~Eager, S.~Franco and K.~Schaeffer, ``Dimer Models and Integrable Systems,''
  JHEP {\bf 1206}, 106 (2012)
  [arXiv:1107.1244 [hep-th]].
 
 \bibitem{mori}
  S.~Moriyama and T.~Nosaka, ``ABJM Membrane Instanton from Pole Cancellation Mechanism,''
  arXiv:1410.4918 [hep-th]; ``Exact Instanton Expansion of Superconformal Chern-Simons Theories from Topological Strings,''   JHEP {\bf 1505}, 022 (2015)
  [arXiv:1412.6243 [hep-th]].
 
 \bibitem{hhm}
  Y.~Hatsuda, M.~Honda and K.~Okuyama, ``Large N non-perturbative effects in $\mathcal{N}=4$ superconformal Chern-Simons theories,''
  arXiv:1505.07120 [hep-th].

\bibitem{wzh} 
X.~Wang, G.~Zhang and M.~x.~Huang, ``New Exact Quantization Condition for Toric Calabi-Yau Geometries,''
  Phys.\ Rev.\ Lett.\  {\bf 115}, 121601 (2015)
  [arXiv:1505.05360 [hep-th]].
   
  \bibitem{hkt}
 S.~Hosono, A.~Klemm and S.~Theisen, ``Lectures on mirror symmetry,''
  Lect.\ Notes Phys.\  {\bf 436}, 235 (1994)
  [hep-th/9403096].



 
\end{thebibliography}
\end{document}